\documentclass[a4paper,11pt]{article}
\pdfoutput=1
\usepackage{jheppub}
\usepackage{amsmath,amssymb,amsfonts,bm,amscd,slashed}
\usepackage{graphicx}
\usepackage{mathtools}
\usepackage{mathrsfs}

\newcommand{\bea}{\begin{eqnarray}}
\newcommand{\eea}{\end{eqnarray}}
\newcommand{\be}{\begin{equation}}
\newcommand{\ee}{\end{equation}}

\newcommand{\Z}{{\mathbb Z}}
\newcommand{\R}{{\mathbb R}}
\newcommand{\C}{{\mathbb C}}

\newcommand{\cZ}{{\mathcal{Z}}}
\newcommand{\cF}{{\mathcal{F}}}
\newcommand{\cN}{{\mathcal{N}}}

\newcommand{\pD}{{\mathscr D}}
\newcommand{\bB}{{\mathscr B}}
\newcommand{\dd}{{\rm d}}

\def\z{\zeta}

\def\tilde{\widetilde}
\def\hat{\widehat}
\def\bar{\overline}

\def\cA{{\mathcal A}}
\def\cB{{\mathcal B}}
\def\cC{{\mathcal C}}
\def\cD{{\mathcal D}}

\def\cF{{\mathcal F}}

\def\cH{{\mathcal H}}

\def\cJ{{\mathcal J}}

\def\cL{{\mathcal L}}
\def\cM{{\mathcal M}}
\def\cN{{\mathcal N}}
\def\cO{{\mathcal O}}
\def\cP{{\mathcal P}}
\def\cQ{{\mathcal Q}}
\def\cR{{\mathcal R}}
\def\cS{{\mathcal S}}

\def\cV{{\mathcal V}}
\def\cW{{\mathcal W}}

\def\cZ{{\mathcal Z}}

\def\qft{{\text{QFT}}}

\renewcommand{\bar}{\overline}
\renewcommand{\hat}{\widehat}

\preprint{CALT-TH 2018-025}
\title{Gluing II: Boundary Localization and Gluing Formulas}

\author[1]{Mykola Dedushenko}

\affiliation[1]{Walter Burke Institute for Theoretical Physics,\\ California Institute of Technology,\\ Pasadena, CA 91125, USA}

\emailAdd{dedushenko@gmail.com}

\abstract{We describe applications of the gluing formalism discussed in the companion paper. When a $d$-dimensional local theory $\qft_d$ is supersymmetric, and if we can find a supersymmetric polarization for $\qft_d$ quantized on a $(d-1)$-manifold $W$, gluing along $W$ is described by a non-local $\qft_{d-1}$ that has an induced supersymmetry. Applying supersymmetric localization to $\qft_{d-1}$, which we refer to as the boundary localization, allows in some cases to represent gluing by finite-dimensional integrals over appropriate spaces of supersymmetric boundary conditions. We follow this strategy to derive a number of ``gluing formulas'' in various dimensions, some of which are new and some of which have been previously conjectured.
First we show how gluing in supersymmetric quantum mechanics can reduce to a sum over a finite set of boundary conditions. Then we derive two gluing formulas for 3D $\cN=4$ theories on spheres: one providing the Coulomb branch representation of gluing, and another providing the Higgs branch representation. This allows to study various properties of their $(2,2)$-preserving boundary conditions in relation to Mirror Symmetry. After that we derive a gluing formula in 4D $\cN=2$ theories on spheres, both squashed and round. First we apply it to predict the hemisphere partition function, then we apply it to the study of boundary conditions and domain walls in these theories. Finally, we mention how to glue half-indices of 4D $\cN=2$ theories.}

\begin{document}
	\maketitle
	\flushbottom
%\cornersize{1}
%scalars
%%%%%%%%%%%%%%%%%%%%%%%%%%%%%%%%%%%%%%%%%%%%%%%%%%%%%%%%%%%%%%%%%%%%%

%%%%%%%%%%%%%%%%%%%%%%%%%%%%%%%%%%%%%%%%%%%%%%%%%%%%%%%%%%%%%%%%%%%%%%%%%%%%%%%%%%%%%%%%%%%%%%

\section{Introduction}
The fundamental property of any local QFT is the cutting and gluing law stating that QFT on a spacetime manifold can be glued from smaller pieces by composing their boundary states \cite{dirac1967principles, Atiyah1988, segal_2004, segal_roles}. Let us focus on gluing Riemannian $d$-manifolds $M$ and $N$ (with QFT defined on them) along their common boundary $W$ such that the glued metric on $M\cup_W N$ is smooth. This procedure is reviewed in the companion paper \cite{glue1}, where we emphasize the following properties of gluing in Lagrangian theories. First of all, to glue along $W$, we have to choose a polarization $\cP$ on the phase space  $X[W]$ of our theory on $W\times \R$. Then gluing along $W$ is represented as a path integral over polarized boundary conditions. Such boundary conditions are given by leaves of a Lagrangian foliation on $X[W]$, i.e. integral submanifolds of the real polarization $\cP\subset T X[W]$. More generally, one can also consider gluing using complex polarizations, however we postpone this to future work.

A point of view advocated in the companion paper \cite{glue1} is that one should think of this gluing path integral as a $(d-1)$-dimensional field theory, called $\qft_{d-1}$, canonically associated to a $d$-dimensional local theory $\qft_d$ and depending on the following choices: a $(d-1)$-manifold $W$ (along with its infinitesimal neighborhood, or germ, $[W]$ inside the $d$-dimensional spacetime, as well as the data of how to define $\qft_d$ in $[W]$), a polarization $\cP\subset TX[W]$ on the phase space $X[W]$, and a pair of states $|\Psi_2\rangle\in\cH_W$ and $\langle\Psi_1|\in\cH_W^\vee$ that we wish to glue. As long as $\qft_d$ has a symmetry that preserves polarization $\cP$, the gluing theory $\qft_{d-1}$ acquires an induced symmetry, whenever the boundary states that we glue are annihilated by the symmetry generator $Q$, i.e. $\langle\Psi_1|Q=Q|\Psi_2\rangle=0$.

In the current paper, we would like to apply this to the case when $Q$ is a supersymmetry. This means that in a supersymmetric theory $\qft_d$, we have to find a polarization on $X[W]$ that is preserved by the chosen SUSY $Q$. When such supersymmetric polarization exists, the gluing theory (describing gluing of $Q$-closed states) becomes supersymmetric. If we are at luck and can efficiently apply supersymmetric localization to this gluing theory, the path integral over polarized boundary conditions may be reduced to the finite-dimensional integral over some space of supersymmetric boundary conditions. For this reason, we often refer to the polarization-preserving $Q$ and $Q_{\rm Loc}$. When this program can be completed, it produces an interesting supersymmetric gluing formula. We refer to the supersymmetric localization in the gluing theory (with respect to $Q_{\rm Loc}$) as the boundary localization. The intuition behind this name is that gluing $N$ to $M$ along $W\subset \partial M$, from the point of view of theory on $M$, can be thought of as imposing a special boundary condition along $W$. This boundary conditions is described by a covector $\langle\Psi_1|$ encoding the result of dynamics taking place in the bulk of $N$. By performing localization of the gluing theory on $W$---the boundary localization---we derive a simple description of this boundary condition.

Notice also that the supersymmetric polarization provides an example of a supersymmetric family of boundary conditions. It often turns out more natural in physics and math to study families rather than individual objects. Boundary conditions are no exception, and a supersymmetric family of boundary conditions is a natural object to consider. Notice that this is \emph{not} the same as a family of supersymmetric boundary conditions, which is another natural object. In a supersymmetric family of boundary conditions, each individual boundary condition does not have to be supersymmetric: after applying SUSY, we obtain another boundary condition within the same family. It means that such a family is parametrized by a supermanifold on which SUSY acts through an odd vector field. If this vector field has fixed points, they correspond to genuine supersymmetric boundary conditions, but in principle there might be no fixed points at all, i.e. no supersymmetric boundary conditions in the family. It seems natural to study such objects. Here we are only interested in supersymmetric polarizations, which is a particular instance of supersymmetric families of boundary conditions. Notice that in space-time dimension $D=2$ and greater, such families are parametrized by infinite-dimensional supermanifolds (hence gluing is represented by a path integral, at least before we apply localization). These are certainly not the simplest supersymmetric families, but they are the ones relevant for gluing, which is our primary interest in the current work.

We will demonstrate this approach by deriving several interesting gluing formulas in dimensions $1$, $3$ and $4$. Some of them will be completely new, while others can be found in the literature, usually as conjectures based on various computational hints and/or naturalness.

We should note that the idea to express gluing as a non-local effective field theory on the interface between $M$ and $N$ has previously appeared in the context of two-dimensional free scalar field theories on Riemann surfaces in \cite{Morozov:1988xk, Losev:1988ea, Morozov:1988gj, Morozov:1988bu} (in relation to bosonic strings). Recently, this analysis was extended to free gauge theories and two-dimensional Yang-Mills theory in \cite{Blommaert:2018oue}. Various versions of supersymmetric gluing have played roles in \cite{Drukker:2010jp, Gaiotto:2014gha, Dimofte:2011py, Beem:2012mb, Gang:2012ff, Gadde:2013wq, Pasquetti:2016dyl, Bullimore:2014nla, Hori:2013ika, Gadde:2013sca, Honda:2013uca, Cabo-Bizet:2016ars, Gava:2016oep, Gukov:2017kmk, LeFloch:2017lbt, Bawane:2017gjf, Dimofte:2017tpi}, with the corresponding gluing rules conjectured based on the physical properties of the system under consideration. We will single out concrete references below whenever we derive a gluing formula that has previously appeared in the literature.

\subsection{Overview and the structure of this paper}
Let us summarize the main results of this paper.

\begin{itemize}
\item We start with an $\cN=2$ supersymmetric quantum mechanics. In this case, the boundaries and interfaces are zero-dimensional, hence the gluing is already represented by a finite-dimensional integral, i.e. a $\qft_0$. To illustrate our approach, we choose a supersymmetric polarization and find that this $\qft_0$ is supersymmetric. Under certain favorable conditions, its finite-dimensional ``path integral'' can be localized to a finite sum over the zeros of potential. Thus the gluing can be represented by a finite sum, see Section \ref{sec:SUSYQM} for more details.

\item In Section \ref{sec:3DN4} we study 3D $\cN=4$ gauge theories quantized on $S^2$. We describe two supersymmetric polarizations invariant under half of the supersymmetry. One of them preserves $\mathfrak{su}(2|1)_A$ on $S^2$, and another preserves $\mathfrak{su}(2|1)_B$, which are the two known variants of $\cN=(2,2)$ physical (as opposed to topological) SUSY algebra on $S^2$ \cite{Benini:2012ui,Doroud:2012xw,Gomis:2012wy,Benini:2016qnm}. Localization on $S^2$ then implies two gluing formulas that allow to sew manifolds along $S^2$, as long as the boundary states preserve the corresponding SUSY.
Such formulas can be used to glue hemisphere partition functions into the full sphere, or attach a cylinder to the boundary of the hemisphere, etc.

\item In Section \ref{sec:su21A} we study the $\mathfrak{su}(2|1)_A$--preserving case, and derive the gluing formula that has previously appeared in \cite{Dedushenko:2017avn}. It takes the form of the Coulomb branch localization answer on $S^2$:
\begin{equation}
\langle\Psi_2|\Psi_1\rangle= \sum_{B\in\Lambda_{\rm cochar}}\frac1{|\cW(H_B)|}\int_{\mathfrak{t}}\dd\sigma\, \mu(\sigma, B) \langle\Psi_2|\sigma,B\rangle \langle\sigma,B|\Psi_1\rangle,
\end{equation}
where $|\sigma,B\rangle$ denotes the half-BPS ($\mathfrak{su}(2|1)_A$-preserving) boundary condition which is of Dirichlet type for the gauge field, imposing magnetic flux $B$ through $S^2$, and of Dirichlet type for one of the scalars of the vector multiplet, fixing its value to a constant $\sigma$ at the boundary, while the rest of the boundary conditions can be determined by SUSY and are given in \eqref{bndryS2}. The coefficient $\mu(\sigma,B)$ is a one-loop determinant for the 2D localization; it depends on the field content of the theory and is given in \eqref{muSuA}. Summation goes over the cocharacter lattice of $G$, $\Lambda_{\rm cochar}$, which is the lattice of allowed magnetic charges through $S^2$. The factor $|\cW(H_B)|$ is the order of the Weyl group of $H_B\subset G$ that is left unbroken by the flux. The integration goes over the Cartan subalgebra $\mathfrak{t}\subset \mathfrak{g}$ of the gauge algebra.

\item In Section \ref{sec:su21B}, we study the $\mathfrak{su}(2|1)_B$--invariant polarization. The gluing formula we obtain is new and takes the form:
\begin{equation}
\langle\Psi_2|\Psi_1\rangle = \int_{\cM}{\rm Vol}_\cM\, Z_{\rm 1-loop}(Y, \bar{Y}) \langle\Psi_2|Y,\bar{Y}\rangle \langle Y,\bar{Y}|\Psi_1\rangle,
\end{equation}
where $|Y,\bar{Y}\rangle$ is another, $\mathfrak{su}(2|1)_B$-preserving, half-BPS boundary condition at $S^2$. It imposes Dirichlet boundary conditions for the vectormultiplets again (with vanishing boundary values of fields or their normal derivatives, unlike in the A case). Hypermultiplets are given half-Dirichlet/half-Neumann boundary conditions: one complex scalars in the hypermultiplet receives a constant value $Y$ at the boundary, while the other complex scalar is given the Neumann boundary condition (i.e., its normal derivative vanishes). The boundary fields are described in \eqref{bndrysuB}. The factor $Z_{\rm 1-loop}(Y, \bar{Y})$ originates from the localization on $S^2$ and is given, in the case of abelian gauge theory, in \eqref{OneLoopB}. Here the integration goes over the ``Higgs branch'' defined in \eqref{QHLocLoc}, and the volume form ${\rm Vol}_\cM$ is constructed in \eqref{VolFormB}.

\item In Section \ref{sec:bndry3DN4} we discuss how the above two gluing formulas are relevant for the study of $(2,2)$--preserving boundary conditions \cite{Chung:2016pgt,Bullimore:2016nji} and domain walls. Imposing a boundary condition at $S^2$ can be described by gluing a thin cylinder, $S^2\times (0,\ell)$, $\ell\to0$, to the boundary, with $S^2\times \{0\}$ supporting a boundary condition and $S^2\times \{\ell\}$ treated as the ``gluing surface''. This implies that the boundary condition can by characterized by its ``wave function'', which is simply the $\ell\to0$ limit of the $S^2\times (0,\ell)$ partition function. In the $\mathfrak{su}(2|1)_A$--invariant case, the ``wave function'' is formally written as $\Psi_\cB(\sigma, B)=\langle\cB|\sigma,B\rangle$, while in the $\mathfrak{su}(2|1)_B$ case it is $\Psi_\cB(Y, \bar{Y})=\langle\cB|Y,\bar{Y}\rangle$. As we will explain, this wave function captures the $Q_{\rm Loc}$--cohomology class of the state created by the boundary condition, where $Q_{\rm Loc}$ is the supercharge used in the boundary localization. 

If the boundary condition $\cB$ is described by coupling to some boundary theory $T$, then the wave function $\Psi_\cB$ is proportional to the partition function of $T$. In some cases, there might exist a non-trivial proportionality factor related to the 3D degrees of freedom ``trapped'' between the two boundaries of $S^2\times (0,\ell)$ and becoming effectively two-dimensional in the $\ell\to 0$ limit. Similarly, domain walls are described by integral kernels, which are also given by $S^2\times (0,\ell)$ partition functions, but with both boundaries treated as the ``gluing surfaces'' (and the actual domain wall sitting at some $S^2\times \{x\}$, $x\in(0,\ell)$, which we usually take to be $S^2\times \{\ell/2\}$).

\item In Section \ref{sec:Mirror3d} we spell out the relation to the Symplectic duality program in 3D $\cN=4$ theories \cite{Bullimore:2016nji}. We show that $\mathfrak{su}(2|1)_A$ and $\mathfrak{su}(2|1)_B$ invariant boundary conditions at the boundary of a hemisphere $HS^3$ give $\cA_C$--bimodules and $\cA_H$--bimodules respectively, where $\cA_C$ and $\cA_H$ are quantized Coulomb and Higgs branch algebras of the corresponding 3D $\cN=4$ gauge theory\footnote{The quantization in \cite{Yagi:2014toa,Bullimore:2015lsa} is achieved by placing a theory in Omega-background; such approach goes back to the work of \cite{Nekrasov:2009rc} where the four-dimensional Omega background \cite{Lossev:1997bz,Nekrasov:2002qd} was used. On the other hand, the quantization of Higgs and Coulomb branches in \cite{Dedushenko:2016jxl,Dedushenko:2017avn} was achieved by placing a theory on $S^3$ background, which is related by conformal map to the quantization in SCFT \cite{Beem:2016cbd,Chester:2014mea,Beem:2013sza}.} \cite{Bullimore:2015lsa,Dedushenko:2016jxl,Dedushenko:2017avn}. The new observation here is that the boundary conditions at $S^2$ have the $\cA$--bimodule, or $\cA\otimes \cA^{\rm op}$--module structure, while in the original flat space approach of \cite{Bullimore:2016nji}, they were one-sided modules.

\item Section \ref{sec:freehyp} discusses the simplest example of 3D $\cN=4$ Mirror Symmetry \cite{Intriligator:1996ex,deBoer:1996mp,deBoer:1996ck,Kapustin:1999ha,Assel:2015oxa,Assel:2017hck,Assel:2018exy}, namely that of a free hyper dual to the $U(1)$ gauge theory with one flavor. We explicitly describe bimodules generated by $\cA_C$ and $\cA_H$ acting on the boundary wave functions. In the former case, the action is through the shift operators, while in the latter case it is given by certain differential operators resulting in a D-module. We also observe that the Mirror Symmetry, i.e. isomorphisms between the $\cA_H$--bimodules on the free hyper side with the $\cA_C$--bimodules on the gauge theory side, is realized by the Fourier-Mellin transform acting on the boundary wave function.

\item In Section \ref{sec:mirrorwall} we outline how this Fourier-Mellin transform is related to the mirror symmetry defect \cite{Bullimore:2016nji}, whose collision with the boundary implements the Mirror Symmetry transformation of boundary conditions. We argue that the kernel of Fourier-Mellin transform simply equals the partition function of the mirror symmetry defect (or the mirror wall, for short).

\item In Section \ref{sec:N2onS3} we derive the formula applicable to 4D $\cN=2$ gauge theories glued along $S^3$. We identify the supersymmetric polarization and write the gluing path integral, which in this case preserves 3D $\cN=2$ SUSY on $S^3$, and then by using the known localization results on $S^3$, we derive the gluing formula:
\begin{equation}
\label{glue_along_s3_review}
\langle\Psi_2|\Psi_1\rangle = \frac{1}{|\cW|}\int_{\mathfrak t} \dd^r a\, \mu(a)\langle\Psi_2|a\rangle\langle a|\Psi_1\rangle,
\end{equation}
where $|a\rangle$ is a half-BPS boundary condition parametrized by a single variable $a\in\mathfrak{t}$, a value of the real part of the vector multiplet scalar at the boundary, while the gauge field is given Dirichlet boundary condition, and the rest is fixed by SUSY. The factor $\mu(a)$, referred to as ``the gluing measure,'' is given by the localization 1-loop determinant on $S^3$. It takes the form $\mu(a)=Z_{\rm v}^{S^3}(a)Z_{\rm ch}^{S^3}(a)$, where $Z_{\rm v}^{S^3}(a)=\det{}'_{\rm Adj}\left[2\sinh\pi a \right]$ is the $\cN=2$ vector multiplet contribution, while $Z_{\rm ch}^{S^3}(a)$ given in \eqref{3dchiral} is the $S^3$ determinant for chiral multiplets of R-charge $1$. If the matter is valued in a self-conjugate representation, $Z_{\rm ch}^{S^3}(a)=1$. This formula allows to glue four-manifolds along the $S^3$ boundary (as long as the boundary state preserves SUSY that was required for the 3D localization). Some special cases of this formula have been conjectured throughout the literature, see \cite{Gava:2016oep,Bullimore:2014nla}. Other references containing the similar-looking expressions include the AGT-related literature \cite{Drukker:2010jp,Hosomichi:2010vh,Terashima:2011qi}), as well as papers on Liouville theory \cite{Teschner:2003at,Teschner:2002vx,Teschner:2003em,Teschner:2005bz}.

\item We also describe deformations of this gluing formula by masses and squashing (that turns $S^3$ into an ellipsoid). The latter is described in Section \ref{sec:ellips}.

\item A simple application of the gluing formula in 4D is to find the hemisphere and squashed hemisphere partition functions. This is briefly described in Sections \ref{sec:HS4Z} and \ref{sec:HEllipsZ}.

\item Similarly to the 3D case, one can use the gluing formula to study boundary conditions and domain walls \cite{Gaiotto:2008sa,Gaiotto:2008ak,Dimofte:2011jd,Dimofte:2011ju,Dimofte:2013lba} through their ``wave functions'' and ``integral kernels'' respectively, which is the subject of Section \ref{sec:4dBCandD}. In this case, the ``wave functions'' are certain Weyl-invariant functions $\Psi_\cB(a)$ on the Cartan subalgebra $\mathfrak{t}\subset \mathfrak{g}$, whose interpretation is again as the partition function on $S^3\times (0,\ell)$ in the $\ell\to0$ limit. We give a few examples and show how the ``trapped'' degrees of freedom on $S^3\times (0,\ell)$ might, in certain cases, be crucial to obtain the right answer. One consequence of this discussion is that we can compute the hemisphere partition function with an arbitrary theory $T$ living at the boundary almost for free. For that we simply have to compute the overlap $\langle \Psi_\cB| HS^4 \rangle$ using \eqref{glue_along_s3_review}, where $\langle a|HS^4\rangle$ is the hemisphere partition function with Dirichlet boundary conditions parametrized by $a\in\mathfrak{t}$, and $\langle \Psi_\cB|a\rangle \equiv \Psi_\cB(a)$ is the boundary ``wave function'' given by the partition function of $T$ with mass $a$ turned on, multiplied by the partition function of trapped degrees of freedom, if present.

\item In Section \ref{sec:SDuality} we further elaborate on this by showing how, in the case of 4D $\cN=4$ (and $\cN=2^*$) theories, S-duality acts on the wave functions of boundary conditions. An example of the gauge group $G=SU(2)$ is discussed in Section \ref{sec:SU2}. Note that the boundary conditions studied here are only supposed to preserve a single supercharge $Q_{\rm Loc}$ -- the supercharge used in boundary localization to derive the gluing formula. However, each $Q_{\rm Loc}$-cohomology class described by the wave function $\Psi_\cB(a)$ also contains a half-BPS boundary condition.

\item In Section \ref{sec:4dindex} we consider 4D $\cN=2$ theories glued along $S^2\times S^1$. The polarization is completely analogous to the one of Section \ref{sec:N2onS3}. Localization works out differently since the gluing path integral on $S^2\times S^1$ is very similar to the one computing indices of 3D $\cN=2$ theories. The gluing measure in this case takes the form of a 3D $\cN=2$ index, and the localization locus is parametrized by gauge holonomies along $S^1$ and gauge fluxes through $S^2$. The gluing formula we obtain in this case has been previously conjectured (in less general form) in \cite{Dimofte:2011py,Gang:2012ff}.

\item Finally, in Section \ref{sec:Discuss} we conclude and provide a list of future directions. In Appendix, we describe some preliminary results on gluing of 2D $\cN=(2,2)$ gauge theories along $S^1$, as well as comment on 3D $\cN=2$ half-indices (that were first introduced in \cite{Gadde:2013wq}, see \cite{Sugishita:2013jca,Okazaki:2013kaa,Yoshida:2014ssa} for some other studies of 3D $\cN=2$ theories with boundaries).

\end{itemize}

\section{Supersymmetric applications: preparatory remarks}
We are going to apply the formalism developed in the companion paper \cite{glue1} to supersymmetric theories, so the phase space $(X,\omega)$ is a supermanifold with even symplectic structure $\omega$. One can extend the notion of polarization and polarized boundary conditions to this setting. We still remain restricted to the case of real polarizations, which in the supersymmetric case is understood as a condition on even (bosonic) directions only. In other words, the underlying (reduced) even manifold $(X_{\rm red},\omega_{\rm red})$ is a usual symplectic manifold, and the polarization $P$ on $X$ induces a polarization $P_{\rm red}$ on $X_{\rm red}$, which is assumed to be real. The fermionic part of the polarization can be complex.\footnote{In principle, it should be possible to extend our formalism to complex bosonic polarizations. It would require more work because for bosonic fields, one has to care about the convergence of the path integral and the choice of integration cycle.} 
The Main Lemma of \cite{glue1} implies that when we can find a supersymmetric polarization, the gluing theory becomes supersymmetric by itself. All SUSYs preserved by the polarization becomes SUSYs of the lower-dimensional gluing theory.

The main idea is that further application of supersymmetric localization to this lower-dimensional theory has a potential of simplifying the gluing procedure. Namely, the infinite-dimensional path integral over polarized boundary conditions might reduce to a finite-dimensional integral over a certain space of supersymmetric boundary conditions. In this way we can derive a number of interesting ``gluing formulas''.

We start, like in the general discussion of \cite{glue1}, from quantum mechanics, namely the $\cN=2$ supersymmetric quantum mechanics, and then move on to higher-dimensional examples.

\subsection{A useful point of view}\label{space_sol}
Recall that as a part of our construction, we have to know how a symmetry that we are interested in acts in the phase space. In particular, for applications to supersymmetric theories, we have to know how supersymmetry acts in the phase space. Of course, in principle it is always possible to compute the supercharge $Q$ and express it in terms of the canonical variables: then it generates the required transformations. In practice, we often start with the Lagrangian description of the SUSY theory, and it may be technically beneficial to take a slightly different approach.

A useful way to think about a phase space of a $d$-dimensional field theory corresponding to a $(d-1)$-manifold $W$ is as the space of all possible solutions to the classical equations of motion (EOM) on $\R \times W$ (modulo gauge transformations). More precisely, only germs of such solutions at $W$ really matter. Then any symmetry, by definition, transforms one solution of the classical EOMs into another. Since each solution corresponds to a point of the phase space $X$, this gives a way to describe how symmetry acts in $X$. Such a description becomes handy when we need to find how SUSY acts on a momentum variable, say $p=\dot{q}$. If we know that $\delta q = \epsilon\psi$, then on the one hand we can write $\delta p = \epsilon\dot{\psi}$, but we know that for fermions (when their EOMs are of the first order), $\dot\psi$ is not interpreted as one of the phase space coordinates. Therefore, we are supposed to express $\dot{\psi}$ in a different way, using the equations of motion. If its EOM were $\dot\psi + m\psi=0$, for example, we would replace $\dot{\psi}$ by $-m\psi$ and conclude that $\delta p = -\epsilon m\psi$.

So in describing how SUSY acts in the phase space, and in particular for understanding boundary conditions, we are allowed (and often have) to use the classical equations of motion. An important point to understand (which sometimes causes confusion in the literature) is that this is legal even when we quantize the theory with path integrals. The reason is that the \emph{classical} phase space with \emph{classical} symmetries serves as an input for the quantization procedure. In particular, when we describe boundary conditions, we need to first choose classically consistent boundary conditions given by Lagrangian submanifolds in the Hamiltonian formalism, and only then quantize the theory. Roughly speaking, the rule one can keep in mind for dealing with such issues is: when in doubt, go back to the phase space path integral. Of course all these statements have to be carefully reexamined in theories that do not admit semiclassical description.

\section{Supersymmetric quantum mechanics}\label{sec:SUSYQM}
Our first class of examples is $\cN=2$ supersymmetric quantum mechanics of \cite{Hori:2014tda} (see also \cite{Witten:1993yc, Hori:2000ck,  Herbst:2008jq}). Being one-dimensional QFTs, gluing in these theories is governed by zero-dimensional QFTs, i.e., ordinary finite-dimensional integrals over even and odd variables. We would like to show that one can use, under certain conditions, localization to further reduce these finite-dimensional integrals to finite sums.

These theories have two supercharges $Q$ and $\bar{Q}$, whose anticommutator is:
\begin{equation}
\{Q,\bar{Q}\} = H=-i\partial_t,
\end{equation}
and their structure mimics that of 2D $\cN=(0,2)$ theories, from which they are obtained by dimensional reduction. The basic multiplets are:
\begin{itemize}
	\item Vector multiplet $\cV=(A_t, \sigma, \lambda, \bar\lambda, D)$, where $A_t$ is a gauge field, $\sigma$ is a real scalar, $\lambda, \bar\lambda$ are fermions and $D$ is an auxiliary field. With $D_t=\partial_t+i A_t$ and $D_t^\pm = D_t \pm i\sigma$, the SUSY variations are given by:
	\begin{align}
	\label{vec1D}
	\delta\sigma&=-\delta A_t = -\frac{i}2\epsilon\bar\lambda - \frac{i}2\bar\epsilon \lambda,\cr
	\delta\lambda &= \epsilon(D_t\sigma+iD),\cr
	\delta D &= \frac12 \epsilon D_t^+ \bar\lambda -\frac12 \bar\epsilon D_t^+\lambda.
	\end{align}
Supersymmetric Lagrangian includes the kinetic term:
	\begin{equation}
	L_g=\frac1{2e^2} {\rm Tr}\left[ (D_t\sigma)^2 +i\bar\lambda D_t^+\lambda +D^2 \right],
	\end{equation}
	as well as the F.I. term:
	\begin{equation}
	L_{FI}=-\zeta(D).
	\end{equation}
	
	\item Chiral multiplet $\cC=(\phi, \bar\phi, \psi, \bar\psi)$ with the SUSY:
	\begin{align}
	\label{chir1D}
	\delta\phi &= -\epsilon\psi,\cr
	\delta\psi &= i\bar\epsilon D_t^+\phi.
	\end{align}
	The Lagrangian is:
	\begin{equation}
	L_{ch}=D_t\bar\phi D_t\phi + i\bar\psi D_t^- \psi + \bar\phi (D - \sigma^2)\phi - i\bar\phi \lambda\psi + i\bar\psi \bar\lambda \phi.
	\end{equation}
	\item Fermi multiplet $\cF=(\eta, \bar\eta, F, \bar{F})$ with the SUSY:
	\begin{align}
	\delta\eta &= \epsilon F + \bar\epsilon E(\phi),\cr
	\delta F &= \bar\epsilon\left( -i D_t^+ \eta + \psi^i \partial_i E(\phi) \right),
	\end{align}
	which couples to the chiral multiplet through the holomorphic superpotential $E(\phi)$. The Lagrangian is:
	\begin{equation}
	L_f = i\bar\eta D_t^+ \eta +\bar{F}F -\bar{E(\phi)}E(\phi) - \bar\eta \partial_i E(\phi) \psi^i - \bar\psi^{\bar i}\partial_{\bar i}\bar{E(\phi)}\eta.
	\end{equation}
\end{itemize}

In addition, there is a $J$-type superpotential determined by a holomorphic function $J(\phi)$:
\begin{equation}
L_J = \psi^i\partial_i J(\phi)\eta - J(\phi)F + c.c.
\end{equation}

In general, chiral and fermi multiplets take values in certain representations of the gauge group, denoted $V_{\rm ch}$ and $V_{\rm f}$ respectively. Then $J$ and $E$ superpotentials are given by $G$-equivariant maps $E: V_{\rm ch} \to V_{\rm f}$ and $J: V_{\rm ch} \to V_{\rm f}^\vee$ satisfying $J(\phi)E(\phi)=0$. (Here $V_{\rm f}^\vee$ is the dual to $V_{\rm f}$.) The anomaly free condition states that $\det (V_{\rm ch} + V_{\rm f})$ has a square root. More details can be found in \cite{Hori:2014tda}. In particular, the authors of \cite{Hori:2014tda} describe one more supersymmetric interaction in $\cN=2$ quantum mechanics given by a Wilson line ${\rm Tr}_\rho {\rm P}\exp\left(-i\int \cA_t \dd t\right)$, where $\rho: G \to U(M)$ is a representation of the gauge group on some $\Z_2$-graded vector space $M$. This allows to relax the anomaly free condition, and instead of requiring that $\det^{1/2} (V_{\rm ch} + V_{\rm f})$ is a well-defined representation of $G$, it is enough to assume that $\det^{1/2} (V_{\rm ch} + V_{\rm f})\otimes M$ is well-defined. The explicit form of the Wilson loop interaction (and even whether it is present in the path integral or not) will not be important to us: in the rest of this section, we simply assume that the theory is well-defined, and everything we do follows from the SUSY variations only.

We would like to define a useful supersymmetric polarization in a general GLSM described above, where by \emph{useful} we mean that it helps to simplify the gluing integral, e.g., by reducing it to a finite sum. We do so by providing the appropriate family of boundary conditions: this amounts to deciding which fields we freeze at the boundary, their boundary values parameterizing this family and called the boundary fields. Note that this differs from the terminology where the name ``boundary fields'' refers to the boundary degrees of freedom at the fixed boundary conditions. The fact that they define a polarization is ensured by vanishing of Poisson brackets between the boundary fields. The fact that this polarization is supersymmetric means that the boundary fields form SUSY multiplets, without mixing with fields that are allowed to fluctuate at the boundary. This is the prerequisite for applying the Main Lemma of \cite{glue1} and concluding that the gluing integral is supersymmetric. Notice that in the case of quantum mechanics, everything is completely rigorous because the gluing integral is finite-dimensional.

It is clear that we cannot preserve both $Q$ and $\bar{Q}$ because cutting a line introduces a point boundary that breaks time-translations. Suppose we choose to preserve $\bar{Q}$. Let us start by fixing $\phi$ and $\bar\phi$ at the boundary:
\begin{equation}
\phi\big| = \varphi,\quad \bar\phi\big| = \bar\varphi.
\end{equation}
Since $\{\bar{Q},\phi\}=0$ and $\{\bar{Q}, \bar\phi\}=-i\bar\psi$, to define supersymmetric polarization, we have to also fix $\bar\psi$ at the boundary, while leaving $\psi$ unconstrained. Since $\{\bar{Q}, \bar\psi\}=0$, this is consistent and closed under SUSY. So we impose the boundary condition:
\begin{equation}
\bar\psi \big| = \bar\chi.
\end{equation}
The boundary fields (which are just numbers, since the boundary is a point) $\varphi, \bar\varphi, \bar\chi$ close to a boundary multiplet under $\bar{Q}$, and they clearly define a real polarization for the chiral multiplet.

Before discussing Fermi and vector multiplets, two remarks are in order. We understand boundary conditions as Lagrangian submanifolds in the phase space, and we learned in subsection \ref{space_sol} that we can use equations of motion when describing the way SUSY acts in the phase space. Thus, we are allowed to use equations of motion when we act with SUSY on the boundary conditions. Another remark is that auxiliary fields, being non-dynamical, do not require any boundary conditions at all. Hence, before discussing boundary conditions and boundary SUSY, we could simply integrate auxiliary fields out first. At the end, we can always reintroduce them if needed so for some reason. In what follows, we will sometimes keep auxiliary fields in the boundary conditions, with the understanding that it is their on-shell value that really enters the expression.

For a Fermi multiplet, the only dynamical fields are $\eta$ and $\bar\eta$, which are canonically conjugate variables. We can choose to fix an arbitrary combination of $\eta$ and $\bar\eta$ at the boundary to define a consistent boundary condition. It is also consistent with SUSY. In particular, choosing $\eta$ is consistent with SUSY because $\{\bar{Q}, \eta\}=-i E(\phi)$, and $\phi$ is already fixed at the boundary. Choosing $\bar\eta$ is also consistent because $\{\bar{Q}, \bar\eta\}=-i \bar{F}^{\rm on-shell}=-i J(\phi)$. It is the matter of convenience whether we pick $\eta$, $\bar\eta$, or their linear combination to define the boundary condition. The most convenient choice depends on what $J$ and $E$ superpotentials look like. Let us assume that we study a model in which $J(\phi)$ has only isolated zeros (modulo gauge transformations). Then because of $\{\bar{Q},\bar\eta\}=-iJ(\phi)$, it is convenient to fix $\bar\eta$ at the boundary:
\begin{equation}
\bar\eta\big| = \bar\rho,
\end{equation}
defining the boundary field $\bar\rho$, and this describes polarization for the Fermi multiplets.

Finally, for the vector multiplet, one option is to start by freezing $\sigma$ at the boundary. Since $\{\bar{Q}, \sigma\}=-\frac12 \lambda$ and $\{\bar{Q}, \lambda\}=0$, this implies that we should also fix $\lambda$ at the boundary, and $(\sigma, \lambda)$ form a boundary multiplet under $\bar{Q}$. This boundary multiplet is a possible choice of supersymmetric polarization for the vector multiplet. One could expect that we also need some sort of boundary condition on $A_t$. However, as we know from the general discussion in \cite{glue1}, such boundary condition does not carry any physical data, and simply corresponds to (partial) gauge-fixing: boundary wave functions do not depend on the value of $A_\perp$, which is $A_t$ in our case. Also, boundary condition on $A_t$ depends on the gauge-fixing condition in the bulk, as was recently pointed out in \cite{Witten:2018lgb}. If we work in Lorenz gauge in the bulk, that is $\partial_t A_t=0$, this automatically enforces $\partial_t A_t\big|=0$. We could also choose to work in temporal gauge, $A_t=0$, in which case $A_t\big|=0$ is the right choice.\footnote{After a SUSY variation, the partial gauge condition, whether it is $A_t\big|=0$ or $\partial_t A_t\big|=0$, breaks down. One needs to perform a compensating gauge transformation with parameter $\kappa$ such that $\kappa\big|=0$, so that it would restore the gauge condition without affecting boundary values of other field.} 

This describes one possible supersymmetric polarization, with the set of boundary fields being $(\varphi, \bar\varphi, \bar\chi)$ for the chiral multiplet, $\bar\rho$ for the Fermi multiplet, and $(\sigma, \lambda)$ for the vector multiplet, where we, due to the limited supply of letters, used the same notation for the bulk fields $\sigma$, $\lambda$ and their boundary values. The gluing can be represented as an integral:
\begin{equation}
\label{QM_glue_gen}
\langle\psi_2|\psi_1\rangle = \frac1{{\rm Vol}(G)} \int \dd\bB\, \langle\psi_2|\bB\rangle \langle\bB|\psi_1\rangle,
\end{equation}
where $\bB=(\varphi, \bar\varphi, \bar\chi, \bar\rho, \sigma, \lambda)$ and $\dd\bB=\dd^2\varphi\, \dd\sigma\, \dd\bar\chi\, \dd\bar\rho\, \dd\lambda$, with $\dd^2\varphi$ a $G$-invariant measure on $V_{\rm ch}$, and $\dd\sigma\, \dd\lambda$ a $G$-invariant measure on $\mathfrak{g}\oplus \Pi\mathfrak{g}$, while $\dd\bar\chi$ and $\dd\bar\rho$ are fermionic measures on $V_{\rm ch}^*$ and $V_{\rm f}^*$ that transform as $\det V_{\rm ch}$ and $\det V_{\rm f}$ respectively. We also include ${\rm Vol}(G)$ because this gluing integral is in general a 0D gauge theory. The SUSY it admits is:
\begin{align}
\label{QM_polar1}
[\bar{Q}, \varphi]&=0,\quad [\bar{Q}, \bar\varphi]=-i\bar\chi, \quad \{\bar{Q}, \bar\chi\}=0,\cr
\{\bar{Q}, \bar\rho\}&=-i J(\varphi),\cr
[\bar{Q}, \sigma]&=-\frac12 \lambda,\quad \{\bar{Q}, \lambda\}=0.
\end{align}

Let us start by looking at a Landau-Ginzburg type theory which has no gauge multiplets, equal number of chiral and Fermi multiplets, only isolated zeros of $J(\phi)$, and the matrix $\partial_i J_j(\phi_0)$ is a non-degenerate square matrix at those zeros. For example, this could be an $\cN=4$ model of chiral multiplets with the superpotential $W(\phi)$ that has only isolated non-degenerate critical points, in which case $J_i(\phi)=\partial W(\phi)/\partial \phi^i$. But more general $J_i(\phi)$, as long as the numbers of Fermi and chiral multiplets are equal, and $\partial_i J_j(\phi)$ is a non-degenerate square matrix at zeros of $J_i(\phi)$, works as well. We apply supersymmetric localization to the gluing integral by inserting the following deformation:
\begin{equation}
\label{QM_deform1}
e^{- t\{\bar{Q},\Xi\}},\quad\text{where}\,\, \Xi=i J_i(\bar{\varphi})\bar\rho^i,
\end{equation}
so:
\begin{equation}
\{\bar{Q},\Xi\} = \sum_i |J_i(\varphi)|^2 + \partial_j J_i(\bar\varphi)\bar\chi^j \bar\rho^i.
\end{equation}
With such a deformation in the limit $t\to\infty$, the integral \eqref{QM_glue_gen} localizes to the sum over zeros of $J_i$, or critical points of $W(\phi)$. Assuming that these are non-degenerate, the one-loop determinant is given by:
\begin{equation}
\pi^{N_{ch}}\frac{\det \partial_i J_i(\bar\varphi)}{|\det \partial_i J_i(\bar\varphi)|^2}=\frac{\pi^{N_{ch}}}{\det \partial_i J_i(\varphi)}.
\end{equation}
The gluing is then represented as:
\begin{equation}
\int \dd\bB\, \langle \psi_2| \bB\rangle \langle \bB|\psi_1\rangle =\sum_{J(\varphi_0)=0} \frac{\pi^{N_{ch}}}{\det \partial J(\varphi_0)} \langle \psi_2| \bB_{\varphi_0}\rangle \langle \bB_{\varphi_0}|\psi_1\rangle,
\end{equation}
where $\bB_{\varphi_0}$ denotes boundary conditions with $\phi|=\varphi_0$ and $\bar\psi|=\bar\eta|=0$. This is a useful gluing formula in the sense explained above: it allows to completely replace the gluing integral by a sum over zeros of $J$ (if such zeros are isolated and non-degenerate, of course).

Next include vector multiplets gauging some symmetry $G$ of the system of chirals. The zeros of $J(\phi)$ now come in gauge orbits that can have positive dimensions, and we assume that these orbits are isolated, meaning that modulo gauge equivalences, the zeros of $J(\phi)$ are still isolated. In other words, each connected component of the space $\{\phi|\, J(\phi)=0\}$ should be a single gauge orbit. In this case, we can use the same localizing deformation \eqref{QM_deform1}. Under the similar non-degeneracy assumption about $\partial_i J_j(\phi)$, but only in the directions orthogonal to the gauge orbit---which in the case when $J_i=\partial W(\phi)/\partial \phi^i$ means that $W(\phi)$ is a Morse-Bott function---we can compute fluctuation determinants and find:
\begin{equation}
\label{Coul0d}
\int \frac{\dd\bB}{{\rm Vol}(G)}\, \langle \psi_2| \bB\rangle \langle \bB|\psi_1\rangle =\int_{\mathfrak{g}} \frac{\dd\sigma\, \dd\lambda}{{\rm Vol}(G)} \sum_{F\subset \{J(\varphi)=0\}}\int_{F} [\dd^2\varphi_0\, \dd\bar\chi_0\, \dd\bar\rho_0] \frac{\pi^{N_{ch}}}{\det \partial J(F)} \langle \psi_2| \tilde\bB\rangle \langle \tilde\bB|\psi_1\rangle.
\end{equation}

Let us unpack this formula. First of all, this is an application of the Atiyah-Bott localization theorem, and the sum on the right goes over $F$'s, connected components of $\{\phi|\, J(\phi)=0\}$. The determinant $\det \partial J(F)$, known as an equivariant Euler class, goes over fluctuations transversal to $F$, because those are the non-zero eigenvectors of the matrix $\partial_i J_j(\phi)$. The integration measure is written as $[\dd^2\varphi_0\, \dd\bar\chi_0\, \dd\bar\rho_0]$ to elucidate the fact that the bosonic integral $\int \dd^2\varphi_0$ goes only over $F$, while the fermionic integral $\int \dd\bar\chi_0\, \dd\bar\rho_0$ is taken over the zero modes of $\partial_i J_j$ only, i.e., the directions tangent to $F$. Finally, the integral over $\dd\sigma\, \dd\lambda$ (both variables are valued in $\mathfrak{g}$) did not disappear, it was not affected by localization at all. As for the boundary conditions $\tilde\bB$ on the right, they take the following form:
\begin{align}
\phi\big| &= \varphi_0 \in F,\quad \bar\psi\big| = \bar\chi_0,\quad
\bar\eta\big| = \bar\rho_0,\cr
\sigma\big| &= \sigma,\quad \lambda\big| = \lambda,
\end{align}
where the last two look somewhat tautological since we use the same letter for the bulk and boundary values of those fields, while $\varphi_0$, $\bar\chi_0$, $\bar\rho_0$ stand for the zero modes along $F$.

The latter representation of gluing can be called the ``Coulomb branch localization'' formula, since the variable $\sigma$ did not disappear, and integration over it can be interpreted as integration over the Coulomb branch. It is not extremely useful for the quantum mechanics case, though, in the sense explained before. Indeed, we did not manage to completely eliminate integration, even though some of the integrals were replaced by finite sums. In what follows, we will describe a different polarization, which has a chance to produce a better result.

An alternative polarization will only differ in the vector multiplet sector, while chiral and fermi multiplets are treated exactly as before. We start by freezing $\bar\lambda$ at the boundary:
\begin{equation}
\bar\lambda\Big| = \theta.
\end{equation}
Then applying SUSY gives $\{ \bar{Q}, \bar\lambda^A \}=-i(D_t\sigma^A - iD^{A, {\rm on-shell}})=-i(D_t\sigma^A + ie^2 \bar\phi t^A \phi - i e^2 \zeta^A)$, where $A$ is an adjoint index, and $\zeta^A$ is an F.I. term that can be non-zero only for those $A$ that correspond to abelian factors of the gauge group. Since we already decided to fix $\phi$ and $\bar\phi$ at the boundary, the expression for $\{\bar{Q}, \bar\lambda^A\}$ suggests that we should also freeze $D_t\sigma$:
\begin{equation}
D_t \sigma\big| = x.
\end{equation}
Then we compute $\{\bar{Q}, D_t \sigma\}=\frac{i}2 [\lambda,\sigma] -\frac12 D_t\lambda = -\frac12 D_t^+ \lambda$, which can be expressed, using the equation of motion,
\begin{equation}
D_t^+\lambda^A - 2e^2\bar\psi t^A \phi=0,
\end{equation}
as $\{\bar{Q}, D_t \sigma^A\} = -e^2 \bar\psi t^A\phi$,
where $t^A$ stands for the Lie algebra generator. Therefore, we have managed to close the $\bar{Q}$ algebra for the vector multiplet on a different set of boundary fields, namely $(x, \theta)$. This completes description of the alternative polarization, which happens to be more useful as we will see.

To summarize, now we describe the wave function as:
\begin{equation}
\psi(\varphi, \bar\varphi, \bar\chi, \bar\rho, x, \theta) \equiv \psi(\bB),
\end{equation}
where $\varphi \in V_{\rm ch}$, $\bar\varphi, \bar\chi \in V_{\rm ch}^*$, $\bar\rho\in V_{\rm f}^*$, $x, \theta\in\mathfrak{g}$, and we have introduced the collective notation $\bB$ for our polarized boundary conditions, as before. The measure in \eqref{QM_glue_gen} becomes $\dd\bB=\dd^2\varphi\, \dd\bar\chi\, \dd\bar\rho\, \dd x\, \dd\theta$. The physical wave functions like this are required to satisfy gauge-invariance in the form of the Gauss law constraint, i.e., they are annihilated by the charge corresponding to global gauge transformations. However, one should keep in mind that $\psi(\bB)$ as a function does not have to be invariant under gauge transformations. Said differently, even though it is annihilated by an operator implementing the Gauss law constraint, it does not have to be annihilated by the Lie derivative along the gauge orbit. 

Notice that the measure $\dd\bB$ in general is not gauge-invariant: indeed, while $\dd^2\varphi$ and $\dd x\, \dd\theta$ are $G$-invariant, $\dd\bar\chi\, \dd\bar\rho$ transforms as $\det V_{\rm ch} \otimes \det V_{\rm f}=\det (V_{\rm ch}\oplus V_{\rm f})$. So according to our general theory, $\langle\psi_2|\bB\rangle\langle\bB|\psi_1\rangle$ has no other choice but to transform in $\det (V_{\rm ch}\oplus V_{\rm f})^{-1}$ and cancel the non-invariance of $\dd\bB$.  If there are no Wilson loops in the QM path integral, $\langle\psi_2|\bB\rangle$ and $\langle\bB|\psi_1\rangle$ should each transform as $\det (V_{\rm ch}\oplus V_{\rm f})^{-1/2}$ -- notice that the existence of this square root is precisely the anomaly-free condition of the parent one-dimensional theory in the absence of Wilson loops \cite{Hori:2014tda}. If the Wilson loop in representation $M$ is present, $\langle\psi_2|\bB\rangle$ and $\langle\bB|\psi_1\rangle$ transform in $\det (V_{\rm ch}\oplus V_{\rm f})^{-1/2}\otimes M^\vee$ and $\det (V_{\rm ch}\oplus V_{\rm f})^{-1/2}\otimes M$ respectively, where $M^\vee$ is the dual of $M$. Recall that the absence of 1D global anomaly in this case requires that $\det (V_{\rm ch}\oplus V_{\rm f})^{-1/2}\otimes M$ is a well-defined representation \cite{Hori:2014tda}.

The boundary SUSY is now given by:
\begin{align}
\label{0dsusy}
[\bar{Q}, \varphi]&=0,\quad [\bar{Q}, \bar\varphi]=-i\bar\chi, \quad \{\bar{Q}, \bar\chi\}=0,\cr
\{\bar{Q}, \bar\rho\}&=-i J(\varphi),\cr
[\bar{Q}, x^A]&=-e^2 \chi t^A\varphi,\quad \{\bar{Q}, \theta^A\}=-i(x^A + ie^2 \bar\varphi t^A \varphi - i e^2 \zeta^A).
\end{align}

Generically, the gluing integral with this new polarization will localize to a lower-dimensional subspace. Its bosonic part is determined by the equations that can be read from \eqref{0dsusy}:
\begin{equation}
\label{BPS0d}
J(\varphi)=0,\quad x^A=0,\quad \bar\varphi t^A\varphi = e^2\zeta^A.
\end{equation}
For example, if we consider a $U(1)$ GLSM flowing to the $\C P^{N-1}$ model, which is one of the examples studied in \cite{Hori:2014tda}, the gluing integral reduces to an even integral over $\C P^{N-1}$ and an odd integral over $N-1$ fermions transforming as holomorphic tangent vectors to $\C P^{N-1}$. While this is a lower-dimensional integral compared to the original one \eqref{QM_glue_gen}, it is not a significant simplification, i.e., it is not useful in the sense defined before. We like to reduce path integrals to finite-dimensional integrals and finite-dimensional integrals to sums; this is when the localization proves to be the most efficient.

To obtain a more useful gluing formula, let us make a further assumption. Every non-zero solution $\varphi_0$ of \eqref{BPS0d} comes with its gauge orbit. Similar to what we had before in the derivation of \eqref{Coul0d}, assume that each such orbit is isolated in the space of solutions to \eqref{BPS0d}, so that solutions to \eqref{BPS0d} modulo gauge equivalences form a set of isolated points. If our theory has no E superpotential, these points are actually isolated (massive) vacua on the Higgs branch, because the potential is $|\sigma\phi|^2+\frac{e^2}{2}(\phi\bar\phi - \zeta)^2 + |J(\phi)|^2$, and the vacua with $\sigma=0$ are precisely solutions to \eqref{BPS0d}. Therefore, we can call it a zero-dimensional analog of the Higgs branch localization. Upon small modification, the latter continues to hold if for every Fermi multiplet, there is only either J or E superpotential present, but not both of them: in this case, we pick $\bar\eta$ to determine the polarization for every Fermi multiplet that has only a J superpotential, and $\eta$ -- for every Fermi multiplet that has only an E superpotential. Then in equations \eqref{BPS0d}, one replaces J by E whenever $J=0$ but $E\ne 0$.

As an illustration, consider the $\C P^{N-1}$ GLSM (i.e., $U(1)$ gauge theory with $N$ charge-one chirals and a positive F.I. parameter) enriched by $N-1$ Fermi multiplets of charges $q^a$, $a=1,\dots, N-1$, and also include $N-1$ homogeneous J-type superpotentials for those, such that equations $J_a(\phi)=0$, with $a=1\dots N-1$, have isolated solutions on $\C P^{N-1}$. These superpotentials satisfy:
\begin{equation}
\label{homogen}
J_a(e^{i\alpha}\phi)=e^{-i q^a \alpha} J_a(\phi).
\end{equation}
D-term relations determine the $\C P^{N-1}$, as in the original GLSM, and the $J$ superpotentials further pick out a finite set of points on it. In this situation, our Higgs branch localization works well.
The localizing term is constructed as:
\begin{align}
\label{locterm0d}
\bar{Q}\left[i J_a(\bar\varphi)\bar\rho^i + i\theta \left(x - ie^2(\sum_i \bar\varphi^i \varphi^i - \zeta)\right)\right]&=|J(\varphi)|^2 + x^2 + e^4(\sum_i\bar\varphi^i \varphi^i -\zeta)^2 \cr
&+\bar\chi^i \bar\rho^a \partial_i \bar{J}_a(\bar\varphi) -2ie^2 \bar\chi^i \varphi^i \theta.
\end{align}
Zeros of this positive definite localizing term are given by points $\varphi_0\in \C^N//U(1)\equiv \C P^{N-1}$ that further satisfy $J(\varphi_0)=0$. More precisely, $\C P^{N-1}$ comes as a base of the generalized Hopf fibration, whose total space $S^{2N-1}$ is determined by the D-term relation $\sum_i \bar\varphi^i \varphi^i = \zeta$. The isolated points on the base determined by the equations $J_a(\varphi_0)=0$ correspond to the isolated $U(1)$ fibers of $S^{2N-1} \to \C P^{N-1}$. The set of such isolated fibers is the zero locus of the localizing term \eqref{locterm0d}. Thus our gluing integral localizes to the sum over such isolated fibers and an integral over each of them. The corresponding integrand is constant along fibers, i.e., is gauge-invariant. Therefore, the integration along fibers simply cancels the factor of $\frac{1}{{\rm Vol}(G)}$, and we can evaluate the integrand at any point along the fiber where $J_a$'s vanish. The integrand involves a one-loop determinant that can be schematically written as $1/\det\partial J$, which should be understood as the appropriate equivariant Euler class. We would like to evaluate it more explicitly in the rest of this subsection.

Let $\varphi_0$ be one such point, i.e., $\sum_{i=1}^N \bar\varphi^i_0 \varphi^i_0=\zeta$ and $J_a(\varphi_0)=0$. Let us introduce $N-1$ orthonormal vectors $v^i_a$, $a=1,\dots, N-1$, that also satisfy:
\begin{equation}
\sum_i \bar\varphi_0^i v^i_a=0,\, a=1,\dots,N-1.
\end{equation}
We can now parametrize fluctuations around $\varphi_0$ as:
\begin{equation}
\label{phichange}
\varphi^i = \varphi^i_0 + \frac{\varphi^i_0}{\sqrt\zeta} y + \sum_{a=1}^{N-1} v^i_a z^a,
\end{equation}
where $y$ corresponds to fluctuations proportional to $\varphi_0$, while $z^a$ are fluctuations tangent to $S^{2N-1}$ in the directions normal to the $U(1)$ orbit. We also make the following coordinate change for $\chi^i$:
\begin{equation}
\label{chichange}
\bar\chi^i = \frac{\bar\varphi^i_0}{\sqrt \zeta}\bar\chi_0 + \sum_{a=1}^{N-1} \bar{v}^i_a \bar\omega^a,
\end{equation}
where $\bar\chi_0$ and $\bar\omega^a$ are new fermions, and $\bar{v}^i_a$ is a complex conjugate of $v^i_a$. After these manipulations, we expand the localizing term up to the quadratic order in fluctuations around $\varphi_0$. It is convenient to introduce a matrix $V=(v^i_a)$ of size $N \times (N-1)$ and to think of $\partial J$ as an $(N-1)\times N$ matrix $(\partial J)^a_i = \partial_i J_a(\varphi_0)$, and also write $\partial \bar{J} = (\partial J)^\dagger$ for the Hermitian conjugate matrix of $\partial J$. We also have $(N-1)$-component columns $z$, $\bar\omega$, $\bar\rho$. Then the quadratic term is:
\begin{align}
\bar{z}^\dagger \left(V^\dagger \partial \bar{J} \partial J V\right) z + x^2 + e^4 \zeta (y+\bar{y})^2 + \bar{\omega}\left( V^\dagger \partial\bar{J}\right)\bar\rho - 2ie^2\sqrt{\zeta} \bar\chi_0 \theta.
\end{align}
Integrating out $x, {\rm Re}(y), \bar\chi_0, \theta$ is trivial and produces a factor of $i\pi$.  It is clear that integrating out $z, \bar{z}$ and $\bar\omega, \bar\rho$ produces determinants that partially cancel and give:
\begin{equation}
\frac{\pi^{N-1}}{\det(\partial J V)}.
\end{equation}
We also have to take into account the Jacobian arising from the coordinate changes \eqref{phichange} and \eqref{chichange}. If we group the column $\frac{\varphi_0^i}{\sqrt{\zeta}}$ together with the matrix $V$ to form a new $N \times N$ matrix $U$, then \eqref{phichange} and \eqref{chichange} become:
\begin{equation}
\varphi = \varphi_0 + U \left(\begin{matrix}
y\\ z^1\\ z^2\\.\\.\\.\\z^{N-1}
\end{matrix} \right),\quad \bar\chi = (\bar\chi_0\,\, \bar\omega^1\,\,\bar\omega^2\,\,\dots\,\,\bar\omega^{N-1}) U^\dagger.
\end{equation}
We further notice that $U$ is a unitary matrix, so the $d^2\varphi$ measure produces a Jacobian $\det(U^\dagger U)=1$, while $d\bar\chi$ produces a non-trivial Jacobian $\det U$.

Finally, we see that integration over ${\rm Im}(y)$ is not suppressed as it is the gauge direction. The gauge orbit contributes a factor of ${\rm Vol}(G)$ which cancels against the similar factor in the gluing integral, as we have already remarked before. For this to be true, the integrand has to be gauge invariant, which is the case as we will see momentarily.

Combining all the pieces together, we obtain:
\begin{equation}
\label{QM_glu_gauge}
\int \frac{\dd\bB}{{\rm Vol}(G)}\, \langle\psi_2|\bB\rangle\langle\bB|\psi_1\rangle = i\sum_{\varphi_0} \frac{\pi^N \det U(\varphi_0) }{\det \partial J(\varphi_0) V(\varphi_0)}\langle\psi_2|\bB_{\varphi_0}\rangle \langle\bB_{\varphi_0}|\psi_1\rangle,
\end{equation}
where the boundary conditions $\bB_{\varphi_0}$ in \eqref{QM_glu_gauge} are: $\bar\psi^i\big|=\bar\eta^a\big|=\bar\lambda\big|=D_t\sigma\big|=0$ and $\phi\big|=\varphi_0$.

In the above expression, the sum goes over the isolated $U(1)$ fibers of the $S^{2N-1}\to \C P^{N-1}$ fibration where $J_a$'s vanish, and in each such fiber we arbitrarily pick $\varphi_0$. Also, for each $\varphi_0$, we pick $v_a^i$, which then determine matrices $V(\varphi_0)$ and $U(\varphi_0)$. It is not hard to see that the answer is independent of these arbitrary choices. Picking another orthonormal set of vectors $v^i_a$ is equivalent to replacing $V \to V T$, where $T$ is an $(N-1)\times (N-1)$ unitary matrix. This also induces a modification $U \to U\times (1\oplus T)$. Therefore, it produces factors of $\det T$ both in the numerator and in the denominator of \eqref{QM_glu_gauge}, which cancel. To prove independence of the choice of $\varphi_0$ in a given fiber, we have to prove that each term on the right of \eqref{QM_glu_gauge} is gauge-invariant. We can assume that $V$ is gauge-invariant, because we have just seen that nothing depends on the choice of $V$. Using \eqref{homogen}, we find that under $\varphi_0\to e^{i\alpha}\varphi_0$, the denominator of \eqref{QM_glu_gauge} contributes a factor of $e^{i\alpha\sum_a(q^a+1)}=e^{i\alpha\sum_a q^a + i\alpha N - i\alpha}$. The numerator $\det U$ simply contributes $e^{i\alpha}$, so the ratio of determinants is multiplied by $e^{i\alpha\sum_a q^a + i\alpha N}$. On the other hand, we know that $\langle\psi_2|\bB_{\varphi_0}\rangle \langle\bB_{\varphi_0}|\psi_1\rangle$ transforms as $\det(V_{\rm ch}\oplus V_{\rm f})^{-1}$, and so it contributes a factor of $e^{-i\alpha N - i\alpha \sum_a q_a}$ that precisely cancels the factor coming from the determinants.

In this way, we see that the answer given in \eqref{QM_glu_gauge} is indeed gauge-invariant and independent of the arbitrary choices. 

\section{3D $\cN=4$ theories quantized on $S^2$}\label{sec:3DN4}
Let us move up in dimension and consider three-dimensional applications, namely to 3D $\cN=4$ gauge theories. We are skipping two-dimensional examples because they are slightly more subtle, see Appendix \ref{openended} for some preliminary results.

The most canonical way to proceed would be to put a theory on $S^2\times \R_+$, but any manifold with the $S^2$ boundary would work, as long as the necessary SUSY is preserved: close to the boundary $S^2$, the normal derivative plays the role of the time derivative on $S^2\times \R_+$. We find it convenient to follow \cite{Dedushenko:2017avn} and study theories on a round hemisphere $HS^3$.

The application of gluing techniques to $\cN=4$ gauge theories on $S^3$ has recently proved to be extremely useful in \cite{Dedushenko:2017avn}, where it allowed, together with the supersymmetric localization in the bulk, to formulate an elegant and simple description of quantized Coulomb branches in those theories, as well as relate them to correlators of the IR SCFTs. (Only abelian theories were considered in \cite{Dedushenko:2017avn}, see \cite{In_progress} for non-abelian generalizations.)

The $\cN=4$ gauge theories on $S^3$, at least their versions studied in \cite{Dedushenko:2017avn}, are based on the superalgebra $\mathfrak{su}(2|1)_\ell\oplus \mathfrak{su}(2|1)_r$, with the bosonic subalgebra $\mathfrak{su}(2)_\ell \oplus \mathfrak{su}(2)_r$ containing isometries of $S^3$ and the bosonic subalgebra $\mathfrak{u}(1)_\ell \oplus \mathfrak{u}(1)_r$ of R-symmetries. This algebra is described as a subalgebra of 3D $\cN=4$ superconformal algebra $\mathfrak{osp}(4|4)$  whose supersymmetries are parametrized in terms of conformal Killing spinors $\xi_{a\dot a}$, with $a=1,2$ and $\dot{a}=1,2$, satisfying conformal Killing spinor equations:
\begin{equation}
\nabla_\mu \xi_{a\dot a} =\gamma_\mu \xi_{a\dot a}',\quad \nabla_\mu \xi_{a\dot a}' =-\frac1{4r^2}\gamma_\mu \xi_{a\dot a}.
\end{equation}
while those belonging to the $\mathfrak{su}(2|1)_\ell\oplus \mathfrak{su}(2|1)_r$ subalgebra satisfy an additional constraint:
\begin{equation}
\xi_{a\dot a}'=\frac{i}{2r}h_a{}^b \xi_{b\dot b}\bar{h}^{\dot b}{}_{\dot a}.
\end{equation}
We follow conventions of \cite{Dedushenko:2017avn} and choose $h=-\tau^2$, $\bar{h}=-\tau^3$, where $\tau^i$ are Pauli matrices. Denoting the supercharges in $\mathfrak{su}(2|1)_\ell$ by $\cQ_\alpha^{(\ell\pm)}$, where $\alpha$ is a spinor index and $\pm$ reflects the $\mathfrak{u}(1)_\ell$ R-charge, and analogously those in $\mathfrak{su}(2|1)_r$ by $\cQ_\alpha^{(r\pm)}$, one has \cite{Dedushenko:2017avn}:
\begin{equation}
\{\cQ_\alpha^{(\ell +)},\cQ_\beta^{(\ell -)}\}=-\frac{4i}{r}\left(  J_{\alpha\beta}^{(\ell)}+\frac12 \varepsilon_{\alpha\beta} R_\ell\right),
\end{equation}
and similarly for $\cQ_\alpha^{(r\pm)}$, where $J$ denotes generators of the corresponding $\mathfrak{su}(2)$ isometry of $S^3$, $R$ is the R-symmetry generator, and $r$ is the radius of $S^3$.

Let us focus, like in \cite{Dedushenko:2017avn}, on theories built from the vectormultiplet $(A_\mu, \Phi_{\dot{a}\dot{b}}, \lambda_{a\dot{a}}, D_{ab})$ in the adjoint of gauge group $G$, and the hypermultiplet $(q_a, \tilde{q}^a, \psi_{\dot a}, \tilde\psi_{\dot a})$ in some representation $\cR\oplus\bar\cR$. Components of the vector multiplet are the gauge field $A_\mu$, the triplet of real scalars $\Phi_{\dot{a}\dot{b}}$, gaugini $\lambda_{a\dot a}$ and the triplet of auxiliary fields $D_{ab}$. (All ``triplets'' are with respect to the appropriate R-symmetries, $SU(2)_H$ or $SU(2)_C$, that would be present in flat space or at the SCFT point, but are not part of $\mathfrak{su}(2|1)_\ell\oplus \mathfrak{su}(2|1)_r$.) Components of the hypermultiplet are the doublet of complex scalars $q_a$, $a=1,2$, with $\tilde{q}^a = (q_a)^*$, and their superpartners $\psi_{\dot a}$, $\tilde\psi_{\dot a}$, $\dot{a}=1,2$. For brevity, we are not going to write 3D $\cN=4$ actions and SUSY transformations here, interested reader can find them in \cite{Dedushenko:2016jxl,Dedushenko:2017avn}. Explicit form of those is needed for the computations results of which are presented in the following two subsections.

We cut $S^3$ in two halves, $HS^3_+$ and $HS^3_-$, to obtain a theory on $HS^3$. We can then glue them back, or glue some other manifold with the $S^2$ boundary to the hemisphere. According to our formalism, there can be many ways to represent this as the gluing path integral on $S^2$, corresponding to different choices of polarization. Let us focus on the supersymmetric polarizations preserving $\cN=(2,2)$ supersymmetry in the gluing theory. As is well-known, physical (i.e., not topological) $\cN=(2,2)$ SUSY on $S^2$ is characterized by the superalgebra $\mathfrak{su}(2|1)$, and there are two versions of it, usually denoted by $\mathfrak{su}(2|1)_A$ and $\mathfrak{su}(2|1)_B$, depending on how they are embedded in the 2D superconformal algebra \cite{Doroud:2013pka}. The former corresponds to choosing the vector R-symmetry of the SCFT, while the latter picks out the axial R-symmetry of the SCFT as the $U(1)$ R-symmetry in the $\mathfrak{su}(2|1)$ subalgebra.

We describe two interesting supersymmetric polarizations on $S^2=\partial (HS^3)$: one -- preserving $\mathfrak{su}(2|1)_A$ (this has appeared in \cite{Dedushenko:2017avn}), and another -- preserving $\mathfrak{su}(2|1)_B$ (this one is novel). We will first review the former and then describe the latter. Both polarizations lead to nice gluing formulas, one relevant for the Coulomb and another -- for the Higgs branches of the gauge theory.

Our strategy in building the supersymmetric polarization is similar to the one we followed for SUSY quantum mechanics in the previous section. First we identify the subalgebra we wish to preserve at the boundary ($\mathfrak{su}(2|1)_A$ or $\mathfrak{su}(2|1)_B$). Then we make a choice to fix some hypermultiplet scalar at the boundary, say $q_1$ or $q_+=q_1 + iq_2$, i.e., postulate Dirichlet boundary conditions for it, as well as choose Dirichlet boundary conditions for the gauge field. After that we act on these chosen boundary conditions with the SUSY that we wish to preserve on $HS^3$ to identify the remaining boundary conditions. Comparing with the SUSY transformations on $S^2$, which are known from \cite{Benini:2012ui,Doroud:2012xw,Gomis:2012wy,Benini:2016qnm}, we identify boundary multiplets. Then we check that their Poisson brackets vanish, i.e., that they indeed determine a polarization.

\subsection{The $\mathfrak{su}(2|1)_A$-invariant polarization}\label{sec:su21A}
As we cut $S^3$ in two halves, we break the isometry down to $\mathfrak{su}(2)$ that, up to conjugation, can be identified with ${\rm diag}\left[ \mathfrak{su}(2)_\ell \oplus \mathfrak{su}(2)_r \right]$. The most straightforward choice for the corresponding supersymmetry that can be preserved by the cut is the diagonal subalgebra ${\rm diag}\left[ \mathfrak{su}(2|1)_\ell\oplus \mathfrak{su}(2|1)_r \right]$. Upon going to the SCFT point, where the SUSY algebra on $S^3$ gets enhanced to the superconformal one, we can identify this ${\rm diag}\left[ \mathfrak{su}(2|1)_\ell\oplus \mathfrak{su}(2|1)_r \right]$ as the $\mathfrak{su}(2|1)_A$ SUSY on $S^2$. Whether we actually obtain the diagonal subalgebra or the one related to it by conjugation depends on where we choose to make a cut. We are going to follow conventions of \cite{Dedushenko:2017avn}, where the $S^3$ was described in coordinates $(\theta, \varphi, \tau)$, with $\theta\in [0,\pi/2]$, $\varphi\in [-\pi,\pi)$ and $\tau\in[0,2\pi)$, with the metric :
\begin{equation}
\dd s^2= r^2\left(d\theta^2 + \sin^2\theta d\varphi^2 + \cos^2\theta d\tau^2 \right).
\end{equation}
The cut is performed along the $S^2$ located at $\varphi=0$ and $\varphi=\pi$. This corresponds to picking a conjugated diagonal subalgebra, i.e., $A\times {\rm diag}\left[ \mathfrak{su}(2|1)_\ell\oplus \mathfrak{su}(2|1)_r \right]\times A^{-1}$, where $A$ is a certain rotation of $S^3$ (a diagonal subalgebra would correspond to a cut along $\varphi=\pm \pi/2$), with the preserved supercharges:
\begin{align}
\label{su2AQ}
Q_1^+ &= \cQ_1^{(\ell +)} + \cQ_1^{(r+)},\quad Q_2^+ = \cQ_2^{(\ell +)}-\cQ_2^{(r+)},\cr
Q_1^- &= \cQ_1^{(\ell-)}-\cQ_1^{(r-)},\quad Q_2^-= \cQ_2^{(\ell -)} + \cQ_2^{(r-)}.
\end{align}
One can identify precisely which conformal Killing spinors $\xi_{a\dot a}$ on $S^3$ correspond to these supercharges, and describe SUSY transformations of vector and hyper multiplets that they generate. For brevity, we are not going to do it here, -- all the details can be found in Appendix A.2 of reference \cite{Dedushenko:2017avn}. In particular, the surviving 2D $(2,2)$ SUSY on $S^2$ is parametrized in terms of two-component conformal Killing spinors $\epsilon$ and $\bar\epsilon$ on $S^2$ satisfying:
\begin{align}
\label{KiSpiS2}
\nabla_\mu \epsilon = \frac1{2r}\gamma_\mu \gamma^3 \epsilon,\quad \nabla_\mu \bar\epsilon = -\frac1{2r}\gamma_\mu \gamma^3 \bar\epsilon.
\end{align}
The boundary fields are defined by the following relations \cite{Dedushenko:2017avn} (below $q_\pm=q_1 \pm i q_2$ and $\tilde{q}_\pm = \tilde{q}_1 \pm i\tilde{q}_2$):
\begin{align}
\label{pol22A}
\phi&=q_+\big|,\quad \bar\phi=i\tilde{q}_-\big|,\cr
\chi&=(\psi_{\dot 1} - \sigma_3 \psi_{\dot 2})\big|,\quad \bar\chi=i(\tilde\psi_{\dot 1} + \sigma_3 \tilde\psi_{\dot 2})\big|,\cr
f&=-\left( \cD_\perp q_- +\frac{\Phi_{\dot1\dot1}-\Phi_{\dot2\dot2}}{2}q_- \right)\big|,\quad \bar{f}=-i\left( \cD_\perp \tilde{q}_+ +\frac{\Phi_{\dot1\dot1}-\Phi_{\dot2\dot2}}{2}\tilde{q}_+ \right)\big|,\cr
a&= A_\parallel\big|,\quad s_1=\frac{\Phi_{\dot1\dot1}+\Phi_{\dot2\dot2}}{2i}\big|,\quad s_2=-\Phi_{\dot1\dot2}\big|,\cr
\lambda&=-\frac12 (\lambda_{1\dot2}-i\lambda_{2\dot2}+\sigma_3(\lambda_{1\dot1}-i\lambda_{2\dot1}))\big|,\quad \bar\lambda=-\frac12(\lambda_{1\dot2}+i\lambda_{2\dot2}-\sigma_3(\lambda_{1\dot1}+i\lambda_{2\dot1}))\big|,\cr
D_{\rm 2d}&=\left(-\frac{\Phi_{\dot1\dot2}}{r}+\frac{i}2 (D_{11}^{\rm on-shell}+D_{22}^{\rm on-shell})+i\cD_{\perp}\left(\frac{\Phi_{\dot1\dot1}-\Phi_{\dot2\dot2}}{2} \right) \right)\big|,
\end{align}
where $D_{ab}^{\rm on-shell}$ denotes the on-shell value of the auxiliary field $D_{ab}$ (recall that the vertical line on the right always means restriction to the boundary $S^2$). We use the notation $A_\parallel$ to denote the component of $A$ parallel to the boundary, while $\cD_\perp$ denotes the normal covariant derivative.

The SUSY transformations of these boundary fields, deduced by restricting the bulk SUSY to the boundary, are given in the Appendix A.2.2 of \cite{Dedushenko:2017avn}. By comparing with the known description of $\cN=(2,2)$ theories on $S^2$ \cite{Benini:2012ui,Doroud:2012xw,Gomis:2012wy,Benini:2016qnm}, one clearly sees the 2D multiplets:
\begin{align}
(\phi, \bar\phi, \chi, \bar\chi, f, \bar{f}):\quad &\text{2D }\cN=(2,2)\text{ chiral multiplet of R-charge 1},\cr
(a_\mu, s_1, s_2, \lambda, \bar\lambda, D_{\rm 2d}):\quad &\text{2D }\cN=(2,2)\text{ vector multiplet},
\end{align} 
with $f$ playing the role of a 2D complex auxiliary field.

Notice that $D_{\rm 2d}$ as defined in \eqref{pol22A} is not real, whereas the 2D auxiliary field in the vector multiplet should be real for convergence of the 2D path integral (with the possible bounded imaginary shift). This discrepancy is the consequence of a relation between Euclidean and Lorentzian signature, which was emphasized in \cite{glue1}. The gluing procedure is more naturally described when the direction normal to the boundary is time-like. If we were in such a situation, we would have $\cD_\perp$ replaced by $-i \cD_0$, in which case the $D_{\rm 2d}$ field at the boundary would be defined as:
\begin{equation}
D_{\rm 2d}=\left(-\frac{\Phi_{\dot1\dot2}}{r}+\frac{i}2 (D_{11}+D_{22})+\cD_0\left(\frac{\Phi_{\dot1\dot1}-\Phi_{\dot2\dot2}}{2} \right) \right)\big|.
\end{equation}
This $D_{\rm 2d}$ is real (and only acquires an imaginary part upon putting $D_{ab}$ on shell, which shifts the integration cycle of $D_{\rm 2d}$ in the imaginary direction but does not change its slope, hence the convergence is preserved), so the boundary ``gluing'' multiplets have the conventional reality properties. The appearance of ``$i$'' in front of $\cD_\perp$ in the definition of $D_{2d}$ in \eqref{pol22A} is thus a manifestation of the analytic continuation discussed in Section 2.5 of \cite{glue1}, and it should not cause any troubles. To remind the conclusions of that analysis, the more precise way to think about this is as follows: the gluing is determined in Lorentzian signature with the boundary fields defined as in \eqref{pol22A}, except that $\cD_\perp$ is replaced by $-i\cD_0$. The wave functions naturally entering the gluing path integral are the Lorentzian time wave functions, and they are obtained from the Euclidean time wave functions by Wick rotation of the momentum variables (whenever the wave function depends on them). The Euclidean time wave functions are evaluated with the boundary conditions \eqref{pol22A}.

It remains to show that \eqref{pol22A} actually determine a polarization, i.e., a maximal Poisson-commuting subset of variables in the phase space.\footnote{Recall that, in a slight abuse of terminology, by ``polarization'' we mean both the Lagrangian distribution on the phase space and the maximal Poisson-commuting set of coordinates that parametrize the space of leaves of this distribution.} For this we need to know all the kinetic terms, as well as the expression for $D_{ab}^{\rm on-shell}$ in terms of propagating fields. Those can be read off from the actions described in \cite{Dedushenko:2017avn}. The kinetic terms are:
\begin{align}
\label{kineticS3}
\cL_{\rm kin}=\cD^\mu \tilde{q}^a\cD_\mu q_a - i\tilde\psi^{\dot a}\slashed{\cD}\psi_{\dot a} + \frac1{g_{YM}^2}{\rm Tr}\left( F^{\mu\nu}F_{\mu\nu} - \cD^\mu\Phi^{\dot a\dot b}\cD_\mu \Phi_{\dot a\dot b} + i\lambda^{a\dot a}\slashed{\cD}\lambda_{a\dot a} \right).
\end{align}
The on-shell values of the auxiliary fields can also be read off from the full action. If we write $D_{ab}=D_{ab}^A T^A$, where $T^A$ are the gauge generators, those are given by:
\begin{equation}
\label{onshellD}
D_{ab}^A{}^{\rm on-shell}=-\frac{ig_{YM}^2}{2}\tilde{q}_{(a}T^A q_{b)} -\frac1{2r}h_{ab} \bar{h}^{\dot a}{}_{\dot b}\Phi^{A,\dot{b}}{}_{\dot a} - \frac{i}{2} g_{YM}^2 \zeta^A h_{ab},
\end{equation}
where $\zeta^A$ is a possible F.I. term that can be present if $T^A$ generates a $U(1)$ factor. 

It is a straightforward exercise to read off the Poisson brackets of various fields from the (Euclidean) action \eqref{kineticS3}. One should note that, because $D_{ab}^{\rm on-shell}$ is expressed in terms of the propagating fields in \eqref{onshellD}, it has non-trivial Poisson brackets with other fields of the theory. One can then check that indeed, all the fields defined in \eqref{pol22A} have vanishing Poisson brackets with each other, and moreover, they determine a middle-dimensional subspace in the space of fields. Closing under the $\cN=(2,2)$ SUSY, they thus form a supersymmetric polarization. With a collective notation, as before, $\bB_A=(\phi, \bar\phi, \chi, \bar\chi, f, \bar{f}, a, s_1, s_2, \lambda, \bar\lambda, D_{\rm 2d})$, we find that the gluing path integral,
\begin{equation}
\label{glueHS2A}
\langle\psi_2|\psi_1\rangle=\int \pD \bB_A\, \langle\psi_2|\bB\rangle\langle\bB|\psi_1\rangle,
\end{equation}
can be thought of as an $\cN=(2,2)$ supersymmetric QFT on $S^2$. 
%\begin{equation}
%D_{\rm 2d}^A=\frac{g_{YM}^2}{8}(\tilde{q}_+ T^Aq_- + \tilde{q}_- T^A q_+) + i\cD_\perp \left(\frac{\Phi_{\dot1\dot1}^A - \Phi_{\dot2\dot2}^A}{2} \right) - \frac{i}{2}g_{YM}^2\zeta^A.
%\end{equation}

Now we would like to localize this path integral. In order to do that, we have to include a $Q$-exact deformation in \eqref{glueHS2A} and compute the 1-loop determinants that it produces. We can use the standard deformation known in the literature (e.g., the kinetic actions for 2D vectors and chirals) and simply copy the results from \cite{Benini:2012ui,Doroud:2012xw,Gomis:2012wy,Benini:2016qnm}, where localization of $\cN=(2,2)$ theories on $S^2$ was thoroughly studied. We mostly follow \cite{Doroud:2012xw}. Using the Coulomb branch representation of the localization formula in 2D, the localization locus is described by the following vevs for the bosonic fields in $\bB_A$:
\begin{align}
\label{CoulBpsSol}
a=\pm\frac{B}{2}(\sin\theta -1)\dd\tau,\quad D_{\rm 2d}=0,\quad s_1=\frac{B}{2r},\quad s_2=-\frac{\sigma}{r},\quad \phi=0,\quad f=0,
\end{align}
while all fermions in $\bB_A$ vanish. Here $\sigma$ is a constant matrix valued in the gauge Lie algebra $\mathfrak{g}$, while $B\in \mathfrak{t}$ is the magnetic flux through $S^2$. Localization equations imply that $\sigma$ and $B$ commute, thus one can make them both lie in the Cartan subalgebra $\mathfrak{t}$ (with the appropriate Vandermonde included).

This solution to the localization equations allows to define the bra- and -ket (co)vectors $\langle\sigma, B|\in\cH_{S^2}^\vee$ and $|\sigma, B\rangle\in\cH_{S^2}$ imposing the half-BPS boundary conditions at $S^2=\partial (HS^3)$. Such boundary conditions are given by inserting the solution \eqref{CoulBpsSol} into the definitions \eqref{pol22A}:
\begin{align}
\label{bndryS2}
&q_+\big|=0,\quad \tilde{q}_-\big| =0,\quad \left( \cD_\perp q_- +\frac{\Phi_{\dot1\dot1}-\Phi_{\dot2\dot2}}{2}q_- \right)\big|=0,\quad \left( \cD_\perp \tilde{q}_+ +\frac{\Phi_{\dot1\dot1}-\Phi_{\dot2\dot2}}{2}\tilde{q}_+ \right)\big|=0,\cr
&(\psi_{\dot 1} - \sigma_3 \psi_{\dot 2})\big|=0,\quad (\tilde\psi_{\dot 1} + \sigma_3 \tilde\psi_{\dot 2})\big|=0,\cr
&A_\parallel\big| =\pm\frac{B}{2}(\sin\theta -1)\dd\tau,\quad \frac{\Phi_{\dot1\dot1}+\Phi_{\dot2\dot2}}{2i}\big|=\frac{B}{2r},\quad \Phi_{\dot1\dot2}\big| =\frac{\sigma}{r},\cr
&(\lambda_{1\dot2}-i\lambda_{2\dot2}+\sigma_3(\lambda_{1\dot1}-i\lambda_{2\dot1}))\big|=0,\quad (\lambda_{1\dot2}+i\lambda_{2\dot2}-\sigma_3(\lambda_{1\dot1}+i\lambda_{2\dot1}))\big|=0,\cr
&\frac{g_{YM}^2}{8i}(\tilde{q}_+ T^Aq_- + \tilde{q}_- T^A q_+) + \cD_\perp \left(\frac{\Phi_{\dot1\dot1}^A - \Phi_{\dot2\dot2}^A}{2} \right) - \frac{1}{2}g_{YM}^2\zeta^A\big|=0,
\end{align}
where in the last line we have substituted the on-shell values $D_{ab}^{\rm on-shell}$ into the definition of $D_{\rm 2d}$. These boundary conditions preserve, quite expectedly, $\cN=(2,2)$ SUSY. By writing the state $\psi\in\cH_{S^2}$ evaluated at \eqref{bndryS2} as $\langle\sigma,B|\psi\rangle$, we find that the gluing path integral \eqref{glueHS2A} reduces after the ``boundary localization'' to:
\begin{equation}
\label{gluingf1}
\langle\psi_2|\psi_1\rangle = \sum_{B\in \Lambda_{\rm cochar}}\frac1{|\cW(H_B)|}\int_{\mathfrak{t}}\dd\sigma\, \mu(\sigma,B)\, \langle\psi_2|\sigma,B\rangle \langle\sigma,B|\psi_1\rangle,
\end{equation}
where one sums over the lattice $\Lambda_{\rm cochar}$ of allowed magnetic fluxes through $S^2$ (cocharacters of $G$), $|\cW(H_B)|$ is the order of the Weyl group of $H_B\subset G$ that is left unbroken by the magnetic flux $B$, and $\mu(\sigma,B)$ is the 1-loop determinant of the $S^2$ localization (that can be found in \cite{Doroud:2012xw}) multiplied by the appropriate Vandermonde (since we integrate $\sigma$ only over the Cartan $\mathfrak{t}$). The coefficient $\mu(\sigma,B)$ has the following structure:
\begin{align}
\label{muSuA}
\mu(\sigma, B)=Z_{\rm one-loop}^{\rm v.m.}(\sigma,B) Z_{\rm one-loop}^{\rm c.m.}(\sigma,B) \cJ(\sigma,B),
\end{align}
where the last factor includes the Vandermonde:
\begin{align}
Z_{\rm one-loop}^{\rm v.m.}(\sigma,B)\cJ(\sigma, B)&= \prod_{\alpha\in\Delta^+} (-1)^{\alpha\cdot B}\left[\left(\frac{\alpha\cdot B}{2r}\right)^2 + \left(\frac{\alpha\cdot\sigma}{r}\right)^2 \right],\cr
Z_{\rm one-loop}^{\rm c.m.}(\sigma,B)&=\prod_{w\in\cR}(-1)^{\frac{|w\cdot B|-w\cdot B}{2}}\frac{\Gamma\left( \frac{1}{2} +i w\cdot \sigma +\frac{|w\cdot B|}{2}\right)}{\Gamma\left( \frac12 - i w\cdot\sigma + \frac{|w\cdot B|}{2} \right)},
\end{align}
where $\Delta^+$ are the positive roots of $\mathfrak{g}$, and the vector and chiral multiplet 1-loop determinants $Z_{\rm one-loop}^{\rm v.m.}$ and $Z_{\rm one-loop}^{\rm c.m.}$ were copied from \cite{Doroud:2012xw}. In the second equation, the product goes over the weights of $\cR$, and we have taken into account that the gluing chiral multiplets (originating from the 3D hypers) have the R-charge 1.

Equation \eqref{gluingf1} is another example of a useful gluing formula. It was previously derived and successfully applied in \cite{Dedushenko:2017avn} as a tool to study Coulomb branches of $\cN=4$ gauge theories on $S^3$. The main non-trivial content of this formula is that the infinite-dimensional integration in \eqref{glueHS2A} has been replaced in \eqref{gluingf1} by a finite-dimensional integration-summation over the family of half-BPS boundary conditions of a certain kind.

The gluing formula \eqref{gluingf1} applies as long as the states $\psi_1$ and $\psi_2$ preserve the supercharge used for the boundary localization. As one can see by comparing our conventions with \cite{Doroud:2012xw}, this supercharge is $Q_1^+ + Q_2^-$ if written in terms of the linear combinations defined in \eqref{su2AQ}. Note that this is the same supercharge as $\cQ^C$ that was used in \cite{Dedushenko:2017avn} to study Coulomb branches on $S^3$. Most (and probably all) $\cQ^C$-invariant states $\psi_1,\psi_2\in\cH_{S^2}$ can be generated at the boundary of a hemisphere $HS^3$ with arbitrary (not necessarily local) $\cQ^C$-closed operators inserted inside the hemispheres. Then the gluing formula \eqref{gluingf1} applies and computes the $S^3$ partition function (with insertions) obtained by sewing two hemispheres.

\subsection{The $\mathfrak{su}(2|1)_B$-invariant polarization}\label{sec:su21B}
As noted in \cite{Dedushenko:2017avn}, the algebra $\mathfrak{su}(2|1)_\ell\oplus \mathfrak{su}(2|1)_r$ has an outer automorphism $\mathfrak{a}$ which leaves most of the generators unchanged, except that it flips the right R-charge, $\mathfrak{a}(R_r)=-R_r$, and correspondingly $\mathfrak{a}(\cQ_\alpha^{r\pm})=\cQ_\alpha^{r\mp}$. This is the $S^3$ version of the 3d mirror symmetry automorphism, and it maps the $\mathfrak{su}(2|1)_A$ subalgebra to a different subalgebra $\mathfrak{a}(\mathfrak{su}(2|1)_A)$, which can also be preserved in a theory on the hemisphere $HS^3$. One can identify $\mathfrak{a}(\mathfrak{su}(2|1)_A)$ as $\mathfrak{su}(2|1)_B$: while the R-symmetry entering $\mathfrak{su}(2|1)_A$ is $R_H$, the one in $\mathfrak{su}(2|1)_B$ is $R_C$. 

As pointed out in \cite{Doroud:2013pka}, the distinction between the two versions of $\cN=(2,2)$ SUSY on $S^2$ is somewhat formal: a theory of vector and chiral multiplets with $\mathfrak{su}(2|1)_B$ SUSY is isomorphic to the theory of twisted vector and twisted chiral multiplets based on $\mathfrak{su}(2|1)_A$ SUSY, and vice versa.  In fact, soon we will describe the $\mathfrak{su}(2|1)_B$-symmetric gluing theory as an $\mathfrak{su}(2|1)_A$-symmetric gluing theory of twisted multiplets. Nevertheless, to draw a clear distinctions between the two cases, we refer to the latter as the $\mathfrak{su}(2|1)_B$-invariant theory, as opposed to $\mathfrak{su}(2|1)_A$-invariant one from the previous subsection. In what follows, we are going to identify the $\mathfrak{su}(2|1)_B$-invariant polarization and the corresponding boundary multiplets. The supercharges in $\mathfrak{su}(2|1)_B$ are:
\begin{align}
\label{su21BQ}
Q_1^+&=\cQ_1^{\ell+} + \cQ_1^{r-},\quad Q_2^+=\cQ_2^{\ell+} - \cQ_2^{r-},\cr
Q_1^-&=\cQ_1^{\ell-} - \cQ_1^{r+},\quad Q_2^-=\cQ_2^{\ell-} + \cQ_2^{r+}.
\end{align}
The supersymmetries they generate on $S^2$ are parametrized in terms of the same $\epsilon$, $\bar\epsilon$ satisfying the same conformal Killing spinor equations \eqref{KiSpiS2} as in the $\mathfrak{su}(2|1)_A$ case. However, because the $\mathfrak{su}(2|1)_B\subset\mathfrak{su}(2|1)_\ell\oplus \mathfrak{su}(2|1)_r$ embedding is different from the $\mathfrak{su}(2|1)_A$ case, the structure of boundary multiplets that originate from the bulk vectors and hypers is going to be different.

We find the boundary multiplets as before, starting with one reference boundary condition and acting on it with the SUSY that we wish to preserve. In this way, we identify the following boundary fields:
\begin{align}
\label{bndrysuB}
Y &= q_1\big|, \quad \bar{Y}=\tilde{q}_2\big|,\cr
\zeta_\alpha &= \left( \begin{matrix}
\psi_{1\dot1}\cr
-\psi_{2\dot2}\cr
\end{matrix} \right)\Big|,\quad \bar\zeta_\alpha=\left( \begin{matrix}
-\tilde\psi_{1\dot2}\cr
-\tilde\psi_{2\dot1}\cr
\end{matrix} \right)\Big|,\cr
G&=(i\cD_\perp q_2 - i\Phi_{12}q_2)\big|,\quad \bar{G}=(-i\cD_\perp \tilde{q}_1 + i\tilde{q}_1\Phi_{12})\big|,\cr
\sigma&=-\Phi_{11}\big|, \quad \bar\sigma=\Phi_{22}\big|,\quad a=A_{\parallel}\big|,\cr
\eta_\alpha&=\left( \begin{matrix}
-\lambda_{11\dot1}\cr
\lambda_{22\dot1}\cr
\end{matrix} \right)\Big|,\quad \bar\eta_\alpha=\left( \begin{matrix}
\lambda_{12\dot2}\cr
\lambda_{21\dot2}\cr
\end{matrix} \right)\Big|,\cr
D_{2d}&=(D_{12}-i\cD_\perp \Phi_{\dot1\dot2})\big|.
\end{align}
Their SUSY transformations are given by:
\begin{align}
\delta\sigma&=\bar\epsilon \eta,\quad \delta\bar\sigma=\epsilon\bar\eta,\cr
\delta a_\mu&= \frac{i}2 (\epsilon\gamma_3\gamma_\mu \eta - \bar\epsilon\gamma_3\gamma_\mu\bar\eta ),\cr
\delta\eta&=i\slashed{D}(\sigma\epsilon) +\bar\epsilon(D_{\rm 2d}+iF)-\frac{i}2 \gamma_3\bar\epsilon [\sigma,\bar\sigma],\cr
\delta\bar\eta&=i\slashed{D}(\bar\sigma\bar\epsilon) + \epsilon(D_{\rm 2d}-iF)-\frac{i}2\gamma_3\epsilon[\sigma,\bar\sigma],\cr
\delta D_{\rm 2d}&=\frac{i}2\left(\cD_\mu(\epsilon\gamma^\mu\eta)-[\sigma,\epsilon\gamma_3\bar\eta]\right)+\frac{i}2\left(\cD_\mu(\bar\epsilon\gamma^\mu\bar\eta)+[\bar\sigma,\bar\epsilon\gamma_3\eta]\right),\cr
\end{align}
and:
\begin{align}
\delta Y&=(\bar\epsilon\gamma_- -\epsilon\gamma_+)\zeta,\quad\delta\bar{Y}=(\bar\epsilon\gamma_+-\epsilon\gamma_-)\bar\zeta,\cr
\delta\zeta_+&=-\gamma_+(i\slashed{D}Y-(G+\frac{i}{2r} Y))\bar\epsilon + i\gamma_+\epsilon \sigma Y,\cr
\delta\zeta_-&=\gamma_-(i\slashed{D}Y - (G+\frac{i}{2r}Y))\epsilon-i\gamma_-\bar\epsilon \bar\sigma Y,\cr
\delta\bar\zeta_+&=\gamma_+(i\slashed{D}\bar{Y}-(\bar{G}-\frac{i}{2r}\bar{Y}))\epsilon-i\gamma_+\bar\epsilon \bar{Y}\bar\sigma,\cr
\delta\bar\zeta_-&=-\gamma_-(i\slashed{D}\bar{Y}-(\bar{G}-\frac{i}{2r}\bar{Y}))\bar\epsilon + i\gamma_-\epsilon\bar{Y}\sigma,\cr
\delta G&=i\epsilon\gamma_-(\slashed{D}\zeta - Y\eta -\sigma \zeta) -i\bar\epsilon\gamma_+(\slashed{D}\zeta+Y\bar\eta-\bar\sigma\zeta)+\frac{i}{2r}(\epsilon\gamma_+-\bar\epsilon\gamma_-)\zeta,\cr
\delta \bar{G}&=i\epsilon\gamma_+(\slashed{D}\bar\zeta -\bar{Y}\eta-\sigma \bar\zeta) - i\bar\epsilon\gamma_-(\slashed{D}\bar\zeta + \bar{Y}\bar\eta-\bar\sigma\bar\zeta) + \frac{i}{2r}(\bar\epsilon\gamma_+-\epsilon\gamma_-)\bar\zeta.
\end{align}
By comparison with \cite{Gomis:2012wy,Doroud:2013pka}, one can easily recognize $(a_\mu, \sigma, \bar\sigma, \eta, \bar\eta, D_{\rm 2d})$ as a twisted vector multiplet and $(Y, \bar{Y}, \zeta, \bar\zeta, G, \bar{G})$ a twisted chiral multiplet of R-charge $q=1$ \cite{Gates:1983py,Gates:1984nk,Gates:1994gx}. Denoting them collectively as $\bB_B$, the gluing is represented as a path integral over $\pD \bB_B$ that has $\mathfrak{su}(2|1)_A$ supersymmetry (because we are describing the $\mathfrak{su}(2|1)_B$-invariant theory in terms of the mirror variables), and the next step is to localize it. 

Localization of twisted vectors that gauge arbitrary systems of twisted chirals on $S^2$ was discussed in \cite{Doroud:2013pka}. On the localization locus, most of the 2D fields vanish, with the exception of $Y$, $\bar{Y}$, which are constant on $S^2$, and the auxiliary fields, which take values $G=-\frac{i}{2r}Y$, $\bar{G}=\frac{i}{2r}\bar{Y}$. The values of $Y, \bar{Y}$ parametrize the following space:
\begin{equation}
\label{QHLocLoc}
\cM = \{Y|\, Y\bar{Y}=\chi\}/G,
\end{equation}
where $G$ is the gauge group, and $\chi$ is an arbitrary F.I. parameter for each $U(1)$ factor of $G$ (nothing depends on its precise value). The equation describing the localization locus can be also written, in a slightly more detailed form, as:
\begin{equation}
\bar{Y} T^A Y=\chi^A,
\end{equation}
where $T^A$ are generators of the gauge group, and $\chi^A$ are non-zero only for the $U(1)$ factors. This $\cM$ is of course the K\"ahler quotient:
\begin{equation}
\cM = \cR // G,
\end{equation}
where $\cR=\C^{|\cR|}$ is the complex representation of the gauge group in which $Y$ takes values.

Proceeding in this way, we arrive at another representation of gluing that is invariant under $\mathfrak{su}(2|1)_B$, and the localization on $S^2$ results in a new gluing formula. Denoting the boundary condition/state $|\bB_B\rangle$ restricted to the localization locus \eqref{QHLocLoc} as $|Y,\bar{Y}\rangle$, we obtain:
\begin{equation}
\label{QHgluingForm}
\langle\Psi_2|\Psi_1\rangle = \int_{\cM} {\rm Vol}_\cM \, Z_{1-\text{loop}}(Y,\bar{Y})\, \langle\Psi_2|Y,\bar{Y}\rangle \langle Y,\bar{Y}|\Psi_1\rangle,
\end{equation}
where ${\rm Vol}_\cM$ is a volume form on $\cM$, and $Z_{1-\text{loop}}$ is the one-loop determinant for the localization on $S^2$. The volume form ${\rm Vol}_\cM$ can be determined in two steps: first we localize on the locus of constant $Y, \bar{Y}$, which is simply $\C^{|\cR|}$ with the natural volume form
\begin{equation}
\frac{\dd^{N_f}Y\wedge \dd^{N_f}\bar{Y}}{{\rm Vol}(G_{\rm global})},
\end{equation}
where we mod out by constant gauge transformations. Then we observe that the localizing deformation of \cite{Doroud:2013pka} also includes a term $(Y\bar{Y} - \chi)^2$, thus $e^{-t (Y\bar{Y} - \chi)^2}$ (with the appropriate $t$-dependent prefactor coming from integrating out other fields) simply generates a delta-function in the $t\to\infty$ limit. Therefore, the measure can be written as:
\begin{equation}
\label{VolFormB}
{\rm Vol}_\cM = \frac{\dd^{N_f}Y\wedge \dd^{N_f}\bar{Y}}{{\rm Vol}(G_{\rm global})} \delta(Y\bar{Y} - \chi),
\end{equation}
where it is understood that integrals over $\cM$ should be written as integrals over $\C^{|\cR|}$ with this delta-functional measure ${\rm Vol}_\cM$.

It is not hard to specialize to the gauge group $U(1)^{N_c}$. If $Q_I^a$ are charges of chirals $Y_I$, $I=1,\dots, N_f$, the measure becomes:
\begin{equation}
{\rm Vol}_\cM = \frac{\dd^{N_f}Y\wedge \dd^{N_f}\bar{Y}}{(2\pi)^{N_c}} \prod_{a=1}^{N_c} \delta(F_a),\quad \text{where } F_a = \sum_{I=1}^{N_f} Q_I^a|Y_I|^2 - \chi_a.
\end{equation}

The authors of \cite{Doroud:2013pka} also evaluate the 1-loop determinant for the gauge group $U(1)^{N_c}$. They introduce matrices:
\begin{equation}
M_I{}^a = Q_I^a Y_I,\quad M_a^\dagger{}^I=Q_I^a \bar{Y}_I,
\end{equation}
and find that
\begin{equation}
\label{OneLoopB}
Z_{1-\text{loop}}=\det(M^\dagger M).
\end{equation}

The new gluing formula \eqref{QHgluingForm} holds as long as the boundary states $\Psi_1$, $\Psi_2$ preserve the supercharge used for the localization. As can be found in \cite{Doroud:2013pka}, this supercharge, written in terms of linear combinations defined in \eqref{su21BQ}, is $Q_1^+ + Q_2^-$. This also coincides with the Higgs branch supercharge $\cQ^H$ used in \cite{Dedushenko:2016jxl} to study quantized Higgs branches on $S^3$.

Notice that the 2D localization allows to insert arbitrary functions of $Y$ at the North pole of $S^2$ and arbitrary functions of $\bar{Y}$ at the South pole of $S^2$, which then enter the localized integral over $\cM$ in the obvious way. These are $q_1(0)=Q(0)$ and $\tilde{q}_2(\pi)=\tilde{Q}(\pi)$ of \cite{Dedushenko:2016jxl}, which in fact belong to the cohomology of $\cQ^H$. From the 3D point of view, for example on the hemisphere $HS^3$, one can also insert $\tilde{q}_1(0)=\tilde{Q}(0)$ at the north pole of the boundary $S^2$ and $q_2(\pi)=Q(\pi)$ at the south pole of the boundary, simply because they preserve the localizing supercharge $\cQ^H$. From the 2D point of view, these fields do not belong to the boundary multiplets defined in \eqref{bndrysuB} (so their boundary values are not fixed by the boundary conditions), so their action on the boundary state is not obvious. However, from the 3D point of view, it is not only obvious that we are allowed to insert these fields, but it is also easy to determine how they act on the boundary states. From \cite{Dedushenko:2016jxl}, one can deduce the commutator $[\tilde{q}_1(0), q_1(0)]=-\frac1{4\pi r}$, which implies that because $q_1(0)$ acts as a multiplication by $Y$, $\tilde{q}_1$ should act as $-\frac1{4\pi r}\frac{\partial}{\partial Y} + (\dots)$. In Section \ref{sec:freehyp} we will find that it simply acts by  $-\frac1{4\pi r}\frac{\partial}{\partial Y}$. Analogously, we find that because $\tilde{q}_2$ is a multiplication by $\bar{Y}$, $q_2$ should act as $-\frac1{4\pi r}\frac{\partial}{\partial\bar Y}$. We will return to this point shortly.

%It would be very interesting to find the role of gluing formula \eqref{QHgluingForm} in the study of mirror symmetry of 3D $\cN=4$ theories. At the moment, the cutting and gluing technique has found its role only in the description of Coulomb branches in \cite{Dedushenko:2017avn}, while the Higgs branch side of \cite{Dedushenko:2016jxl} is described in a completely different way. Presumably, the gluing formula \eqref{QHgluingForm} could provide new insights into this relation.

\subsection{Arbitrary supersymmetric boundary conditions}\label{sec:bndry3DN4}
We can use the $\mathfrak{su}(2|1)_A$ or $\mathfrak{su}(2|1)_B$ invariant gluing formulas to classify arbitrary boundary conditions that preserve these algebras (or, more generally, preserve $\cQ^C$ or $\cQ^H$). The idea is straightforward: a boundary condition $\cB$ is completely characterized by the state $\langle\cB|$ it creates, and we can impose $\cB$ simply by gluing $\langle\cB|$ to the boundary. Furthermore, the gluing formulas imply that $\cQ^C$-invariant boundary conditions $\cB_A$ are completely characterized, up to $\cQ^C$-exact terms, by their ``wave functions'':
\begin{eqnarray}
Z_{\cB_A}(\sigma,B) := \langle\cB_A|\sigma, B\rangle,
\end{eqnarray}
and $\cQ^H$-invariant boundary conditions $\cB_B$ are similarly characterized, up to $\cQ^H$-exact terms, by their ``wave functions'':
\begin{equation}
Z_{\cB_B}(Y, \bar{Y}):=\langle \cB_B|Y, \bar{Y}\rangle.
\end{equation}
When we say ``up to $\cQ$-exact terms,'' we mean that the corresponding ``wave function'' determines the $\cQ$-cohomology class of the physical state created by the boundary condition. This also means that, as long as the bulk theory preserves the corresponding $\cQ$, the function $Z_\cB$ is all we need to know about the boundary condition in order to compute correlation functions of any $\cQ$-closed observables. As a reminder of the gluing formulas, note that if the bulk dynamics creates at $S^2$ a boundary state $|\Psi\rangle$ annihilated either by $\cQ^C$ or $\cQ^H$, then imposing an $\mathfrak{su}(2|1)_A$ or $\mathfrak{su}(2|1)_B$ invariant boundary condition respectively is described by:
\begin{align}
\langle \cB_A|\Psi_+\rangle &= \sum_{B\in \Lambda_{\rm cochar}}\frac1{|\cW(H_B)|}\int_{\mathfrak{t}}\dd\sigma\,\mu(\sigma,B)\, Z_{\cB_A}(\sigma,B) \langle\sigma,B|\Psi_+\rangle,\cr
\langle \cB_B|\Psi_+\rangle &= \int_{\cM}{\rm Vol}_\cM\, Z_{\rm 1-loop}(Y,\bar{Y})Z_{\cB_B}(Y,\bar{Y}) \langle Y, \bar{Y}|\Psi_+\rangle.
\end{align}

\begin{figure}[t!]
	\centering
	\includegraphics[width=0.5\textwidth]{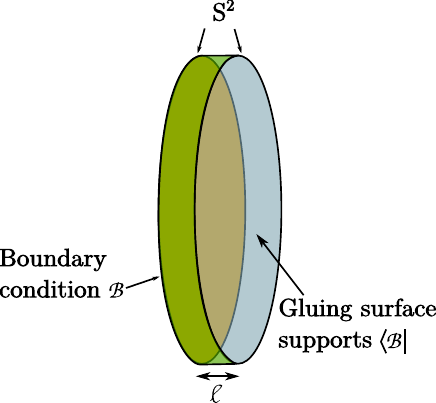}
	\caption{\label{fig:bndry} Imposing a boundary conditions $\cB$ can be thought of as gluing in a cylinder $S^2 \times I$ of length $\ell\to 0$ supporting a state $\langle\cB|$ at the gluing surface that was created by the boundary condition $\cB$ imposed at the opposite end of $I$. The boundary condition is characterized by $Z_\cB$, the $\ell\to 0$ limit of the partition function on this cylinder.}
\end{figure}

To be slightly more pedantic, since we want to impose a single boundary condition $\cB$, while gluing involves integration over families of boundary conditions, it is convenient to start with a cylinder $S^2\times I$, where $I=(0,\ell)$ is an interval of length $\ell$. We impose the boundary condition $\cB$ at $S^2\times \{0\}$, while the other boundary component, $S^2\times \{\ell\}$, serves as the ``gluing surface'' -- we glue it to the $S^2$ boundary of the three-manifold we are working with, such as the hemisphere $HS^3$ from the previous subsections. See Figure \ref{fig:bndry} for the illustration. In the limit $\ell\to 0$, we can think of this as simply imposing the boundary condition $\cB$ at the boundary of $HS^3$. If the boundary condition $\cB$ preserves $\mathfrak{su}(2|1)_A$ (or, more generally, just $\cQ^C$), we can use the $\mathfrak{su}(2|1)_A$-invariant polarization for gluing, and the boundary condition at the gluing surface is parametrized by $(\sigma, B)$. It makes obvious that $Z_\cB(\sigma, B)$ is simply the partition function on $S^2\times I$ in the limit $\ell\to 0$, with $\cB$ imposed at $S^2\times \{0\}$ and $(\sigma, B)$ imposed at $S^2\times \{\ell\}$. Similarly, if $\cB$ preserves $\mathfrak{su}(2|1)_B$ (or, more generally, just $\cQ^H$ ), we use the $\mathfrak{su}(2|1)_B$-invariant polarization for gluing. Then the boundary condition at the gluing surface $S^2\times \{\ell\}$ is parametrized by $(Y, \bar{Y})$, while the boundary condition $\cB$ is imposed at $S^2\times \{0\}$, so the $\ell\to 0$ limit of the $S^2\times I$ partition function in this case determines $Z_\cB(Y,\bar{Y})$.

\paragraph{Coupling to a boundary theory.} Quite a general class of boundary conditions can be described by coupling the bulk fields that can fluctuate at the boundary to some boundary theory. Suppose that $\cB$ is of this type: first we impose Neumann boundary conditions on the gauge fields (completed to the half-BPS Neumann boundary conditions on the vector multiplet), we also impose some boundary conditions on the hypers, and then we couple the unconstrained fields to some $\mathfrak{su}(2|1)$-invariant boundary theory $T$. In particular, the boundary values of the bulk vector multiplet gauge some global symmetry of $T$ at the boundary. 

We can couple $T$ to the 3D bulk either in an $\mathfrak{su}(2|1)_A$ or $\mathfrak{su}(2|1)_B$ invariant way, identifying the $\mathfrak{su}(2|1)$ SUSY of $T$ either with $\mathfrak{su}(2|1)_A$ or $\mathfrak{su}(2|1)_B$ preserved at the boundary. Let us assume that $T$ has the flavor symmetry $G_F$ acting on chiral multiplets -- we refer to it as the ``chiral'' flavor symmetry as opposed to the ``twisted chiral'' flavor symmetry acting on twisted chirals. When we describe the $\mathfrak{su}(2|1)_A$ invariant coupling, the 3D vector and hyper multiplets give rise to background 2D vector and chiral multiplets at the boundary (as follows from Section \ref{sec:su21A}). They couple to $T$, in particular background vectors gauge the subgroup $G\subset G_F$ of the chiral flavor symmetry, as well as enter the twisted superpotential (through their field strength superfield); background chirals can appear in the superpotential. For the $\mathfrak{su}(2|1)_B$-invariant coupling, the 3D vectors and hypers also give background vector and chiral multiplets at the boundary (if we use the $\mathfrak{su}(2|1)_B$ description!) that can couple to $T$ by gauging $G\subset G_F$ and through the superpotentials. Since we use the $\mathfrak{su}(2|1)_A$ language for the $B$ case in Section \ref{sec:su21B}, the boundary background multiplets are actually twisted vectors and twisted chirals, and they should be coupled to $\tilde{T}$, the mirror dual of $T$. This $\tilde{T}$ has a twisted chiral flavor symmetry $G_F$, and $G\subset G_F$ is gauged by the background twisted vectors. Note also that when we couple $\tilde{T}$, the regular and twisted superpotentials are the twisted and the regular ones of $T$. If any of the readers got confused by this discussion, we want to reassure them that we are going to provide more details in the next subsection.

When we take the $\ell\to 0$ limit, some modes in the bulk of $S^2\times (0,\ell)$ become ``frozen'', while some might survive as ``trapped'' effectively two-dimensional degrees of freedom. This of course depends on the boundary conditions: if $S^2\times \{0\}$ and $S^2\times \{\ell\}$ support ``complementary'' or ``transversal'' boundary conditions (which are parts of transversal polarizations in the sense of \cite{glue1}), then no degrees of freedom survive the $\ell\to0$ limit. The basic example is when every field has a Dirichlet boundary condition on one end and Neumann on another: in this case, boundary conditions completely fix the background value for this field on the cylinder, while fluctuations require energy $\frac1{\ell}$ and hence are suppressed (or ``frozen''). If both ends support Dirichlet boundary conditions, they are not transversal any more, however the freezing property still holds: no degrees of freedom survive the $\ell\to0$ limit. On the other hand, if we have Neumann boundary conditions on both ends, the modes that are constant in the $(0,\ell)$ direction survive and become effective ``trapped'' 2D degrees of freedom in the $\ell\to0$ limit.

In the next section, in the discussion of 4D $\cN=2$ theories and their domain walls, we will have more to say about such ``trapped'' modes. In particular, the boundary conditions ``wave function'' $Z_\cB$ is related to the boundary theory partition function $Z_T$ and the trapped modes partition function $Z_{\rm trapped}$ through: 
\begin{equation}
Z_\cB = Z_T \times Z_{\rm trapped}.
\end{equation}

For now, let us assume that the boundary conditions at $S^2\times \{0\}$ and $S^2\times \{\ell\}$ are transversal (or complementary), so all bulk degrees of freedom are frozen in the $\ell\to0$ limit. In the $\mathfrak{su}(2|1)_A$ case, the hypermultiplet fields on $S^2\times I$ vanish in the $\ell\to 0$ limit, while the vector multiplet is frozen to the background characterized by a constant scalar vev $\sigma$ and a constant magnetic flux $B$. From the point of view of the boundary theory $T$, these are vevs of the background vector multiplet (gauging $G$ subgroup of the flavor symmetry): $\sigma$ plays a role of the twisted mass, while $B$ is the background flux. In the $\mathfrak{su}(2|1)_B$ case, the vector multiplet fields vanish in the $\ell\to 0$ limit, while the hypers acquire constant background values given by $Y$ and $\bar{Y}$. The boundary theory $\tilde{T}$ feels them as the background twisted chiral superfields with vevs $Y, \bar{Y}$, which can only enter its twisted superpotential.

Therefore $Z_\cB$, the $\ell\to 0$ limit of the $S^2\times I$ partition function, is simply the partition function of $T$ on $S^2$ in the appropriate background. In the $\mathfrak{su}(2|1)_A$ case, this is $T$ coupled to the background flux $B$ and the twisted masses $\sigma$:
\begin{equation}
Z_\cB(\sigma, B)=Z_T(\sigma, B).
\end{equation}
In the $\mathfrak{su}(2|1)_B$ case, this is $\tilde{T}$, the mirror of $T$, coupled to the background twisted chiral superfields:
\begin{equation}
Z_\cB(Y,\bar{Y})=Z_{\tilde T}(Y, \bar{Y}).
\end{equation}

\subsubsection{Connection to Mirror Symmetry and Symplectic Duality}\label{sec:Mirror3d}
Some readers might have noticed parallels with the work of \cite{Bullimore:2016nji}, so let us make them more precise.

Recall from \cite{Dedushenko:2016jxl,Dedushenko:2017avn} that on $S^3$, the $\cQ^H$--cohomology of local operators is an associative algebra $\cA_H$ quantizing the Higgs branch chiral ring $\C[\cM_H]$. Similarly, the $\cQ^C$ cohomology forms $\cA_C$, quantization of the Coulomb branch chiral ring $\C[\cM_C]$. On the hemisphere, if the boundary condition preserves $\cQ^H$, one can bring bulk operators to the boundary, therefore acting with $\cA_H$ on the boundary condition and defining an $\cA_H$--module. This module is described by how $\cA_H$ acts on the boundary condition wave function $Z_\cB(Y, \bar{Y})$. Analogously, the $\cQ^C$-invariant boundary condition encoded in $Z_\cB(\sigma, B)$ determines an $\cA_C$--module. This is very similar to \cite{Bullimore:2016nji}, with the difference that in their case, the quantization was achieved by placing a theory in the Omega-background, while in our case it is done by putting it on $S^3$.
\begin{figure}[t!]
	\centering
	\includegraphics[width=0.5\textwidth]{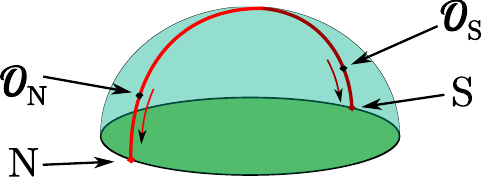}
	\caption{\label{fig:module} Bringing operators in the $\cQ^C$ (or $\cQ^H$) cohomology to the North and South poles of the boundary defines the action of $\cA_C\otimes\cA_C^{\rm op}$ (or $\cA_H\otimes\cA_H^{\rm op}$) on boundary conditions.}
\end{figure}

However, if we look closer, the actual picture is slightly richer. Operators in the $\cQ^H$ or $\cQ^C$ cohomology can be inserted along the great circle $S^1\subset S^3$ \cite{Dedushenko:2016jxl,Dedushenko:2017avn}, which on the hemisphere $HS^3$ becomes a great semicircle normal to the boundary. It meets boundary at two antipodal points -- the North and the South poles. Hence every bulk operator can act on the boundary condition in two distinct ways: through the North and the South pole actions, see Figure \ref{fig:module}. Since going around the semicircle reverses the order in which operators hit the boundary, the second action is in fact opposite to the first. Therefore, the complete correspondence looks as follows:
\begin{align}
\cQ^C\text{-invariant boudnary condition }\cB_C,\, Z_{\cB_C}(\sigma, B) &\longrightarrow \cA_C\otimes \cA_C^{\rm op}-\text{module},\cr
\cQ^H\text{-invariant boudnary condition }\cB_H,\, Z_{\cB_H}(Y, \bar{Y}) &\longrightarrow \cA_H\otimes \cA_H^{\rm op}-\text{module}.
\end{align}

One can contrast this to \cite{Bullimore:2016nji}, where the $(2,2)$-supersymmetric boundary conditions of 3D $\cN=4$ theories were studied in flat space (with Omega-deformation), as opposed to our $S^2$ boundaries. In their case, the boundary conditions had the structure of one-sided $\cA_H$ and $\cA_C$ modules. We would recover such a structure by forgetting the $\cA^{\rm op}$--action, i.e. only focusing on the operators hitting the North pole on Figure \ref{fig:module}. We explicitly observe the bimodule structure later in an example.

Besides building modules, \cite{Bullimore:2016nji} also describe a construction of pairs of $\cA_H$ and $\cA_C$ modules which are proposed to give a physical realization of the symplectic duality. They consider a $(2,2)$--preserving boundary condition $\cB$ in the UV theory and study its IR images on the Higgs and Coulomb branches as a pair $(\cB_H,\cB_C)$ of the $\cA_H$--module $\cB_H$ and the $\cA_C$--module $\cB_C$. Can we do the same? The naive answer would be no, simply because to describe $\cA_H$-modules, we need to preserve $\cQ^H$, while $\cA_C$--modules are described by $\cQ^C$--preserving boundary conditions. However, the anticommutator $\{\cQ^H, \cQ^C\}$ involves isometries of $S^3$ that do not preserve the boundary of $HS^3$, therefore $\cQ^H$ and $\cQ^C$ cannot be preserved simultaneously. This seems to imply that we cannot generate pairs of modules from a single UV boundary condition in our case.

However, a closer look reveals that we can, and our discussion in the previous subsection on how to couple the boundary theory $T$ to the bulk was a preparation for this. The first point we need to make is that the flat space limit (radius $r\to\infty$) of $\cQ^C$--invariant boundary conditions \eqref{bndryS2} is related the the flat space limit of $\cQ^H$--invariant boundary conditions from Section \ref{sec:su21B} by an $SU(2)_H\times SU(2)_C$ R-symmetry rotation. If we were studying the $\cQ^H$ case using the $\mathfrak{su}(2|1)_B$ multiplets, this would also include a 3D mirror symmetry operation, but since we use the $\mathfrak{su}(2|1)_A$ language, it is a simple R-symmetry rotation. For completeness, we provide the matrices of such a rotation in our conventions:
\begin{align}
U_a{}^b=\left(\begin{matrix}
\frac1{\sqrt 2} & \frac{i}{\sqrt 2}\\
-\frac{i}{\sqrt 2}& -\frac{1}{\sqrt 2}
\end{matrix} \right),\quad V_{\dot a}{}^{\dot b}=\left(\begin{matrix}
\frac1{\sqrt 2} & -\frac{1}{\sqrt 2}\\
\frac{1}{\sqrt 2}& \frac{1}{\sqrt 2}
\end{matrix} \right).
\end{align}
This rotation is available in SCFT or in flat space, but not in a non-conformal theory on $S^3$. If we think of our $\cQ^C$ and $\cQ^H$ preserving boundary conditions as deformations of the flat space ones (with the deformation parameter $r^{-1}$), it gives a natural prescription of how ``the same'' boundary conditions can be imposed in a $\cQ^C$ or $\cQ^H$ invariant way. Indeed, if we have some $\cQ^C$--preserving boundary condition at $S^2$, we can take its $r\to\infty$ limit, and by performing the R-symmetry rotation obtain an $r\to\infty$ limit of some $\cQ^H$--preserving boundary condition. One could worry if it is possible to recover the finite-$r$ version of the latter, but our system is rigid enough due to SUSY which ensures positive answer to such a concern.

This suggests that boundary conditions can be imposed both in $\cQ^C$--invariant and in $\cQ^H$--invariant fashion: this clearly holds for Neumann, Dirichlet, and exceptional Dirichlet of \cite{Bullimore:2016nji}. It is also not hard to see what to do with a more general boundary theory $T$. If we have a $\cQ^C$--preserving boundary condition described by the 2D theory $T$ coupled (by gaugings and superpotentials) to the $\cQ^C$--invariant Neumann boundary of the 3D theory, we construct its $\cQ^H$--invariant counterpart by coupling $\tilde{T}$, the 2D mirror of $T$, to the $\cQ^H$--invariant Neumann boundary. Note that this can be successfully done only when both $U(1)_V$ and $U(1)_A$ are preserved at the boundary, simply because $U(1)_V$ and $U(1)_A$ are parts of $\mathfrak{su}(2|1)_A$ and $\mathfrak{su}(2|1)_B$ respectively. This manifests itself on the $\cQ^H$ side as follows: in the $\mathfrak{su}(2|1)_B$ language, $U(1)_A$ is the R-symmetry on $S^2$ that is part of $\mathfrak{su}(2|1)_B$, and one should impose anomaly cancellation to make the SUSY background consistent. In the $\mathfrak{su}(2|1)_A$ language, the same anomaly cancellation condition (but for $U(1)_V$) appears in terms of charges of the twisted chirals. The authors of \cite{Bullimore:2016nji} were also mostly interested in $U(1)_V\times U(1)_A$--preserving boundary conditions.

To provide slightly more details, notice that when we are on the $\cQ^C$ side, the 2D boundary superpotential $\cW$ is Q-exact and only affects the answer implicitly by determining the R-charges of boundary degrees of freedom. On the other hand, the twisted superpotential $W$ explicitly enters the 2D localization answer \cite{Doroud:2012xw,Doroud:2013pka}. As we switch to the $\cQ^H$ side, the superpotential $\cW$ of $T$ becomes the twisted superpotential of $\tilde{T}$ and now explicitly enters the answer, while $W$ becomes the superpotential and only affects the answer implicitly.

So indeed we find that our construction provides the hemisphere version of the story in \cite{Bullimore:2016nji}, with the main difference being that the boundary conditions give pairs of $\cA_C\otimes\cA_C^{\rm op}$ and $\cA_H\otimes\cA_H^{\rm op}$ modules now. This extra structure immediately raises the question: what are the Hochschild homologies of $\cA_C$ and $\cA_H$ with coefficients in such bimodules?

We can describe the Hochschild complex quite explicitly. If $\cB_H$ is an $\cA_H$--module associated to the boundary condition $\cB$, then the zeroth degree chains are simply the module itself, $C_0(\cA_H, \cB_H)=\cB_H$. We denote elements of $C_0(\cA_H, \cB_H)$ pictorially by a black semicircle (representing the semicircle where we can insert local operators) ending on a red interval representing the boundary condition:
\begin{equation}
	\includegraphics[width=0.08\textwidth]{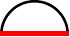}\in C_0(\cA_H, \cB_H).
\end{equation}
The $n$-th degree chains are given by $C_n(\cA_H, \cB_H)=\cB_H\otimes \cA_H^{\otimes n}$, which clearly has the meaning of inserting $n$ ordered observables $\cO_1,\dots,\cO_n\in\cA_H$ on the semicircle. We represent them pictorially by red dots. For example, the elements of $C_3(\cA_H, \cB_H)$ are depicted as:
\begin{equation}
\includegraphics[width=0.08\textwidth]{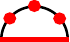}\in C_3(\cA_H, \cB_H).
\end{equation}
The boundary operator is defined by colliding insertions with each other and with the boundary:
\begin{align}
\partial:\quad & \includegraphics[width=0.08\textwidth]{C0}\mapsto 0,\cr
		  & \includegraphics[width=0.08\textwidth]{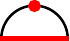}\mapsto \includegraphics[width=0.08\textwidth]{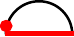}-\includegraphics[width=0.08\textwidth]{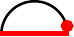}\cr
		  & \includegraphics[width=0.08\textwidth]{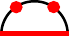}\mapsto \includegraphics[width=0.08\textwidth]{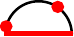}-\includegraphics[width=0.08\textwidth]{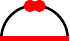}+\includegraphics[width=0.08\textwidth]{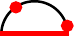}\cr
		  & \includegraphics[width=0.08\textwidth]{C3}\mapsto \includegraphics[width=0.08\textwidth]{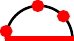}-\includegraphics[width=0.08\textwidth]{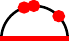}+\includegraphics[width=0.08\textwidth]{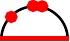}-\includegraphics[width=0.08\textwidth]{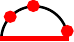}\cr
		  &\text{etc.}
\end{align}
The homology of this complex is $HH_n(\cA_H, \cB_H)$. The $\cA_C$ case is defined analogously.

It would be interesting to understand what property of boundary conditions is captured by the corresponding Hochschild homology. More generally, a further study of the $\cA_C$ and $\cA_H$ bimodules generated by boundary conditions, and in particular understanding their role in symplectic duality, furnishes an interesting direction for further inquiries. For now we will only describe the simplest example of a basic mirror-dual pair.

\subsubsection{An Example}\label{sec:freehyp}
Let us consider the simplest example of a mirror dual pair, in which case we can explicitly see the structure of modules generated by boundary conditions. On one side of duality, we will have a free hyper, on the other side -- a $U(1)$ gauge theory with $N_f=1$ hyper of charge $1$.

Let us start with the free hyper. The Coulomb branch is trivial (consists of a single point), the Coulomb branch chiral ring coincides with its quantization $\cA_C=\C$ and only contains a unit operator. The $\mathfrak{su}(2|1)_A$-invariant polarization described above does not have parameters $\sigma, B$ (because there are no vector multiplets), thus the wave function characterizing the $\cQ^C$-preserving boundary condition is a single number $\psi\in\C$. This boundary condition is a basic Dirichlet boundary condition from \eqref{bndryS2} (with only one hypermultiplet present), and it generates a one-dimensional module $\C$ over $\cA_C=\C$.

The Higgs branch is less trivial: it is equal to $\C^2$, the corresponding chiral ring is $\C[Q,\tilde{Q}]$, and its quantization $\cA_H$ is generated by $Q, \tilde{Q}$ with the relation $[\tilde{Q}, Q]=-\epsilon$, where $\epsilon=\frac1{4\pi r}$ in terms of the radius of the sphere. The $\mathfrak{su}(2|1)_B$-preserving boundary condition is characterized by a wave function $Z(Y,\bar{Y})$, where $Y$ is a complex variable. This $Z(Y,\bar{Y})$ generates an $\cA_H$-bimodule (or $\cA_H\otimes\cA_H^{\rm op}$-module), with the left and right actions of $\cA_H$ given by:
\begin{align}
\label{leftright_Q}
\text{Left action: } Q_L&=Y,\quad \tilde{Q}_L=-\epsilon\overrightarrow{\frac{\partial}{\partial Y}},\cr
\text{Right action: } \tilde{Q}_R&=\bar{Y},\quad Q_R=-\epsilon\overleftarrow{\frac{\partial}{\partial\bar{Y}}},
\end{align}
and we will comment on how this action is determined shortly.

There exists a distinguished boundary condition whose wave function corresponds to the vacuum state of the free hyper SCFT: it is a state generated at the boundary of the empty hemisphere. We call it the vacuum boundary condition. If we have a theory defined on some three-manifold with boundary $S^2$ (say, on a semi-infinite cylinder $S^2\times \R_+$), then imposing the vacuum boundary condition is simply done by gluing an empty hemisphere to this $S^2$. At the level of equations, if we have some boundary state $\Psi$ living on $S^2$, imposing the vacuum boundary conditions is expressed as an overlap:
\begin{equation}
\langle 0|\Psi\rangle = \int \dd Y \dd \bar{Y} Z_0(Y,\bar{Y}) \langle Y,\bar{Y}|\Psi\rangle.
\end{equation}
Here $Z_0(Y,\bar{Y})$ is the empty hemisphere partition function of the free hyper with the $\mathfrak{su}(2|1)_B$-preserving boundary conditions: $q_1\big|=Y$, $\partial_\perp q_2\big|=-\frac{1}{2r}Y$ (plus boundary conditions for the fermions that follow from \eqref{bndrysuB}). It is easily computed to be:
\begin{equation}
Z_0(Y,\bar{Y}) = \sqrt{4r} e^{-4\pi r Y\bar{Y}}=\sqrt{\frac1{\pi\epsilon}}e^{-\frac1{\epsilon}Y\bar{Y}},
\end{equation}
where the normalization is fixed so that the full sphere partition function of the free hyper is $\int\dd Y\dd\bar{Y} (Z_0(Y,\bar{Y}))^2=\frac12$, as expected. 

We can now comment on how \eqref{leftright_Q} was determined. From the boundary localization, we know that $Q_L=Y$ and $\tilde{Q}_R=\bar{Y}$. If we also include requirements that $[\tilde{Q}_L,Q_L]=[\tilde{Q}_R, Q_R]=-\epsilon$, that left and right action commute with each other, and that on the vacuum $Z_0(Y,\bar{Y})$ the left and right actions coincide:
\begin{align}
\label{vac_condition}
Q_L Z_0(Y,\bar{Y})&=Q_R Z_0(Y,\bar{Y}),\cr
\tilde{Q}_L Z_0(Y,\bar{Y})&=\tilde{Q}_R Z_0(Y,\bar{Y}),
\end{align}
this uniquely determines the operators to be as in \eqref{leftright_Q}. The requirement \eqref{vac_condition} follows from the argument as in \cite{Dedushenko:2016jxl,Dedushenko:2017avn} (which has also appeared in many other places, including \cite{Bullimore:2016nji}). We can move $Q$ or $\tilde{Q}$ through the empty hemisphere from one pole of the boundary to the other without ever colliding with other observables; because such operation is $\cQ^H$--exact, the answer should remain the same. This means that left and right actions should agree on the vacuum wave function.

By acting with $Q, \tilde{Q}$, either through the left or the right action, we see that the general element of the module takes the form:
\begin{equation}
\sqrt{\frac1{\pi\epsilon}}P(Y,\bar{Y})e^{-\frac1{\epsilon}Y\bar{Y}},
\end{equation}
where $P(Y,\bar{Y})$ is some polynomial.

Let us now look at the dual $U(1)$ gauge theory with $N_f=1$ charge-one hyper. This theory has a trivial Higgs branch (a single point), and the Higgs branch chiral ring coincides with its own quantization, $\cA_H=\C$. Indeed, the $\mathfrak{su}(2|1)_B$-invariant boundary conditions would be parametrized by:
\begin{equation}
\{Y\bar{Y} = \chi\}/U(1),
\end{equation}
with some $\chi>0$, which is just a single point. Thus the $\mathfrak{su}(2|1)_B$-invariant boundary conditions do not have any parameters, and the corresponding wave function is just a number $\psi\in\C$, which lives in a one-dimensional module over $\cA_H=\C$.

The Coulomb branch of $U(1)$ with $N_f=1$ can be found in \cite{Dedushenko:2017avn,Bullimore:2015lsa}. First of all, it is described by the equation $XY=Z$, so as a variety it is simply $\C^2$. The chiral ring is $\C[X,Y]$, with its quantization determined by $[Y,X]=\epsilon$. The $\mathfrak{su}(2|1)_A$-preserving boundary condition is characterized by the wave function $Z(\sigma,B)$, where $\sigma\in\R$ is the Coulomb branch parameter and $B\in\Z$ is the magnetic flux. This $Z(B,\sigma)$ generates the $\cA_C$-bimodule, with the left and right actions given by:
\begin{align}
\text{Left action: } X_L&=-\frac1{\sqrt{4\pi r}}\left( \frac12 + i\sigma - \frac{B}2 \right)\overrightarrow{e^{-\frac{i}{2}\partial_\sigma - \partial_B}},\quad Y_L=\frac1{\sqrt {4\pi r}} \overrightarrow{e^{\frac{i}{2}\partial_\sigma + \partial_B}},\cr
\text{Right action: } X_R&=\frac1{\sqrt{4\pi r}} \overleftarrow{e^{\frac{i}{2}\partial_\sigma - \partial_B}},\quad Y_R=-\frac1{\sqrt{4\pi r}}\overleftarrow{e^{-\frac{i}{2}\partial_\sigma + \partial_B}}\left( \frac12 + i\sigma + \frac{B}2 \right),
\end{align}
where by $e^{a\partial_x}$ we mean a shift operator acting as $e^{a\partial_x}f(x)=f(x+a)$. The distinguished $\mathfrak{su}(2|1)_A$--invariant boundary condition corresponds to the vacuum wave function, which is computed by the empty hemisphere partition function with the boundary conditions \eqref{bndryS2} (that depend on $\sigma$ and $B$). It was evaluated in \cite{Dedushenko:2017avn} and is given by:
\begin{equation}
Z_0(\sigma,B)=\delta_{B,0}\frac1{\sqrt{2\pi}}\Gamma\left(\frac12-i\sigma\right).
\end{equation}
The normalization is such that the full sphere partition function, equal in this case to $\int\dd\sigma\,\mu(\sigma,B=0)(Z_0(\sigma,0))^2$ with $\mu(\sigma,0)=\frac{\Gamma\left(\frac12 + i\sigma\right)}{\Gamma\left(\frac12 - i\sigma\right)}$ (see \cite{Dedushenko:2017avn}), is $\frac12$ again. By acting either with $X_L$, $Y_L$ or $X_R$, $Y_R$ (or both), we find that all elements of the module built on $Z_0(\sigma, B)$ take the following form:
\begin{align}
\sum_{n\in 2\Z}\delta_{B,n} P_n(\sigma)\Gamma\left(\frac12 - i\sigma\right) + \sum_{n\in 1 + 2\Z}\delta_{B, n} P_n(\sigma)\Gamma(1-i\sigma),
\end{align}
where $P_n(\sigma)$ are polynomials, and each sum has only finitely many non-zero terms.

We see that the two sides of duality look very much alike: the algebras match precisely, and only the vacuum modules do not look completely identical. While the free hyper Higgs vacuum module is generated by differential operators acting on $e^{-4\pi r Y\bar{Y}}$, the $U(1), N_f=1$ Coulomb branch vacuum module is generated by shift operators acting on $\Gamma(1/2-i\sigma)$. As it turns out, there exists a simple integral transform that explicitly establishes their isomorphism. It is known under the name of Fourier-Mellin transform (because it is Fourier in the angular direction and Mellin in the radial direction), and is defined by: 
\begin{equation}
\cM[f(r,\theta)](u,v)=\frac1{2\pi}\int_0^\infty \dd r\, r^{-1-iu}\int_0^{2\pi}\dd\theta\,f(r,\theta)e^{-iv\theta}.
\end{equation}
By adjusting normalization and changing variables, we write it in the form more appropriate for our application:
\begin{equation}
\label{int_tr}
\cM_{\sigma,B}[\Psi(Y,\bar{Y})]\equiv\sqrt{\frac{2r}{\pi}}\int \dd Y\dd\bar{Y} (4\pi r Y\bar{Y})^{-\frac12 -i\sigma} e^{i B\varphi} \Psi(Y,\bar{Y}),
\end{equation}
where $\varphi=\arg(Y)$. We claim that:
\begin{equation}
\Psi(\sigma,B)=\cM_{\sigma,B}[\Psi(Y,\bar{Y})].
\end{equation}
Indeed, it is easy to check that:
\begin{align}
\label{mod_isom}
\cM_{\sigma,B}\left[ Y^k \bar{Y}^{k+n}\sqrt{4r} e^{-4\pi r Y\bar{Y}} \right]=\delta_{B,n}\frac1{\sqrt{2\pi}} \frac1{(4\pi r)^{k+\frac{n}{2}}}\Gamma\left(k+\frac{n}{2}+\frac12 - i\sigma\right),
\end{align}
which in particular transforms vacua $(\pi \epsilon)^{-1/2}e^{-Y\bar{Y}/\epsilon}$ and $(2\pi)^{-1/2}\Gamma(1/2 - i\sigma)$ into each other. By introducing the inverse transform,
\begin{equation}
\label{inv_int}
\tilde{\cM}_{Y, \bar{Y}}[\Psi(\sigma,B)]=\sqrt{\frac{2r}{\pi}}\sum_{B\in\Z} e^{-i B\varphi}\int_{-\infty}^\infty \dd\sigma\, (4\pi r Y\bar{Y})^{-\frac12 +i\sigma} \Psi(\sigma,B),\quad \varphi=\arg(Y),
\end{equation}
we can also establish the relation between the differential operators of the Higgs side and the shift operators of the Coulomb side: 
\begin{align}
\label{diff_shift}
\cM\circ Q_{L,R}\circ\tilde\cM &= Y_{L,R},\cr
\cM \circ\tilde{Q}_{L,R}\circ\tilde\cM &= X_{L,R}.
\end{align}
This equality shows that the Fourier-Mellin transform provides an explicit isomorphism between the $\cA_H$-bimodule on the free hyper side and the $\cA_C$-bimodule on the gauge theory side.

Relations \eqref{diff_shift} mean that this isomorphism also holds for more general modules. If $Z(Y,\bar{Y})$ is some wave function on the Higgs side (not necessarily from the vacuum module), we simply define the dual wave function to be $Z(\sigma,B)=\cM_{\sigma,B}[Z(Y,\bar{Y})]$, and vice versa. Because of \eqref{diff_shift}, such an identification respects the action of $\cA_H$ and $\cA_C$ on the two sides of duality. Therefore, it establishes equivalence of the category of $\cQ^H$--preserving boundary conditions on the free hyper side with the category of $\cQ^C$--preserving boundary conditions on the $U(1)$, $N_f=1$ gauge theory side, and this equivalence respects the bimodule structure.

\subsubsection{Relation to the mirror wall}\label{sec:mirrorwall}
We have demonstrated above how the Fourier-Mellin transform realizes mirror symmetry of boundary conditions at the level of Q-cohomology. A more standard approach to this would be to collide boundary condition with the mirror symmetry interface \cite{Bullimore:2016nji}. To make a connection, we observe that the $\tilde{\cM}_{Y\bar{Y}}$ transform above is expressed as a convolution with the following kernel:
\begin{equation}
\langle Y,\bar{Y}|\hat{M}|\sigma,B\rangle = \frac1{\mu(\sigma,B)} \sqrt{\frac{2r}{\pi}}(4\pi r Y\bar{Y})^{-\frac12 +i\sigma} \left(\frac{\bar{Y}}{Y}\right)^{B/2},
\end{equation}
while the inverse transform is realized by the kernel:
\begin{equation}
\langle\sigma,B|\hat{M}^{-1}|Y,\bar{Y}\rangle =\sqrt{\frac{2r}{\pi}}(4\pi r Y\bar{Y})^{-\frac12-i\sigma} \left(\frac{Y}{\bar{Y}}\right)^{B/2},
\end{equation}
where $\hat{M}$ is the mirror symmetry defect. These kernels must coincide with the supersymmetric partition functions on $S^2\times(0,\epsilon)$, where the mirror symmetry defect is placed at $S^2\times \{\epsilon/2\}$, and the appropriate boundary conditions at the two ends are imposed. In principle, it should be possible to derive this from the description of the mirror wall given in \cite{Bullimore:2016nji}.

\paragraph{Conjectural partition function on the mirror wall}
Here we present a somewhat imprecise, hence conjectural calculation of partition function of the mirror wall that seems to reproduce the above result. Suppose we are given an $\mathfrak{su}(2|1)_A$-invariant boundary condition with the wave function $\psi(\sigma,B)$ in the $U(1), N_f=1$ theory. To compute its mirror, we need to know the kernel $\langle Y,\bar{Y}|\hat{M}|\sigma,B\rangle$. The latter is given by the partition function on $S^2\times (0,\epsilon), \epsilon\to 0$, with the following objects living on it. 

In the spherical layer $S^2\times (0,\epsilon/2)$, we place a $U(1), N_f=1$ gauge theory. It has Dirichlet boundary conditions parametrized by $(\sigma,B)$ at $S^2\times \{0\}$. At the opposite boundary, $S^2\times\{\epsilon/2-0\}$, we impose transversal Neumann boundary conditions. In the limit $\epsilon\to 0$, all degrees of freedom inside this layer are frozen and $(\sigma, B)$ play the role of background fields on the mirror wall (we will return to this in a moment).

In the spherical layer $S^2\times (\epsilon/2, \epsilon)$, we place a free hypermultiplet. Its boundary conditions at $S^2\times \{\epsilon\}$ are as in \eqref{bndrysuB} (with fermions vanishing), i.e. $q_1\big|=Y$, $\partial_\perp q_2\big| =0$, etc. At the opposite surface, namely at $S^2\times \{\epsilon/2 + 0\}$, we impose transversal boundary condition, i.e. $q_2\big| =\partial_\perp q_1\big| =0$ etc. In the limit $\epsilon\to 0$, again, all degrees of freedom inside this layer are frozen and $(Y,\bar{Y})$ play the role of background fields on the wall.

Therefore, we simply have to compute the partition functions of the theory on the wall in a given background parametrized by $(\sigma, B, Y, \bar{Y})$. From the point of view of this theory, $(Y,\bar{Y})$ form a background twisted chiral multiplet. On the other hand, $(\sigma, B)$ parametrize a background chiral multiplet $\sigma + i\frac{B}{2}$ on the wall that in \cite{Bullimore:2016nji} was denoted by $\varphi$ (it originates from the adjoint chiral which is part of the 3D $\cN=4$ vector multiplet, and which receives Neumann boundary conditions at $S^2\times \{\epsilon/2-0\}$). 

In \cite{Bullimore:2016nji}, the theory on the wall was described as a theory of a $\C^*$-valued twisted chiral multiplet $Z$, that has both a twisted superpotential $W(Z)$ and an ordinary superpotential $\cW(\tilde{Z})$, where $\tilde{Z}$ is the T-dual of $Z$. Such a description is not Lagrangian, and appears to pose a problem. To avoid this problem, and compute the partition function, we are going to use the following trick, which is not completely precise and is responsible for ``conjectural'' status of this computation. We are going to use localization for the $\mathfrak{su}(2|1)_A$-invariant theory on $S^2$. It is known \cite{Doroud:2013pka} that with respect to the standard  choice of the supercharge, the superpotential is Q-exact, while twisted superpotential is not. Hence, we argue that in the computation of the partition function, we can drop the superpotential, while the twisted superpotential should be kept. After this, the theory effectively becomes Lagrangian: it is a theory of a single twisted chiral multiplet $Z$ with a given twisted superpotential $W(Z)$. We can read off $W(Z)$ from \cite{Bullimore:2016nji} (adjusting coefficients to get the right answer):
\begin{equation}
W=\sqrt{\frac{4\pi}{r}}YZ - \frac{i}{r}\left(\sigma + i\frac{B}{2}\right)\log Z.
\end{equation}
The partition function is then given by:
\begin{equation}
\cZ=\int \frac{\dd Z\dd \bar{Z}}{\sqrt{Z\bar{Z}}} e^{rW(Y,Z) - r \bar{W}(\bar{Y},\bar{Z})}=\int \frac{\dd Z\dd \bar{Z}}{\sqrt{Z\bar{Z}}} e^{\sqrt{4\pi r}(YZ-\bar{Y}\bar{Z})}(Z\bar{Z})^{-i\sigma}\left( \frac{Z}{\bar{Z}} \right)^{B/2},
\end{equation}
where we have to further assume that the nature of this $\C^*$-valued field is such that the proper measure is $\frac{\dd Z\dd \bar{Z}}{\sqrt{Z\bar{Z}}}$. By writing $Z=\frac1{\sqrt{4\pi r}Y} \rho e^{i\theta}$, we find:
\begin{align}
\cZ=(4\pi rY\bar{Y})^{-\frac12 +i\sigma} \left(\frac{\bar{Y}}{Y} \right)^{B/2}\int \dd\rho\, \dd\theta\, e^{2i\rho\sin\theta}\rho^{-2i\sigma} e^{i\theta B}
\end{align}
The latter integral can be evaluated to be:
\begin{equation}
\int \dd\rho\, \dd\theta\, e^{2i\rho\sin\theta}\rho^{-2i\sigma} e^{i\theta B}=\pi (-1)^B\frac{\Gamma\left(\frac{B+1}{2}-i\sigma\right)}{\Gamma\left( \frac{B+1}{2}+i\sigma\right)}=\frac{\pi}{\mu(\sigma,B)}.
\end{equation}
We see that, up to an overall normalization, $\cZ$ on the wall indeed reproduces $\langle Y,\bar{Y}|\hat{M}|\sigma,B\rangle$. This serves as a good evidence that our integral transform approach agrees with the mirror symmetry wall approach of \cite{Bullimore:2016nji}. However, the above derivation clearly calls for better understanding.

\subsubsection{More boundary conditions.} Finally, we make a few remark on other boundary conditions. The Dirichlet boundary conditions on the gauge theory side are computed by the $\ell\to 0$ limit of the $S^2\times (0,\ell)$ partition function with \eqref{bndryS2} imposed on the two ends (and replacing $\sigma$ by $\sigma_0$ on one end). The corresponding wave function is:
\begin{equation}
\psi_D(\sigma, B)=\frac1{\mu(\sigma, B)} \delta_{B,0}\delta(\sigma-\sigma_0).
\end{equation}
The Neumann boundary condition is computed by the $\ell\to 0$ limit of the $S^2\times (0,\ell)$ partition function with \eqref{bndryS2} imposed on one end, and the complementary (transversal) Neumann boundary condition on another end. The corresponding wave function is:
\begin{equation}
\psi_N(\sigma, B)=1.
\end{equation}
It can be further enriched by the boundary theory $T$, in which case the wave function is simply the partition function of $T$ in the background of $\sigma, B$.

On the free hyper side, the Dirichlet boundary condition setting $q_1\big|=c$ and $\partial_\perp q_2\big| =-c/2r$ is described by the obvious wave function:
\begin{equation}
\psi_D(Y,\bar{Y}) = \delta^{(2)}(Y-c).
\end{equation}
The wave function of the complementary (transversal) boundary condition is computed by the $S^2\times (0,\ell)$ partition function with $q_1\big| =Y,\ \partial_\perp q_2\big| =-Y/2r$ imposed at $S^2\times \{\ell\}$ and $q_2\big| =Z,\ \partial_\perp q_1\big| =-Z/2r$ -- at $S^2\times \{0\}$. It is equal to:
\begin{equation}
\psi_N(Y,\bar{Y})=1.
\end{equation}
We can similarly enrich this boundary condition by the boundary degrees of freedom. If we put a (twisted) chiral multiplet $\phi$ at $S^2\times \{0\}$ and couple it to $q_1$ through the superpotential $W(q_1 ,\phi)$, we obtain a new boundary condition wave function given by:
\begin{equation}
\psi(Y,\bar{Y}) = \int \dd^2\phi\, e^{rW(Y,\phi)-r\bar{W}(\bar{Y},\bar{\phi})}.
\end{equation}
For example, picking $W=\phi(Y-c)$ results in:
\begin{equation}
\psi(Y,\bar{Y}) =\left(\frac{\pi}{r}\right)^2\delta^{(2)}(Y-c).
\end{equation}
Indeed, the boundary superpotential $\phi(Y-c)$ is known to flip the Neumann and Dirichlet boundary conditions for the hypermultiplet components.

\section{4D $\cN=2$ theories quantized on $S^3$}\label{sec:N2onS3}
Let us move up in dimension and consider four-dimensional supersymmetric theories. In this section, we describe gluing along $S^3$ in 4D $\cN=2$ theories (though we will also briefly consider gluing along squashed spheres). For example, we can sew supersymmetric path integrals on the two hemispheres $HS^4_\pm$ to get the full $S^4$ answer, or glue a hemisphere to a half-cylinder $S^3\times \R_+$ and obtain a cigar-like geometry.

To proceed and find a supersymmetric polarization, we need an $\cN=2$ theory formulated on some SUSY-preserving four-manifold with boundary $S^3$, such as $S^3\times \R_+$ or $HS^4$. Just like in 3D case of the previous section, we choose a hemisphere $HS^4$, which is obtained by cutting $S^4$ along the equator, though $S^3\times\R$ would work just fine. Theories with $\cN=2$ SUSY on $S^4$ received a lot of attention in the literature \cite{Pestun:2007rz,Okuda:2010ke,Gomis:2011pf,Hama:2012bg,Nosaka:2013cpa,Gerchkovitz:2014gta,Cabo-Bizet:2014nia}, and we find it unnecessary to review them here, instead we closely follow conventions of a review \cite{Hosomichi:2016flq}. SUSY in these theories is based on conformal Killing spinors $\xi_{A\alpha}$ and $\bar\xi_{A}^{\dot \alpha}$ that satisfy equations (2.1) and constraints (2.2) from \cite{Hosomichi:2016flq}. The field content is the vector multiplet $\cV = (A_m, \phi, \bar\phi, \lambda_{A\alpha}, \bar\lambda_A^{\dot{\alpha}}, D_{AB})$, where $A,B=1,2$ are the $SU(2)$ R-symmetry\footnote{Only the maximal torus of $SU(2)_R$ is preserved on $S^4$.} indices and the gauge indices are implicit, and the hypermultiplet $\cH=(q_{IA}, \psi_{I\alpha}, \bar\psi_I^{\dot \alpha})$, where $I=1,2$. The reality condition on $q$ is:
\begin{equation}
(q_{IA})^\dagger = \varepsilon^{IJ}\varepsilon^{AB}q_{JB}.
\end{equation}
The hypermultiplet $\cH$ takes values in a unitary representation $\cR\oplus\bar\cR$ of the compact gauge group $G$ (this notation means that $q_{11}$ and $q_{12}$ take values in $\cR$, while $q_{21}$ and $q_{22}$ are valued in $\bar\cR$). 

We choose $S^4$ to have radius $\ell=1$, and it is described using the stereographic projection to $\R^4$ with coordinates $(x^1, x^2, x^3, x^4)$. The radius $\ell$ can always be reintroduced on dimensional grounds. We cut $\R^4$ along $x^4=0$, the two half-spaces $\R^3\times \R_{>0}$ and $\R^3 \times \R_{<0}$ corresponding to $S^4 = HS^4_+ \cup HS^4_-$; the boundary $\R^3$ is the stereographic chart of $S^3=\partial (HS^4_+)$. Then we construct a supersymmetric polarizations of the phase space associated to this $S^3$.

\subsection{Supersymmetric Dirichlet polarization and the gluing formula}
One can define the following fields at the boundary:\footnote{Note the unusual reality condition $\bar\phi=- \phi^\dagger$.}
\begin{align}
\label{HS4fields}
\varphi&=q_{11}\big|,\quad \bar\varphi=q_{22}\big|,\cr
\chi_\alpha&=(\psi_{1\alpha} + i\sigma^4_{\alpha\dot{\alpha}} \bar\psi_1{}^{\dot \alpha})\big|,\quad \bar\chi_\alpha=(\psi_{2\alpha} - i\sigma^4_{\alpha\dot{\alpha}} \bar\psi_2{}^{\dot \alpha})\big|,\cr
F&=-(\cD_\perp q_{12} + (\phi-\bar\phi)q_{12})\big|,\quad \bar{F}=(\cD_\perp q_{21} - (\phi-\bar\phi)q_{21})\big|,\cr
\cA&=A_\parallel\big|, \quad \sigma=i(\phi+\bar\phi)\big|,\cr
\lambda_\alpha&=-i(\lambda_{2\alpha} + i\sigma^4_{\alpha\dot \alpha}\bar\lambda_2{}^{\dot \alpha})\big|,\quad \bar\lambda_\alpha=-i(\lambda_{1\alpha} - i\sigma^4_{\alpha\dot \alpha}\bar\lambda_1{}^{\dot \alpha})\big|,\cr
D_{\rm 3d}&=-D_{12} + i\cD_\perp (\phi-\bar\phi)\big|.
\end{align}
Using the kinetic part of the $\cN=2$ Lagrangian on $S^3\times\R$, or alternatively on $S^4$ close to the equator (the $S^4$ Lagrangian is given in \cite{Hosomichi:2016flq} in equations (2.10) and (2.18)), one can check that these boundary fields Poisson-commute\footnote{For the computation of Poisson brackets, one may assume that $A_\perp\big| =0$. However, $A_\perp\big|$ does not carry any physical data, and any condition on it is merely part of gauge fixing. If we work in Lorenz gauge in the bulk, it enforces $\cD_\perp A_\perp\big|=0$ at the boundary, while using temporal gauge in the vicinity of the boundary implies $A_\perp\big| =0$.} with each other, forming a polarization in the phase space on $S^3$. Moreover, they define a supersymmetric polarization that preserves $\cN=2$ SUSY on $S^3$. In particular, fields $(\varphi,\bar\varphi,\chi, \bar\chi, F, \bar{F})$ form a chiral multiplet of R-charge $\Delta=1$, while $(\cA, \sigma, \lambda, \bar\lambda, D_{\rm 3d})$ close into a 3D $\cN=2$ vector multiplet on $S^3$. Note that the $U(1)$ R-symmetry of a 4D theory (whenever it is present, e.g., in the SCFT) is broken by the boundary conditions, while the Cartan of $SU(2)$ R-symmetry (which is the only R-symmetry generically preserved on $S^4$ anyways) becomes the $U(1)$ R-symmetry of the $\cN=2$ SUSY on $S^3$.

Therefore, gluing is represented by a 3D $\cN=2$ path integral over the fields uniquely determined by the 4D field content: each 4D $\cN=2$ vector multiplet gives rise to a 3D $\cN=2$ vector multiplet, and each 4D hypermultiplet gives a 3D $\cN=2$ chiral multiplet of R-charge $\Delta=1$. We can further apply known localization results for gauge theories on $S^3$ to arrive at a simple gluing formula. 

The localization locus of $\cN=2$ theories on $S^3$, in the approach of \cite{Kapustin:2009kz}, is parametrized by a single constant matrix $a\in \mathfrak{t}$, where $\mathfrak{t}$ is a Cartan subalgebra of the gauge algebra. The 3D $\cN=2$ vector multiplet one-loop determinant, combined with the Vandermonde factor due to integration over $\mathfrak{t}$ rather than $\mathfrak{g}$, is given by \cite{Kapustin:2009kz}:
\begin{equation}
\label{3dvector}
Z^{S^3}_{\rm v}(a)=\det{}_{\rm Adj}'[2 \sinh \pi a],
\end{equation}
where $\det_{\rm Adj}'$ denotes determinant over the non-zero roots of $\mathfrak{g}$ only. The one-loop determinant for chiral multiplets of arbitrary R-charge can be found in \cite{Jafferis:2010un, Hama:2010av, Hosomichi:2014hja}, and the answer for R-charge $\Delta=1$ is:
\begin{equation}
\label{3dchiral}
Z^{S^3}_{\rm ch}(a)=\prod_{w\in\cR}\prod_{n>0} \left(\frac{n+iw(a)}{n-iw(a)}\right)^n=\prod_{w\in\cR} s_{b=1}(-w(a)).
\end{equation}
In this way, we derive the following gluing formula:
\begin{equation}
\label{HS4glue}
\langle\Psi_1|\Psi_2\rangle=\frac{1}{|\cW|} \int_{\mathfrak{t}} \dd^r a \, \mu(a) \langle\Psi_1|a\rangle \langle a|\Psi_2\rangle,\quad \mu(a)=Z^{S^3}_{\rm v}(a)Z^{S^3}_{\rm ch}(a).
\end{equation}
where $\langle\Psi_1|a\rangle$ and $\langle a|\Psi_2\rangle$ are the wavefunctions evaluated at the boundary conditions given by \eqref{HS4fields} specialized to the locus where all the 3D fields vanish except for $\sigma = a \in \mathfrak{t}$.:
\begin{align}
\label{HS4bc}
q_{11}\big|&=q_{22}\big|=0,\cr
(\psi_{1\alpha} + i\sigma^4_{\alpha\dot{\alpha}} \bar\psi_1{}^{\dot \alpha})\big|&=(\psi_{2\alpha} - i\sigma^4_{\alpha\dot{\alpha}} \bar\psi_2{}^{\dot \alpha})\big|=0,\cr
(\cD_\perp q_{12} + (\phi-\bar\phi)q_{12})\big|&=(\cD_\perp q_{21} - (\phi-\bar\phi)q_{21})\big|=0,\cr
i(\phi+\bar\phi)\big|&=a,\cr
A_\parallel\big|&=-D_{12} + i\cD_\perp (\phi-\bar\phi)\big|=0,\cr 
(\lambda_{2\alpha} + i\sigma^4_{\alpha\dot \alpha}\bar\lambda_2{}^{\dot \alpha})\big|&=(\lambda_{1\alpha} - i\sigma^4_{\alpha\dot \alpha}\bar\lambda_1{}^{\dot \alpha})\big|=0.\cr
\end{align}

Such boundary conditions are precisely the Dirichlet boundary conditions of \cite{Gava:2016oep}, where they computed the empty (i.e., without insertions) round hemisphere partition function using the supersymmetric localization, and used it to conjectured a formula akin to \eqref{HS4glue}. Notice that similar observation was made much earlier in \cite{Drukker:2010jp} based on the AGT correspondence.\footnote{Note that the 3D chiral multiplet contribution $Z_{\rm ch}(a)$ of \eqref{3dchiral} is often dropped in the literature. This is allowed as long as the matter representation $\cR$ is self-conjugate. Otherwise, it is a non-trivial $a$-dependent phase factor that has to be included.} We have proven this gluing formula in full generality here, which also implies its validity in the presence of arbitrary insertions of observables, as long as the 3D $\cN=2$ localizing supercharge is preserved.

Another remark we should make is that physical wave functions of the gauge theory are always invariant under the Weyl group action on $\mathfrak{t}$:
\begin{equation}
\langle w(a)|\Psi\rangle = \langle a|\Psi\rangle,\quad a\in\mathfrak{t},\ w\in\cW.
\end{equation}

\paragraph{Deformation by masses. } One can easily include masses for the hypermultiplets by turning on background vevs for the vector multiplets gauging flavor symmetries. This affects the gluing measure $\mu(a)$ in a rather obvious way: we should include effects of the extra background vector multiplets into it. If the $i$-th flavor of matter has non-zero mass $m_i$ and dynamical gauge representation $\cR_i$, the 3D chiral contribution to $\mu(a)$ becomes:
\begin{equation}
\label{3dchiralMassive}
Z^{S^3}_{\rm ch}(a,m)=\prod_i\prod_{w_i\in\cR_i}\prod_{n>0} \left(\frac{n+iw_i(a) + im_i}{n-iw_i(a)-im_i}\right)^n=\prod_i\prod_{w_i\in\cR_i} s_{b=1}(-w_i(a)-m_i).
\end{equation}
The vector multiplet contribution $Z_{\rm v}^{S^3}(a)$ is not modified, and we obtain the mass-deformed gluing measure $\mu(a,m)=Z_{\rm v}^{S^3}(a) Z^{S^3}_{\rm ch}(a,m)$.

Notice that for self-conjugate representations, even though $Z^{S^3}_{\rm ch}(a,0)=1$, the massive factor might be non-trivial, $Z^{S^3}_{\rm ch}(a,m)\ne 1$. In order for a massive factor to be trivial, $Z^{S^3}_{\rm ch}(a,m)= 1$, the matter should transform in the self-conjugate representation of the flavor symmetry that has been gauged by the background vector multiplet. For example, if a theory has two hypers transforming in the same self-conjugate representation $\cR$ of the gauge group, the flavor symmetry is $U(2)_F$ and it allows two masses, $m_1, m_2$. For generic values of $m_1$ and $m_2$, 
\begin{equation}
Z^{S^3}_{\rm ch}(a,m)=\prod_{w\in\cR}s_{b=1}(-w(a)-m_1)s_{b=1}(-w(a)-m_2)\ne 1.
\end{equation}
However, if $m_1=-m_2$, this corresponds to turning on the background gauge multiplet only for $SU(2)_F\subset U(2)_F$; hypers transform in the fundamental representation of $SU(2)_F$, which is self-conjugate. In this case, indeed: 
\begin{equation}
Z^{S^3}_{\rm ch}(a,m)=\prod_{w\in\cR}s_{b=1}(-w(a)-m_1)s_{b=1}(-w(a)+m_1)= 1.
\end{equation}

\subsubsection{One-loop determinants on $HS^4$}\label{sec:HS4Z}
As a side remark, one can use the gluing formula to predict the hemisphere partition function and check the results of \cite{Gava:2016oep}. The structure of the hemisphere partition function with Dirichlet boundary conditions parametrized by $a\in\mathfrak{t}$ is: 
\begin{equation}
Z^{HS^4}(a)=Z^{HS^4}_{\rm cl}(a)Z^{HS^4}_{\rm vec, 1-loop}(a)Z^{HS^4}_{\rm hyp, 1-loop}(a)Z_{\rm inst}(a),
\end{equation}
where $Z^{HS^4}_{\rm cl}(a)$ is the classical piece, $Z_{\rm inst}(a)$ is the Nekrasov instanton partition function \cite{Nekrasov:2002qd} (from point-like instantons at the pole), and the remaining are the one-loop determinants. Gluing two copies of these, one with the opposite orientation (which flips the $\theta$-term and thus complex conjugates $Z_{\rm inst}(a)$), we can write the full sphere answer as:
\begin{equation}
Z^{S^4}=\frac1{|\cW|}\int_{\mathfrak{t}}\dd^r a\, \mu(a) \left(Z^{HS^4}_{\rm cl}(a)Z^{HS^4}_{\rm vec, 1-loop}(a)Z^{HS^4}_{\rm hyp, 1-loop}(a)\right)^2 |Z_{\rm inst}(a)|^2.
\end{equation}
Comparing this with the known $S^4$ answer \cite{Pestun:2007rz}:
\begin{equation}
Z^{S^4}=\frac1{|\cW|}\int_{\mathfrak{t}}\dd^r a\, Z^{S^4}_{\rm cl}(a)Z^{S^4}_{\rm vec, 1-loop}(a)Z^{S^4}_{\rm hyp, 1-loop}(a) |Z_{\rm inst}(a)|^2,
\end{equation}
we conclude that there must exist a relation between the 1-loop factors on $S^4$ and $HS^4$ of the form $Z^{S^4}_{\rm vec, 1-loop}(a)Z^{S^4}_{\rm hyp, 1-loop}(a) = \mu(a)\left(Z^{HS^4}_{\rm vec, 1-loop}(a)Z^{HS^4}_{\rm hyp, 1-loop}(a)\right)^2$. Collecting contributions from the hyper and vector multiplets separately from each other, and remembering that $\mu(a)=Z_{\rm v}^{S^3}(a)Z_{\rm ch}^{S^3}(a)$, we find:
\begin{align}
Z^{HS^4}_{\rm vec, 1-loop}(a)&=\left(\frac1{Z^{S^3}_{\rm v}(a)} Z^{S^4}_{\rm vec, 1-loop}(a)\right)^{1/2},\cr
Z^{HS^4}_{\rm hyp, 1-loop}(a)&=\left(\frac1{Z^{S^3}_{\rm ch}(a)} Z^{S^4}_{\rm hyp, 1-loop}(a)\right)^{1/2}.
\end{align}
Using expression \eqref{3dvector} for $Z^{S^3}_{\rm v}(a)$ and taking $Z^{S^3}_{\rm v}(a)$ as in \cite{Pestun:2007rz}, this gives the answer for $Z^{HS^4}_{\rm vec, 1-loop}(a)$ as in \cite{Gava:2016oep}:
\begin{equation}
Z^{HS^4}_{\rm vec, 1-loop}(a)=\prod_{\alpha\in\Delta_+}\frac{\alpha(a)}{\sinh(\pi\alpha(a))}\prod_{n>0}(n+i\alpha(a))^n(n-i\alpha(a))^n.
\end{equation}

However, the result for $Z^{HS^4}_{\rm hyp, 1-loop}(a)$ is slightly different. Using $Z^{S^3}_{\rm ch}(a)$ as given in \eqref{3dchiral}, and $Z_{\rm hyp, 1-loop}^{S^4}(a)=\prod_{w\in\cR}\prod_{n>0}\left(n+iw(a) \right)^{-n}\left(n-iw(a) \right)^{-n}$ taken from \cite{Pestun:2007rz}, we find:
\begin{equation}
\label{HS4HypDet}
Z_{\rm hyp}^{HS^4}(a)=\prod_{w\in\cR}\prod_{n>0}(n + iw(a))^{-n}.
\end{equation}
In general, this answer differs from the one-loop determinant of \cite{Gava:2016oep}, which was written there as $\prod_{w\in\cR}\prod_{n>0}\left[(n + iw(a))^{-n}(n-iw(a))^{-n}\right]^{1/2}$, and agrees with it only for the self-conjugate representations.\footnote{Only for real scalars, the answer can possibly involve square roots of the determinants, such as $\prod_{w\in\cR}\prod_{n>0}\left[(n + iw(a))^{-n}(n-iw(a))^{-n}\right]^{1/2}$. However, the Dirichlet boundary conditions do not violate complex structure of the hypermultiplet scalars, thus both on $HS^4$ and $S^4$ one has to compute determinants for complex scalars only. This can be considered as an argument in favor of \eqref{HS4HypDet}.}
 
\subsection{The ellipsoid}\label{sec:ellips}
One can slightly generalize discussion of the previous subsection by turning on squashing for $S^3$ and $S^4$. Namely, now the $S^4$ is replaced by the following submanifold in $\R^5$:
\begin{equation}
\label{sqashedS4}
\frac{b^{-2}}{\ell^2}(X_1^2 + X_2^2) + \frac{b^{2}}{\ell^2}(X_3^2 + X_4^2) + \frac{X_5^2}{r^2}=1,\quad \text{where } b\geq1.
\end{equation}
By cutting through $X_5=0$, we obtain a 3D slice given by the squashed $S_b^3\subset \R^4$:
\begin{equation}
b^{-2}(X_1^2 + X_2^2) + b^{2}(X_3^2 + X_4^2)=\ell^2,
\end{equation}
and we intend to derive a formula that describes gluing along this $S^3_b$. To make equations less cumbersome, we pick $\ell=1$, but $\ell$ can always be recovered from dimensional analysis.

4D $\cN=2$ theories on the ellipsoid \eqref{sqashedS4} were studied in \cite{Hama:2012bg}. The construction of their Lagrangians is based on the same off-shell 4D $\cN=2$ conformal supergravity as on $S^4$, and differs only by the presence of non-zero vevs for certain background supergravity fields. As a result, one can use precisely the same definitions of the boundary fields as for the round case \eqref{HS4fields}. In parallel with the round sphere case, it is also true that 4D vector multiplets give rise to 3D vector multiplets in the gluing theory on $S^3_b$, while 4D hypermultiplets give 3D chiral multiplets of unit R-charge in the theory on $S^3_b$.

Using the results of \cite{Hama:2011ea} where $\cN=2$ theories on $S^3_b$ were studied, and specializing to the chirals of unit R-charge, we note that the gluing measure has the same structure:
\begin{equation}
\mu_b(a)= Z_{\rm v}^{S^3_b}(a) Z_{\rm ch}^{S^3_b}(a),
\end{equation}
with the chiral multiplet contribution given by:
\begin{align}
Z_{\rm ch}^{S^3_b}(a)=\prod_{w\in\cR}\prod_{n,m\geq 0} \frac{mb + nb^{-1} + \frac{Q}2 + iw(a)}{mb + nb^{-1} +\frac{Q}2 -i w(a)}=\prod_{w\in\cR} s_b\left(-w(a)\right),\text{ where }Q=b + b^{-1},
\end{align}
and the vector multiplet contribution:
\begin{equation}
Z_{\rm vec}^{S^3_b}(a)=\prod_{\alpha\in\Delta_+} 4\sinh\left(\pi b \alpha(a)\right) \sinh\left( \pi b^{-1}\alpha(a) \right).
\end{equation}
Note that the matter one-loop factor $Z_{\rm ch}^{S^3_b}(a)$ is still a pure phase, which also cancels for the self-conjugate $\cR$ like in the $b=1$ case. The gluing formula that we find in this way:
\begin{equation}
\label{S3bGlue}
\langle\Psi_1|\Psi_2\rangle=\frac{1}{|\cW|} \int_{\mathfrak{t}} \dd^r a \, \mu_b(a) \langle\Psi_1|a\rangle \langle a|\Psi_2\rangle,
\end{equation}
should also be compared with the one from the AGT literature \cite{Drukker:2010jp}.

\paragraph{Deformation by masses. } Just like on the round sphere, we can allow hypers to have masses originating from flavor symmetries of the massless theory. In complete analogy with the round case, this only affects the gluing measure $\mu_b(a)$ through the factor $Z_{\rm ch}^{S_b^3}$, which becomes mass-dependent and non-trivial even for self-conjugate $\cR_i$'s:
\begin{equation}
Z_{\rm ch}^{S^3_b}(a, m)=\prod_{i}\prod_{w_i\in\cR_i}\prod_{n,k\geq 0} \frac{kb + nb^{-1} + \frac{Q}2 + i w(a) + im_i}{kb + nb^{-1} +\frac{Q}2 -i w(a) - im_i}=\prod_i\prod_{w_i\in\cR_i}s_b\left(-w_i(a)-m_i\right).
\end{equation}

\subsubsection{Partition function on a squashed hemisphere}\label{sec:HEllipsZ}
Similar to the round case, we can apply the gluing formula to find the partition function of a half of the ellipsoid with Dirichlet boundary conditions. The full ellipsoid answer is known from \cite{Hama:2012bg}. The gluing formula tells us how it factorizes between the two halves. We find that:
\begin{equation}
Z^{HS^4_b}(a)=Z^{HS^4_b}_{\rm cl}(a)Z^{HS^4_b}_{\rm vec, 1-loop}(a)Z^{HS^4_b}_{\rm hyp, 1-loop}(a)Z_{\rm inst}(a,b),
\end{equation}
where
\begin{equation}
Z^{HS^4_b}_{\rm cl}(a)=e^{-\frac{4\pi^2}{g_{YM}^2}{\rm Tr}(a^2)},\quad Z_{\rm inst}(a,b)=Z_{\rm Nekrasov}\left(a, \epsilon_1=b, \epsilon_2=b^{-1}, \tau=\frac{\theta}{2\pi}+\frac{4\pi i}{g_{YM}^2}\right),
\end{equation}
and the one-loop factors are found by taking square roots of answers from \cite{Hama:2012bg} divided by $Z_{\rm vec}^{S^3_b}(a)$ and $Z_{\rm ch}^{S^3_b}(a)$:
\begin{align}
Z^{HS^4_b}_{\rm vec, 1-loop}(a)&=\prod_{\alpha\in\Delta_+}\left(\frac{\Upsilon(i\alpha(a))\Upsilon(-i\alpha(a))}{\sinh(\pi b\alpha(a))\sinh(\pi b^{-1}\alpha(a))}\right)^{1/2},\cr
Z^{HS^4_b}_{\rm hyp, 1-loop}(a)&=\prod_{w\in\cR}\prod_{m,n\geq 0}\left(mb + nb^{-1}+\frac{Q}{2}+iw(a)\right)^{-1},
\end{align}
where $\Upsilon(x)=\prod_{m,n\geq 0}(mb + nb^{-1} + x)(mb + nb^{-1} + Q - x)$.

\subsection{Boundary conditions and domain walls}\label{sec:4dBCandD}
Just like for the case of 3D $\cN=4$ theories studied in the previous section, the supersymmetric gluing formula that we found here can be applied to study supersymmetric boundary conditions \cite{Gaiotto:2008sa,Gaiotto:2008ak}. Namely, we are interested in boundary conditions in 4D $\cN=2$ theories \cite{Dimofte:2011jd,Dimofte:2011ju,Dimofte:2013lba} that preserve the SUSY $Q_{\rm Loc}$ used for the boundary localization. 

An argument similar to the one illustrated on Figure \ref{fig:bndry} implies that a supersymmetric boundary condition $\cB$ at $S^3$ (or $S^3_b$) is completely determined by a single Weyl-invariant function $f_\cB(a)$ (its ``wave function'') on the Cartan subalgebra $\mathfrak{t}\subset \mathfrak{g}$. By analogy with \cite{Hori:2013ika}, we could also call this $f_\cB(a)$ a ``brane factor''. Similar wave functions on the Cartan subalgebra (and ideas somewhat reminding the ones discussed here) have also previously appeared in \cite{Nishioka:2011dq}. 

To be more precise, the boundary conditions $\cB$ is determined by $f_\cB(a)$ only up to $Q$-exact terms, in the same sense as in the previous section. Also, as a reminder, we should note that this classifies only those boundary conditions that preserve the supercharge used for the boundary localization. Alternatively, since for every $a\in\mathfrak{t}$, we have a half-BPS boundary condition \eqref{HS4bc}, we can say that $f_\cB(a)$ determines a half-BPS boundary condition that preserves $\cN=2$ SUSY at the boundary. To impose a boundary condition determined by some $f_\cB(a)$, one should first compute the path integral with boundary conditions \eqref{HS4bc}, and then integrate the result against $\mu(a) f_\cB(a)$, i.e., apply the gluing formula.

If the boundary condition $\cB$ is described by coupling to some boundary theory $T_\cB$, the wave function $f_\cB(a)$ can be thought of as its partition function $Z_{T_\cB}(a)$, however one should be careful with such an interpretation. The picture we have in mind is similar to the one from Figure \ref{fig:bndry}: we consider a thin cylinder $S^3_b\times (0,\ell)$, with the boundary condition $\cB$ imposed at $S^3_b\times \{\ell\}$. Then we can glue this cylinder along the ``gluing surface'' $S^3_b\times \{0\}$ to the boundary $S^3_b=\partial M$ of another manifold. In the limit $\ell\to 0$, this is equivalent to simply imposing the boundary condition $\cB$ at $S^3_b=\partial M$. How does the cylinder partition function behave in the $\ell\to0$ limit? Depending on how the boundary theory $T_\cB$ couples to the bulk, in the limit $\ell\to 0$, some of the bulk degrees of freedom might survive in the thin slice between the two boundaries of $S^3\times (0,\ell)$, and manifest themselves as additional 3D degrees of freedom, -- we call them ``trapped'' degrees of freedom. If no such degrees of freedom survive, i.e., all bulk fields are completely frozen in the $\ell\to0$ limit, then $f_\cB(a)=Z_{T_\cB}(a)$, with $a$ playing the role of masses in the boundary theory. This happens, for example, if each field has Dirichlet boundary condition on one end and Neumann or Dirichlet on another. Then we can take fields that have Neumann boundary condition at $S^3_b\times \{\ell\}$ and couple their boundary values to $T_\cB$. However, if some of the bulk degrees of freedom stay dynamical, i.e., can fluctuate in the $\ell\to 0$ limit, then:
\begin{equation}
f_\cB(a)= Z_{T_\cB}(a) Z_{\rm trapped}(a),
\end{equation}
where $Z_{\rm trapped}(a)$ is the partition function of such ``trapped'' 3d degrees of freedom.

For example,
\begin{equation}
\label{DirS3}
f_\cB(a)=\frac1{\mu_b(a)}\delta_{\mathfrak t}(a,\tilde{a})
\end{equation}
corresponds to the Dirichlet boundary condition \eqref{HS4bc} with $a$ replaced by $\tilde{a}$, where $\delta_{\mathfrak t}$ is a Weyl-invariant delta-function\footnote{One might worry that $\frac1{\mu_b(a)}\delta_{\mathfrak t}(a,\tilde{a})$ is poorly defined at $\tilde{a}=0$, since $\mu_b(a)$ vanishes at $a=0$. In fact, it is well-defined. The wave function on the full algebra $\mathfrak{g}$ would be $\frac{\det_{\rm Adj}' a}{\mu_b(a)}\delta_{\mathfrak g}(a)$, and this is clearly well-defined. Upon passing to integration over $\mathfrak{t}\subset \mathfrak{g}$, the Vandermonde factor $\det{}_{\rm Adj}' a$ disappears. This is analogous to the  delta function on $\R^2$ being well-defined despite having a dangerous-looking expression $\frac1{2\pi r}\delta(r)$ in polar coordinates due to the Jacobian.} on $\mathfrak{t}$:
\begin{equation}
\delta_{\mathfrak t}(a_1, a_2)=\sum_{w\in\cW}\delta^r(a_1 - w(a_2)).
\end{equation} 
It is obvious from our gluing formula \eqref{S3bGlue} that \eqref{DirS3} indeed describes Dirichlet boundary conditions: the $1/\mu_b(a)$ factor cancels the gluing measure, and further integration is saturated by the delta function which simply imposes Dirichlet boundary conditions. On the other hand, we can again interpret \eqref{DirS3} as the partition function on the thin cylinder (in the $\ell\to 0$ limit) with the Dirichlet boundary conditions \eqref{HS4bc} imposed on both ends. Even though we call them ``Dirichlet'' (based on the gauge field having Dirichlet boundary conditions), some of the fields in the multiplets actually have Neumann boundary conditions on both ends of the cylinder. Therefore, in this case we have some ``trapped'' bulk degrees of freedom on $S^3_b\times (0,\ell)$, and the factor $1/\mu_b(a)$ in \eqref{DirS3} can be thought of as their $S^3_b$ partition function. We will understand this factor more precisely by the end of this subsection.

Another natural wave function is given by:,
\begin{equation}
\label{NeumBC}
f_p(a)=e^{4\pi^2 a\cdot p},
\end{equation}
where $a\cdot p$ denotes the invariant inner product between $a, p\in \mathfrak{t}$. Such a wave function describes the half-BPS Neumann boundary conditions obtained from \eqref{HS4bc} by flipping Dirichlet and Neumann for all fields (we call it the complementary boundary condition). In particular, $i(\phi+\bar\phi)\big|=a$ is replaced by $i\cD_\perp(\phi+\bar\phi)\big| =p$. To understand why \eqref{NeumBC} describes such boundary conditions, we follow the same steps as before: $f_p(a)$ is given by the partition function on the thin cylinder $S^3_b\times (0,\ell)$, $\ell\to 0$, with the Dirichlet boundary conditions \eqref{HS4bc} imposed at $S^3_b\times \{\ell\}$ and the complementary Neumann boundary conditions imposed at $S^3_b\times \{0\}$. With such boundary conditions, all fields living on the cylinder become frozen in the $\ell\to 0$ limit, so there are no leftover ``trapped'' degrees of freedom. Hence the partition function is simply given by the classical action in the $\ell\to 0$ limit, which also vanishes except possibly for the boundary terms. Recall from \cite{glue1} that consistently imposing general boundary conditions might require adding special boundary terms to the action. In our case, most boundary terms vanish, except the one for the field $i(\phi + \bar\phi)$. Indeed, this field satisfies:
\begin{align}
i(\phi + \bar\phi)\big|_{S^3_b\times \{\ell\}}=a,\cr
i\cD_\perp(\phi + \bar\phi)\big|_{S^3_b\times \{0\}}=p,
\end{align}
and we need to include a term $-2{\rm Tr}\int_{S^3_b \times \{0\}}i(\phi + \bar\phi)i\cD_\perp(\phi + \bar\phi)$ in the action to make the Neumann boundary condition at $S^3_b\times\{0\}$ consistent. In the limit $\ell\to 0$, this term evaluates to $-4\pi^2a\cdot p$, and the cylinder partition function becomes $e^{4\pi^2 a\cdot p}$. 

We could also choose to flip Dirichlet and Neumann boundary conditions only for a subgroup $H\subset G$ of the gauge group, with the Cartan $\mathfrak{t}_H \subset \mathfrak{t}$. In such a case, the wave function could be guessed as $f_\cB(a)=e^{4\pi^2 a\cdot p}\frac{\mu^H_b(a)}{\mu^G_b(a)}\delta_{\mathfrak{t}_H^\perp}(a-\tilde{a})$, where $p\in\mathfrak{t}_H, \tilde{a}\in\mathfrak{t}_H^\perp$, $\delta_{\mathfrak{t}_H^\perp}$ is the delta-function on the orthogonal complement of $\mathfrak{t}_H$ inside of $\mathfrak{t}$, and $\mu^G_b(a)\equiv \mu_b(\sigma)$ is the original measure factor (with the gauge group $G$), while $\mu^H_b(a)$ is the measure factor written for the same matter content as $\mu_b^G(a)$ but with the gauge group $H$.

\begin{figure}[t!]
	\centering
	\includegraphics[width=0.5\textwidth]{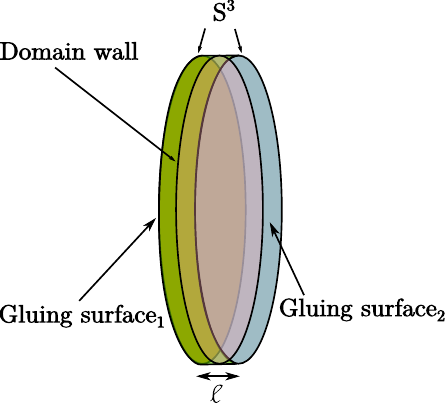}
	\caption{\label{fig:domain} Inserting a domain wall can be thought of as gluing in a cylinder $S^3 \times (0,\ell)$ of length $\ell\to 0$ supporting the domain wall in the middle, at $S^3\times \{\ell/2\}$. The two gluing surfaces are assigned Dirichlet boundary conditions with parameters $a_1$ and $a_2$, so such a cylinder computes the integral kernel $Z(a_1, a_2)$ of the domain wall.}
\end{figure}

It is clear that one can also describe domain walls in this way: adding a domain wall along $S^3_b$ corresponds to making a cut along this $S^3_b$ and gluing in an infinitesimal cylinder $S^3_b\times (0,\ell)$, $\ell\to0$, supporting the domain wall in the middle of $(0,\ell)$, e.g. at $S^3\times \{\ell/2\}$ (see Figure \ref{fig:domain}). It is described by an operator $\hat{Z}$ that completely characterizes the effect of domain wall on the physical states. At the level of $Q_{\rm Loc}$-cohomology, this $\hat{Z}$ is fully captured by the integral kernel $Z(a_1, a_2)=\langle a_1|\hat{Z}|a_2\rangle$, which also enters the supersymmetric gluing formula:
\begin{equation}
\label{dwall}
\langle\Psi_1|\hat{Z}|\Psi_2\rangle=\frac{1}{|\cW|^2} \int_{\mathfrak{t}} \dd^r a_1\, \dd^r a_2 \, \mu_b(a_1)\, \mu_b(a_2)\, \langle\Psi_1|a_1\rangle Z(a_1,a_2)\langle a_2|\Psi_2\rangle.
\end{equation}
Notice that, just like in the case of boundary conditions, the two thin cylindrical layers on both sides of the wall may support ``trapped'' bulk degrees of freedom in the $\ell\to0$ limit, where they become effectively three-dimensional. In the absence of such ``trapped'' degrees of freedom, $Z(a_1, a_2)$ simply equals partition function of the theory living at the wall. More generally, one should also include partition function of trapped modes. Whether or not they are present depends on how the domain wall couples to the bulk. We will have a few notable examples below.

The simplest possible domain wall is the transparent wall, it has $\hat{Z}=1$ and:
\begin{equation}
\label{transpWall}
Z_0(a_1, a_2)=\langle a_1|a_2\rangle=\frac{1}{\mu_b(a_1)}\delta_{\mathfrak t}(a_1,a_2),
\end{equation}
which clearly coincides with the Dirichlet wave function \eqref{DirS3}. With such a wall, equation \eqref{dwall} reduces to the gluing formula \eqref{S3bGlue}. Integral kernel \eqref{transpWall} of the transparent domain wall, just like the Dirichlet wave function, is given by the $\ell\to0$ limit of the empty $S^3_b\times (0,\ell)$ partition function  with boundary conditions \eqref{HS4bc} imposed on both ends. The factor $1/\mu_b(a)$ corresponds to the same ``trapped'' bulk modes (that remain dynamical in the $\ell\to 0$ limit) as for the Dirichlet wave function.

One can actually understand precisely what are these ``trapped'' degrees of freedom. The 4D hypermultiplet consists of two 4D chiral multiplets in conjugate representations, with $q_{11}$ in $\cR$ and $q_{21}$ in $\bar\cR$ as their lowest components. As seen from \eqref{HS4bc}, $q_{11}|_{S^3_b\times \{0\}}= q_{11}|_{S^3_b\times \{\ell\}}=0$, therefore the first chiral multiplet is frozen in the $\ell\to 0$ limit. On the other hand, the boundary condition $\cD_\perp q_{21}|_{S^3_b\times \{0\}}=\cD_\perp q_{21}|_{S^3_b\times \{\ell\}}=0$ simply implies that $q_{21}$ becomes independent of the $(0,\ell)$ direction. This means that the second chiral multiplet effective becomes an $\bar\cR$-valued chiral multiplet on $S^3_b$. Using the property $s_b(x)s_b(-x)=1$, we find that the contribution of this multiplet is $1/Z_{\rm ch}^{S^3_b}(a,m)$. Similar analysis is slightly more subtle for the vector multiplet, but one can find that the surviving degrees of freedom form a 3D adjoint chiral multiplet of R-charge zero. Such a multiplet contributes the inverse of the vector multiplet factor, $1/Z_{\rm vec}^{S^3_b}(a)$. Multiplying the two contributions, we indeed find that the ``trapped'' degrees of freedom on $S^3_b\times (0,\ell)$ produce a factor of $1/\mu_b(a)$.

Our next example is slightly less trivial -- the S-duality domain walls in $\cN=4$ SYM.

\subsubsection{S-duality of BPS boundary conditions in $\cN=4$ SYM}\label{sec:SDuality}
As anther fun application, let us consider boundary conditions in 4D $\cN=4$ SYM. From the 4D $\cN=2$ SUSY point of view, the theory has one adjoint hypermultiplet, and so the boundary fields determined in \eqref{HS4fields}, forming a 3D $\cN=2$ vector multiplet and an adjoint chiral multiplet of unit R-charge, actually build up a 3D $\cN=4$ vector multiplet. Thus the gluing theory is the theory of a single 3D $\cN=4$ vector multiplet. So the basic Dirichlet boundary condition given in \eqref{HS4bc} is half-BPS from the 4D $\cN=4$ point of view: it is not hard to check that it actually coincides with the basic half-BPS Dirichlet boundary condition in 4D $\cN=4$ SYM, as described in \cite{Gaiotto:2008sa}.

However, to apply our formalism, we do not need to preserve the full 3D $\cN=4$ SUSY at the boundary: it is sufficient to preserve the supercharge used in the boundary localization, which is the 3D $\cN=2$ localization on $S^3$ in our case. Throughout this subsection, as before, we refer to such a supercharge as $Q_{\rm Loc}$. In particular, since $Q_{\rm Loc}$ belongs to 3D $\cN=2$ subalgebra, this includes quarter-BPS boundary conditions, but even more general boundary conditions are allowed. Each such boundary condition $\cB$ is completely characterized by the physical state it creates, and the $Q_{\rm Loc}$-cohomology class of this state is completely determined by the Weyl-invariant wave function $f_\cB(a)$. \footnote{Notice that each $Q_{\rm Loc}$-cohomology class determined by $f_\cB(a)$ contains a half-BPS boundary condition. To construct it, impose the Dirichlet boundary condition \eqref{HS4bc} parametrized by $a$, and then integrate the result against $\mu(a)f_\cB(a)$. Therefore, general $Q_{\rm Loc}$-invariant boundary conditions, while probably highly non-trivial in the full theory, are equivalent to half-BPS boundary conditions at the level of $Q_{\rm Loc}$-cohomology.} 

Since we only have an adjoint-valued hyper, and the adjoint representation is self-conjugate, in the absence of mass ($\cN=2^*$) deformation the gluing measure only contains contribution from the vector multiplet:
\begin{equation}
\mu_b(a)=Z_{\rm vec}^{S^3_b}(a)=\prod_{\alpha\in\Delta_+} 4\sinh\left(\pi b \alpha(a)\right) \sinh\left( \pi b^{-1} \alpha(a) \right).
\end{equation}
Note that if we turn on a single possible mass parameter, i.e., consider the $\cN=2^*$ theory, chirals start to contribute non-trivially in the measure:
\begin{equation}
Z_{\rm ch}^{S_b^3}(a,m)=\prod_{\alpha\in\Delta_+}s_b\left(-\alpha(a)-m\right)s_b\left(\alpha(a)-m\right).
\end{equation}
The latter becomes $1$ only for $m=0$ and should not be forgotten at $m\neq 0$. This $Z_{\rm ch}^{S_b^3}(a,m)$ is usually not included in the literature, and in some cases indeed it can be safely skipped. We will comment more on this issue soon. 

Let us find how $f_\cB(a)$ transforms under the $SL(2,\Z)$ duality of the 4D $\cN=4$ theory. The easiest way to do this is by fusing the boundary condition $\cB$ with duality walls. The duality wall corresponding to $T$ transformation differs from the transparent wall described before only by an $\cN=4$ Chern-Simons term living on it. Indeed, recall from \cite{Gaiotto:2008ak} that $T$ transformation of the boundary condition can be realized as a unit shift of the Chern-Simons level in the boundary theory (if the boundary condition is described by such a boundary theory). It is known how this affects the 3D localization computation, with the unit Chern-Simons term resulting in the factor of
\begin{equation}
e^{-i\pi {\rm Tr}a^2}
\end{equation}
in the Kapustin-Willett-Yaakov matrix model \cite{Kapustin:2009kz}, where for classical groups, ${\rm Tr}$ denotes the trace in defining representation. We will see that the action of $T$ transformation might involve an additional phase shift $\varphi_0$, to be determined later from consistency. So the integral kernel of $T$ transformation is given by:
\begin{equation}
T(a_1, a_2)=e^{i\varphi_0-i\pi {\rm Tr}a_1^2} Z_0(a_1, a_2)=e^{i\varphi_0-i\pi {\rm Tr}a_1^2}\frac{1}{\mu_b(a_1)}\delta_{\mathfrak t}(a_1,a_2).
\end{equation}
It simply multiplies the wave function by a phase:
\begin{equation}
\textbf{T}:\ f_\cB(a) \mapsto e^{i\varphi_0-i\pi {\rm Tr}a^2}f_\cB(a).
\end{equation}

Description of the $S$-duality wall is also known in the literature: it supports a special theory usually denoted by $T[G]$ that has a $G\times {}^LG$ global symmetry \cite{Gaiotto:2008ak}. The corresponding integral kernel is proportional to its partition function,
\begin{equation}
\label{SwallProp}
S(a_2, a_1)\propto Z_{T[G]}(a_1, a_2),
\end{equation}
with masses $a_1$ in the Cartan of $G$ and F.I. parameters $a_2$ in the Cartan of ${}^LG$. To establish the exact equality, we should be careful about the possibility to have ``trapped'' degrees of freedom inside $S^3\times (0,\ell)$, see Figure \ref{fig:domain}.

\paragraph{Trapped degrees of freedom near the S-wall.}
To proceed, we have to determine the proportionality coefficient in \eqref{SwallProp}. Let us first review how $T[G]$ couples to the two sides of the wall in flat space, following \cite{Gaiotto:2008ak,Gaiotto:2008sa}. Six real scalars of the 4D $\cN=4$ vector multiplet are separated into two groups, $X^{1,2,3}$ and $Y^{1,2,3}$, acted on by $SO(3)_X\times SO(3)_Y\subset SO(6)_R$, the R-symmetry of the theory. On the one side of the wall, we impose Dirichlet boundary conditions on $Y^{1,2,3}$ and Neumann on $X^{1,2,3}$ and gauge fields, which are completed by fermions to give a half-BPS boundary condition. On the other side of the wall, we similarly impose Neumann boundary conditions on the gauge fields and scalars $Y^{1,2,3}$, while $X^{1,2,3}$ are given Dirichlet boundary conditions. In the flat space case, both boundary conditions preserve full 3D $\cN=4$ SUSY at the boundary. The dynamical degrees of freedom not fixed by boundary conditions form two 3D $\cN=4$ vector multiplets: an ordinary vector multiplet on one side of the wall, and a twisted vector multiplet on another side. The ordinary vector multiplet is taken to gauge the $G\subset G\times {}^LG$ symmetry of $T[G]$ on the wall, while the twisted multiplet is gauging the ${}^LG\subset G\times {}^LG$ subgroup. The two vector multiplets, one acted on by $SO(3)_X$ and another -- by $SO(3)_Y$, exchange their roles under the 3D mirror symmetry.

In our case, the boundary is $S^3$ and the $\cN=4$ SUSY preserved there is $\mathfrak{su}(2|1)_\ell\oplus \mathfrak{su}(2|1)_r$.\footnote{For $S^3_b$, this is broken down to $\mathfrak{su}(1|1)_\ell\oplus \mathfrak{su}(1|1)_r$.} As reviewed in Section \ref{sec:su21B}, the 3D mirror symmetry acts as an automorphism $\mathfrak{a}$ on this algebra which flips the sign of R-charge in $\mathfrak{su}(2|1)_r$ but does not affect $\mathfrak{su}(2|1)_\ell$. In the present context, it is useful to take as the boundary localization supercharge $Q_{\rm Loc}$ the supercharge invariant under the automorphism $\mathfrak{a}$. A convenient choice is the 3D $\cN=2$ supercharge of Kapustin-Willett-Yaakov \cite{Kapustin:2009kz}, $\cQ^{\rm KWY}\in\mathfrak{su}(2|1)_\ell$. With this choice, we know that whenever some boundary condition is $Q_{\rm Loc}$-invariant, so is its mirror dual.

\begin{figure}[t!]
	\centering
	\includegraphics[width=1\textwidth]{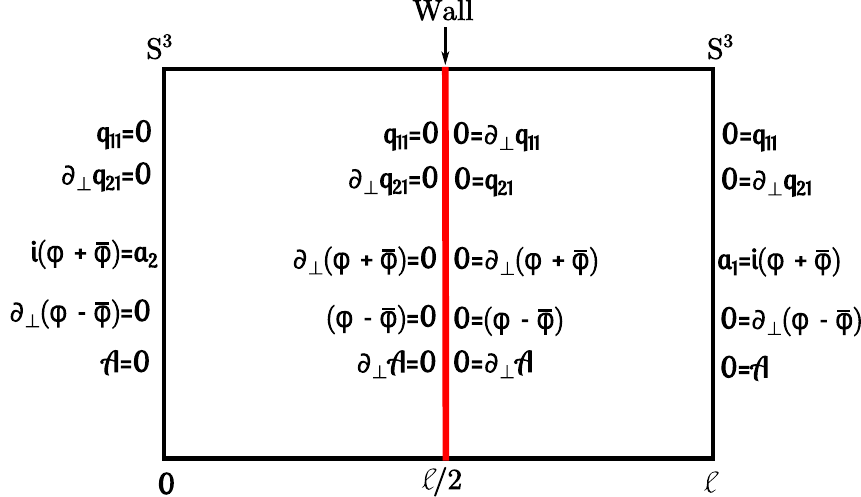}
	\caption{\label{fig:swall} Boundary conditions on the cylinder $S^3\times (0,\ell)$ and on the S-duality wall inserted at $S^3\times \{\ell/2\}$.}
\end{figure}

To describe the kernel $S(a_1,a_2)$, we consider again a thin cylinder $S^3_b\times (0,\ell)$, with boundary conditions of type \eqref{HS4bc} imposed at the two ends, see Figure \ref{fig:swall}. In the center of this cylinder, at $S^3_b\times \{\ell/2\}$, we put a wall with the theory $T[G]$ on it. On the right surface of this wall, we impose a boundary condition that is complementary to \eqref{HS4bc}: each field that has Dirichlet boundary condition in \eqref{HS4bc}, is given Neumann boundary condition on the wall, and vice versa. This is a half-BPS boundary condition, and fields not fixed by it form a 3D $\cN=4$ vector multiplet that gauges $G$ global symmetry of $T[G]$ (or we can describe it as a 3D $\cN=2$ vector multiplet coupled to a 3D $\cN=2$ adjoint chiral). In particular, gauge field has Neumann boundary condition on the wall. On the other surface of the wall, we impose boundary condition that is mirror to this one. It is achieved by flipping Dirichlet and Neumann boundary conditions for the 4D hypers, while boundary conditions for the vector multiplet remain the same. Fields that are not fixed on this surface form a 3D $\cN=4$ twisted vector multiplet that gauges ${}^LG$ symmetry of $T[G]$. We summarize the boundary condition on Figure \ref{fig:swall}. In the limit $\ell\to 0$, the two sides of the wall behave somewhat differently. We see from the Figure \ref{fig:swall} that the boundary conditions freeze all modes to the right from the wall: every field has Dirichlet boundary conditions on one end and Neumann on the other end of that region. Therefore, there are no ``trapped'' degrees of freedom on the right. On the other hand, the left side of the wall has the field $q_{21}$ which is allowed to fluctuate along $S^3_b$, while the vector multiplet is frozen. This gives an adjoint-valued chiral multiplet of unit R-charge trapped on the left of the wall.

The above analysis shows that the proportionality coefficient in \eqref{SwallProp} is given by the partition function of the ``trapped'' adjoint chiral multiplet of R-charge $1$ on $S^3_b$ in the background of a frozen vector multiplet. For ordinary 4D $\cN=4$ theory, this proportionality factor is $1$ -- we already mentioned before that $Z_{\rm ch}^{S^3_b}(a)=1$ for R-charge $1$ chirals in self-conjugate representations, such as the adjoint. In other words:
\begin{equation}
S(a_1, a_2)=Z_{T[G]}(a_1, a_2).
\end{equation}
However, if we consider the $\cN=2^*$ theory, the mass deformation makes such a factor non-trivial, as already explained previously. Recall that $\cN=2^*$ deformation is achieved by turning on twisted mass for the $U(1)$ global symmetry of the pair of chirals forming a hyper. The field $q_{11}$ has charge 1, while $q_{21}$ has charge $-1$ under this symmetry. Therefore, effectively, we are turning on mass $-m$ for the ``trapped'' adjoint chiral on $S^3_b$. This results in a factor of $Z_{\rm ch}^{S^3_b}(a_2,-m)$:
\begin{align}
S(a_1, a_2)&=Z_{T[G, m]}(a_1, a_2)Z_{\rm ch}^{S^3_b}(a_2,-m),\cr
Z_{\rm ch}^{S_b^3}(a,-m)&=\prod_{\alpha\in\Delta_+}s_b\left(-\alpha(a)+m\right)s_b\left(\alpha(a)+m\right),
\end{align}
where $T[G,m]$ is a known $\cN=2$ mass deformation of the 3D $\cN=4$ theory $T[G]$. Note that this $m$ corresponds to a global $U(1)$ symmetry of $T[G]$ which is different from $G\times{}^LG$. Turning on vevs for $G$ and ${}^LG$ background vector multiplets introduces masses and F.I. parameters $a_1$ and $a_2$ respectively that preserve $\cN=4$ SUSY. On the other hand, non-zero $m$ breaks it down to $\cN=2$.

The $S$ transformation then generally acts by:
\begin{equation}
\textbf{S}:\ f_\cB(a) \mapsto \frac1{|\cW|}\int_{\mathfrak{t}} \dd^r a'\, f_\cB(a')\, \mu_b(a',m)\, Z_{T[G, m]}(a', a)Z_{\rm ch}^{S^3_b}(a,-m).
\end{equation}

\subsubsection{Example of $G=SU(2)$.}\label{sec:SU2}

Let us make this explicit for the $\cN=4$ SYM with the gauge group $SU(2)$, first for the round sphere and in the absence of $\cN=2^*$ deformation $m$. The boundary condition is characterized by $f_\cB(a)$, a complex function of a single variable $a\in \R$. Weyl-invariance means that this function is symmetric, $f_\cB(-a)=f_\cB(a)$. The gluing measure is simply: 
\begin{equation}
\mu(a)=4\sinh^2\pi a.
\end{equation}
The $T$ transformation in this case, up to a constant phase, is:
\begin{equation}
\textbf{T}:\ f_\cB(a) \mapsto e^{-\frac{i}2\pi a^2}f_\cB(a).
\end{equation}
The $S$-transform kernel $T[SU(2)]$ is described by a 3D $\cN=4$ gauge theory with gauge group $U(1)$ and two charge-one hypers. Its partition function with the mass $a_1$ and F.I. term $a_2$ is \cite{Benvenuti:2011ga}:
\begin{align}
S(a_2, a_1)&=\frac1{\sqrt 2}\int_\R \dd\sigma\, \frac{e^{-2i\pi a_2 \sigma}}{4\cosh \pi\left(\sigma + \frac{a_1}{2}\right)\cosh \pi\left(\sigma - \frac{a_1}{2}\right)}\cr 
&= \frac1{\sqrt 2}\times\frac{1}{2\sinh\pi a_1}\times \frac{1}{2\sinh\pi a_2}\times 2\sin\pi a_1 a_2,
\end{align}
where $1/\sqrt{2}$ was included for normalization. It is easy to check that $S^2=1$:
\begin{align}
\frac12 \int\dd a\, S(a_1, a)\,\mu(a)\, S(a,a_2)&=\frac1{4\sinh\pi a_1 \sinh\pi a_2}\int\dd a\, \sin(\pi a_1 a)\sin(\pi a a_2)\cr
&=\frac{1}{\mu(a_1)}\delta_{\mathfrak t}(a_1, a_2),
\end{align}
where the latter coincides with the kernel $Z_0(a_1, a_2)$ of the transparent wall.

To make sure that these $S$ and $T$ indeed define the $SL(2,\Z)$ action, one should check that $(ST)^3=1$. Explicit computation shows that $(ST)^3=e^{-\frac{3i\pi}{4}}$. This means that in order to have $(ST)^3=1$, we should actually redefine the $T$ transformation by a constant phase:
\begin{equation}
\label{Ttransform}
\textbf{T}:\ f_\cB(a) \mapsto e^{\frac{i\pi}{4}-\frac{i}2\pi a^2}f_\cB(a).
\end{equation}
With this $T$, we indeed have the full $SL(2,\Z)$ action on the space of symmetric complex-valued functions\footnote{It is not obvious what is the appropriate space of physically allowed functions $f_\cB(a)$. However, it is very natural to conjecture that it can be identified with the $\Z_2$-invariant subspace of complex tempered distributions, $(\cS'(\R)\otimes\C)^{\Z_2}$: 1) we have seen that for Dirichlet boundary conditions, $f_\cB(a)$ is a delta-function; 2) the form of $S$-transformation suggests that $\sinh(\pi a)f_\cB(a)$ should have a Fourier transform.} of a single real variable $a$. 

But what is the origin of this somewhat ad hoc phase $e^{i\pi/4}$? Recall that the $T$-transformation wall is defined by coupling a level-one SUSY Chern-Simons term to the two sides of the wall. In Chern-Simons theory, such phases often appear due to the framing anomaly \cite{Witten:1988hf}, so it is natural to guess that this $e^{i\pi/4}$ is related to framing. On the other hand, we have learned in \cite{glue1} that the gluing procedure is unambiguous once the original theory is well-defined and anomaly-free. Thus, assuming that the action of $SL(2,\Z)$ duality is unambiguous, the $T$ transformation should be uniquely defined as well. This suggests that the framing ambiguity of Chern-Simons on the $T$ wall is resolved, and this phase $e^{i\pi/4}$ must be its fixed left-over value. It would be interesting to understand the precise mechanism behind this.

\paragraph{Two examples of dual boundary conditions.} To illustrate the use of S-kernel, let us look at wave functions of some S-dual boundary conditions. The dual of Dirichlet boundary condition with the wave function $f_\cB(a)=\frac{1}{\mu(a)}\delta_{\mathfrak t}(a)$ is given by:
\begin{equation}
\frac12\int\dd a'\, S(a, a')\mu(a') \frac{1}{\mu(a')}\delta_{\mathfrak t}(a') = S(a,0).
\end{equation}
This can clearly be interpreted as a wave function of the Neumann boundary condition coupled to $T[SU(2)]$, which is known to be S-dual to the Dirichlet boundary condition \cite{Gaiotto:2008ak}. Indeed, the simplest Neumann boundary condition has the wave function $f_{p=0}(a)=1$, as we have learned before, and further coupling it to $T[SU(2)]$ through the mass parameter $a$ simply produces the partition function of $T[SU(2)]$ with mass $a$, that is $S(a,0)$.

Another example is the dual of the basic Neumann boundary condition that is complementary to the basic Dirichlet boundary condition \eqref{HS4bc}. Since it has the wave function $f_{p=0}(a)=1$, the S-dual is given by:
\begin{equation}
(\textbf{S}\cdot 1)(a)= \frac12\int_\R \dd a'\, S(a, a')\mu(a')=\frac1{\sqrt{2}\sinh\pi a} \int_\R \dd a'\, \sinh\pi a' \sin \pi a a'.
\end{equation}
This expression is obviously divergent and ill-defined. Indeed, the S-dual of Neumann boundary conditions is known to be the Nahm pole \cite{Gaiotto:2008ak}, and it is natural to expect that the $S^3\times (0,\ell)$ partition function with the Nahm pole at $S^3\times \{\ell\}$ and the Dirichlet boundary conditions at $S^3\times \{0\}$ should have a subtle behavior in the $\ell\to0$ limit. However, we could proceed formally and evaluate the above integral as:
\begin{equation}
\label{NahmFormal}
(\textbf{S}\cdot 1)(a)=\frac{\sqrt{\sinh^2\pi a}}{i\sqrt{2}\sinh^2\pi a}\delta_{\mathfrak t}(a,i).
\end{equation}
What this means, of course, is that we analytically continue $a$ to be imaginary, and rotate the $\dd a'$ integration cycle by $e^{i\pi/2}$. Furthermore, formally applying S-transform to this $(\textbf{S}\cdot 1)(a)$ gives back the Neumann wave function $1$.

This can be interpreted as the statement that the Nahm pole wave function is related to the Dirichlet one by the analytic continuation $a\to i$ and multiplication by an extra factor $2\sqrt{2}\sin\pi a$. We can check this claim and make the above formal manipulation more precise in the following way. Suppose we study the empty hemisphere wave function $|HS^4\rangle$. With the Dirichlet boundary conditions, we know what it looks like from \cite{Gava:2016oep} and from the remarks made around the equation \eqref{HS4HypDet} earlier in this paper:
\begin{align}
\langle a|HS^4\rangle = Z^{HS^4}_{\rm cl}(a)Z^{HS^4}_{\rm vec}(a)Z^{HS^4}_{\rm hyp}(a)Z_{\rm inst}(a)\equiv Z_D(a),\cr
\end{align}
with $Z_{\rm inst}(a)=1$, since we are looking at the 4D $\cN=4$ theory here \cite{Nekrasov:2002qd,Pestun:2007rz}. All the other components are given in \cite{Gava:2016oep}.\footnote{Since hypermultiplet takes values in the adjoint representation, which is self-conjugate, the answer for $Z^{HS^4}_{\rm hyp}(a)$ in \cite{Gava:2016oep} coincides with \eqref{HS4HypDet} in this case.} They are:
\begin{align}
Z_{\rm cl}^{HS^4}(a)&=e^{-\frac{4\pi^2 {\rm tr}a^2}{g_{YM}^2}},\cr
Z_{\rm hyp}^{HS^4}(a)&=\prod_{n>0}(n^2 + a^2)^{-n}=\frac1{G(1+ia)G(1-ia)},\cr
Z_{\rm vec}^{HS^4}(a)&=\prod_{n>0}(n^2 + a^2)^{n-1}=\frac{a}{\sinh \pi a}G(1+ia)G(1-ia),
\end{align}
so the hemisphere wave function is simply:
\begin{equation}
\label{HS4Neq4}
Z_D(a)=\frac{a}{\sinh(\pi a)} e^{-\frac{2\pi^2 a^2}{g_{YM}^2}}.
\end{equation}
This $Z_D(a)$ decays fast enough at infinity to make the S-transform well-defined. Denoting the Nahm pole boundary condition by $\langle N.p.|$, we would like to compute $\langle N.p.|HS^4\rangle$:
\begin{equation}
\langle N.p.|HS^4\rangle=\langle N.p.|\textbf{SS}|HS^4\rangle=\langle \cN|\textbf{S}|HS^4\rangle,
\end{equation}
where we use the fact that the S-dual of Nahm pole is the Neumann boundary condition, whose state we denoted by $\langle\cN|$. Using the wave functions $\langle a|HS^4\rangle=Z_D(a)$ and $\langle a|\cN\rangle=1$, we then have:
\begin{align}
\langle N.p.|HS^4\rangle&=\frac14 \int_{\R} \dd a_1\,\mu(a_1)\int_\R \dd a_2\, S(a_1, a_2)\mu(a_2)Z_D(a_2)\cr
&=\sqrt{2}\int_{\R} \dd a_1\int_\R \dd a_2\, \sinh(\pi a_1)\sin(\pi a_1 a_2)\sinh(\pi a_2)Z_D(a_2).
\end{align}
With the explicit form of $Z_D(a)$ given in \eqref{HS4Neq4}, we can evaluate this to be:
\begin{equation}
\langle N.p.|HS^4\rangle=2\sqrt{2} e^{\frac{2\pi^2}{g^2_{YM}}},
\end{equation}
which can indeed be obtained from $Z_D(a)$ by the analytic continuation $a\to i$ and further multiplication by $2\sqrt{2}\sin\pi a$. This supports the claim that the Nahm pole boundary condition can be described by the wave function \eqref{NahmFormal}, but it would also be great to check this through the direct localization computation.

\paragraph{Deformations: squashing and $\cN=2^*$.} As it follows from our earlier discussion, the $SL(2,\Z)$ action that we have just studied should admit two deformation parameters: the squashing $b\geq 1$ and the $\cN=2^*$ mass deformation $m$. These deformation do not affect the $T$ transform \eqref{Ttransform}, simply because it does not change the Chern-Simons contribution $e^{-\frac{i}{2}\pi a^2}$, and it is natural to expect that the framing factor $e^{i\pi/4}$ is not affected by continuous deformations of $S^3$. On the other hand, the $S$-transform is non-trivially modified. First, the ``trapped'' adjoint chiral multiplet contributes at non-zero mass, so we have:
\begin{equation}
S(a_1, a_2)=\frac1{\sqrt{2}}\frac{Z_{T[G,m]}(a_1, a_2)}{s_b(-a_2-m)s_b(-m)s_b(a_2-m)}=\frac1{\sqrt{2}}\frac{Z_{T[G,m]}(a_1, a_2)}{Z_{\rm ch}^{S^3_b}(a_2,m)}.
\end{equation}
Second, the $T[G,m]$ partition function has a more complicated form \cite{Hosomichi:2010vh}:
\begin{align}
\label{SkerDeformed}
Z_{T[G,m]}(a_1, a_2)=\frac{1}{s_b(m)}\int_{\R} \dd\sigma\, \frac{s_b\left(\sigma+\frac{a_1}2 +\frac{m}2 +\frac{iQ}4\right)s_b\left(\sigma-\frac{a_1}2 +\frac{m}2 +\frac{iQ}4\right)}{s_b\left(\sigma+\frac{a_1}2-\frac{m}2-\frac{iQ}4\right)s_b\left(\sigma-\frac{a_1}2 -\frac{m}2 -\frac{iQ}4\right)}e^{-2i\pi a_2\sigma},
\end{align}
where $Q=b+b^{-1}$ as usual. The multiplicative factor $1/s_b(m)=s_b(-m)$ comes from the free 3D $\cN=2$ chiral multiplet which at $m=0$ is a part of the 3D $\cN=4$ vector multiplet. To check that $S^2=1$, one would have to use the deformed gluing measure $\mu_b(a,m)=Z_{\rm vec}^{S^3_b}(a)Z_{\rm ch}^{S^3_b}(a,m)$ and compute:
\begin{align}
\frac12 \int\dd a\, S(a_1,a)\mu_b(a,m) S(a,a_2)=\frac1{Z_{\rm ch}^{S^3_b}(a_2,m)}\frac14\int\dd a\, Z_{T[G,m]}(a_1, a) Z_{\rm vec}^{S^3_b}(a) Z_{T[G,m]}(a, a_2).
\end{align}
Notice that the factors of $Z_{\rm ch}^{S_b^3}(a,m)$ canceled inside the integral, and the formula is simply gluing $Z_{T[G,m]}$ with itself using the measure $Z_{\rm vec}^{S^3_b}(a)\dd a=4\sinh(\pi b a)\sinh(\pi b^{-1}a)\dd a$. Such a measure has appeared previously in the AGT literature in various contexts, for example in the discussion of S-duality wall placed at the equator of the sphere (see, e.g., \cite{Hosomichi:2010vh,Terashima:2011qi,Drukker:2010jp}). Moreover, integral expressions like this were matched to the S-duality kernel in Liouville theory \cite{Teschner:2003at,Teschner:2002vx,Teschner:2003em,Teschner:2005bz}. Such matchings implied that $\frac14\int\dd a\, Z_{T[G,m]}(a_1, a) Z_{\rm vec}^{S^3_b}(a) Z_{T[G,m]}(a, a_2)=(Z_{\rm vec}^{S^3_b}(a_1))^{-1}\delta_{\mathfrak t}(a_1,a_2)$ holds, which in our case gives the desired property:
\begin{equation}
\frac12 \int\dd a\, S(a_1,a)\mu_b(a,m) S(a,a_2)=\frac1{\mu_b(a,m)}\delta_{\mathfrak t}(a_1, a_2).
\end{equation}
Equation $(ST)^3=1$ similarly holds by matching with the Liouville theory literature.

\section{4D $\cN=2$ index and half-index}\label{sec:4dindex}
Another straightforward application of the gluing formalism is to a four-dimensional Schur index \cite{Gadde:2011uv}, or rather an $S^1\times S^3$ partition function, which can be glued from the two copies of an $S^1\times D^3$ partition function known as the half-index. Exhaustive explanations provided in the earlier sections of this paper allow us to avoid a lot of technical details here and go directly to the derivation of the gluing formula and its applications. In particular, we are not going to discuss the supersymmetric background for $\cN=2$ theories on $S^3\times S^1$ or $S^3\times \R$, which will be described elsewhere, along with other technical details.

A simple exercise shows that the Dirichlet polarization, defined in the previous section for $\cN=2$ theories quantized on $S^3$, also works for these theories quantized on $S^1\times S^2$, which can be, e.g., a boundary of $S^1\times HS^3$. In particular, we obtain a 3D $\cN=2$ supersymmetric gluing theory at the interface $S^1\times S^2$ whose field content is dictated by the parent 4D theory: 4D $\cN=2$ vector multiplet in the adjoint of $G$ gives rise to a 3D $\cN=2$ vector multiplet in the adjoint of $G$; 4D hypermultiplet in the representation $\cR\oplus \bar\cR$ gives rise to a 3d chiral multiplet of R-charge $\Delta_\Phi=1$ valued in $\cR$. These rules are identical to those found in the previous sections, and more generally, the same pattern holds for all dimensions and geometries. The only difference is that now the 3D multiplets transform according to 3D $\cN=2$ SUSY on $S^2\times S^1$, whereas before it was 3D $\cN=2$ SUSY on $S^3$.

Let %$J_{L,R}$ be the Cartan generators of $SU(2)_L\times SU(2)_R$, the isometry group of $S^3$, 
$R$ be the Cartan generator of the 4d $SU(2)$ R-symmetry, and $F_i$ are Cartan generators of the flavor symmetry. Consider the path integral on $S^3\times S^1$, with $S^3$ of radius $\ell$ and $S^1$ of radius $\beta\ell$, %$\beta=\beta_1 + \beta_2$, $p=e^{-\beta_1}$, 
$x=e^{-\beta}$, and also introduce $t_i=e^{-\gamma_i}$. With the Schur index in mind, we consider the path integral over fields in the twisted sector:
\begin{equation}
\label{twist}
\cF(\tau + \beta\ell) = e^{-\beta R + \gamma_i F_i} \cF(\tau),
\end{equation}
where $\tau$ is the coordinate of $S^1$, and $\cF$ stands for all variables in the theory. Next, we cut $S^3$ into two hemispheres $HS^3_\pm$.

From the point of view of the boundary $S^2\times S^1$, we end up in the setup of \cite{Imamura:2011su}. In particular, the equation \eqref{twist} defining the twisted sector on $S^1\times S^3$ reduces to eqn. (17) of \cite{Imamura:2011su} defining the twisted sector on $S^1\times S^2$ for $\beta_1=\beta_2$.\footnote{In their case, the answer does not depend on $\beta_1$, so we might as well choose $\beta_1=\beta_2$, which leads to $e^{-\beta_1 R}=e^{-\beta \frac{R}{2}}$ in their equation (17). This $R/2$ in 3d is precisely our $R$ in 4d.} In other words, the gluing theory is the 3D $\cN=2$ QFT on $S^1\times S^2$ whose field content and SUSY transformations are precisely those of \cite{Imamura:2011su}

We can now localize the gluing theory by simply applying the results of \cite{Imamura:2011su} for the localization of superconformal indices in 3D. In particular, most of the fields vanish on the localization locus. The non-vanishing fields are parametrized in terms of two constant matrices: $z\in \mathbb{T}$ valued in the maximal torus of the gauge group, and $m\in \mathfrak{t}$ valued in the Cartan subalgebra. This $z$ is the holonomy of the gauge field around $S^1$, which makes it valued in the maximal torus. On the other hand, $m$ is a gauge magnetic flux through $S^2$, so it is quantized as the GNO charge. Another field that has a non-zero vev is $\sigma = \frac{m}{2r}$, a scalar in the 3D $\cN=2$ vector multiplet. Hence integration over the localization locus is represented by a sum over $m\in \Lambda_{\rm cochar}$ (a cocharacter lattice) and an integral over $z\in\mathbb{T}$. Including the 1-loop contributions computed in \cite{Imamura:2011su}, we obtain the gluing formula expressing index in terms of the half-indices:

\begin{align}
I(x, t_i) = \sum_{m\in \Lambda_{\rm cochar}} \frac1{|\cW(G_m)|}\int_{\mathbb{T}} \prod_{j=1}^{{\rm rk}(G)}\frac{\dd z_j}{2\pi i z_j} \, z^{b_0} x^{\epsilon_0} \left(\prod_i t_i^{q_{0i}}\right) &\exp\left[\sum_{n=1}^\infty \frac1{n} f_{\rm tot}(z^{n}, x^n, t_i^n) \right]\cr
&\times\Pi^{+}_m(x, t, z) \Pi^-_m(x, t, z),
\end{align}
where $\Pi^\pm_m(x,t,z)$ are half-indices computed with particular half-BPS boundary conditions determined by $m$ and $z$ (and with fugacities $x$ and $t$ turned on as before). At the boundary $\partial (HS^3\times S^1) = S^2\times S^1$, the gauge field is given Dirichlet boundary condition such that it has holonomy $z$ around $S^1$ and magnetic flux $m$ through $S^2$. Scalar $\sigma$ is given a Dirichlet b.c. with the value $m/2r$. Matter multiplets are given boundary conditions as in the previous section: each hyper decomposes as a pair of chirals, one of which has Dirichlet and one -- Neumann boundary conditions.

To unpack the rest of notations in the above gluing formula, we provide definitions:
\begin{align}
\epsilon_0 &= -\frac{1}{2}\sum_{\alpha\in \Delta(\mathfrak{g})}|\alpha(m)|,\cr
q_{0i}&= -\sum_{w\in \cR}\frac12 |w(m)|F_i,\cr
b_0 &= -\sum_{w\in\cR} \frac12 |w(m)|\rho,\cr
f_{\rm tot}&= f_{\rm vector} + f_{\rm chiral},\cr
f_{\rm vector}(z,x)&= \sum_{\alpha\in \Delta(\mathfrak{g})}\left(-z^{\alpha}x^{|\alpha(m)|}\right),\cr
f_{\rm chiral}(z,x,t_i)&=\sum_{w\in\cR}\frac{x^{|w(m)|+1}}{1-x^2}\left[z^{w}t_i^{F_i} - z^{-w}t_i^{-F_i} \right],
\end{align}
where we have specialized the results of \cite{Imamura:2011su} to the case of $\Delta_\Phi=1$ that is relevant to us, since all the gluing chiral multiplets have R-charge 1. Following \cite{Kapustin:2011jm}, it is also convenient to rewrite the answer in terms of the q-Pochhammers:
\begin{equation}
I(x, t_i) = \sum_{m\in \Lambda_{\rm cochar}} \frac1{|\cW(G_m)|}\int_{\mathbb{T}} \prod_{j=1}^{{\rm rk}(G)}\frac{\dd z_j}{2\pi i z_j}\, Z_{\rm gauge}(z, m; x)Z_\Phi(z, m; t; x)\Pi^{+}_m(x, t, z) \Pi^-_m(x, t, z),
\end{equation}
where
\begin{equation}
Z_{\rm gauge}(z,m;x)=\prod_{\alpha\in \Delta(\mathfrak{g})}x^{-\frac12 |\alpha(m)|} \left(1 - z^{\alpha}x^{|\alpha(m)|} \right)
\end{equation}
is a vector multiplet factor, and the chirals contribute through:
\begin{equation}
\label{chiral_index}
Z_\Phi(z,m;t;x)=\prod_{w\in\cR}\left(\prod_{j=1}^{{\rm rk}(G)} z_j^{-w} \prod_i t_i^{-F_i} \right)^{\frac12 |w(m)|}\frac{(z^{-w}\,t_i^{-F_i}x^{|w(m)|+1};x^2)_\infty}{(z^{w}\,t_i^{F_i}x^{|w(m)|+1};x^2)_\infty}.
\end{equation}
Note again that we specialize to $\Delta_\Phi=1$ here as well, compared to \cite{Kapustin:2011jm}, because this is the R-charge of the chiral multiplet in the gluing theory.

If $\cR$ is a self-conjugate representation, (i.e., it contains weights in pairs, $w$ and $-w$,) and if we ignore flavor fugacity (put $t_i=1$), then $Z_\Phi=1$, and the gluing measure contains only $Z_{\rm gauge}$. This gives the gluing formula as written in \cite{Dimofte:2011py} for 4D $SU(2)$ $\cN=2^*$ theory, and also in \cite{Gang:2012ff} for 4D $\cN=4$ SYM. Indeed, in the 4D $\cN=4$ SYM described in $\cN=2$ language, the hypermultiplet is in the adjoint representation, and we do not turn on any flavor fugacities as it is incompatible with $\cN=4$ SUSY. Thus $Z_\Phi=1$, and the gluing formula reduces to the one found in the above references. 
If we choose to turn on flavor fugacity for the hypermultiplet, this breaks SUSY down to $\cN=2$ (and can be considered as a version of the $\cN=2^*$ theory). In this case, the contribution of $Z_\Phi\ne 1$ given in \eqref{chiral_index} should be included in the gluing formula. This is analogous to our discussion of the $\cN=2^*$ theory in the previous section.

However, the actual $\cN=2^*$ theory obtained by turning on mass for the hypermultiplet has to be examined more thoroughly in this case, which we leave for the future work (or as an exercise for the readers). The reason is that including $\cN=2$ preserving masses in index computations on $S^3\times S^1$ is somewhat subtle and involves turning on unusual supergravity backgrounds \cite{Dumitrescu_in_progress}. In this case the gluing theory is likely to take a different form. In our version of the gluing theory, this problem manifests itself as follows. By analogy with the previous section, one could expect that turning on masses in the parent theory (by giving vevs to the background vector multiplet, when it is possible) corresponds to giving vevs to the background vector multiplet scalar in the gluing theory. It is indeed possible to give vevs to the background vector multiplet fields on $S^1\times S^2$, however they need to satisfy BPS equations. Such BPS equations, among other things, relate the scalar vev to the magnetic flux through $S^2$, $F_{12}=\frac{\sigma}{r}$, see \cite{Imamura:2011su}. Introduction of such vevs for flavor symmetries is certainly possible, and it leads to the notion of generalized index \cite{Kapustin:2011jm}. However, such a discrete background (recall that magnetic flux takes on discrete values) certainly does not describe the mass deformation. It would be very interesting to properly understand $\cN=2$ masses on $S^1\times S^3$, and how the gluing works in their presence.

\subsection{Gluing of more general states}
Note also that, even though we describe our gluing formula as a result about indices and half-indices, it certainly holds for arbitrary states in $\cH_{S^1\times S^2}$ that preserve the localizing supercharge. To write such a general gluing formula, we replace $I(x, t_i)$ by $\langle\Psi_1|\Psi_2 \rangle$, and $\Pi^{+}_m(x, t, z) \Pi^-_m(x, t, z)$ by $\langle \Psi_1|z,m\rangle \langle z,m |\Psi_2\rangle$, where $\langle z,m|$ corresponds to the half-BPS boundary condition parametrized by the boundary gauge holonomy $z$ along $S^1$ and boundary gauge flux $m$ through $S^2$. In this notation, possible fugacities $x$ and $t$ (corresponding to twisted sectors) are suppressed.

\subsection{Boundary conditions and domain walls}
All discussions from the previous section related to the boundary conditions and domain walls supported at $S^3$ can be translated to the current case of gluing along $S^1\times S^2$. One major difference is that while in the $S^3$ case, the boundary conditions were described by wave functions on $\mathfrak{t}\subset \mathfrak{g}$, now they are wave functions on $\mathbb{T}\times\Lambda_{\rm cochar}$, where $\mathbb{T}\subset G$ is the maximal torus of the gauge group. Every such wave function $f_\cB(z,m)$ depends on the boundary holonomy $z\in \mathbb{T}$ and the boundary flux $m\in \Lambda_{\rm cochar}$.

Our goal in the present paper was to introduce the supersymmetric gluing technique and describe several illustrative examples. Careful analysis of boundary conditions, domain walls, trapped degrees of freedom, as well as S-duality of boundary wave functions for the $S^1\times S^2$ case goes beyond the scope of this paper and is left for the future work. This also includes, as we have mentioned, investigating the possibility of mass deformations.

\section{Discussion and future directions}\label{sec:Discuss}
In this paper we have presented a selection of examples where gluing techniques studied in \cite{glue1} combined with the supersymmetric localization imply non-trivial gluing formulas. They are expressed as integrals/sums over spaces of supersymmetric boundary conditions. We have looked at such formulas in spacetime dimensions $D=1,\, 3,\, 4$ only, demonstrating how they can be used to deduce new and old results about theories with high enough SUSY. Many subtleties were pointed out along the way.

The current work does not provide an overwhelming study of all possible aspects of supersymmetric gluing. Quite to the contrary, it opens up a box of many more problems that can and should be addressed along similar lines. Here we give a list of questions that, in our opinion, provide the most immediate and obvious directions of further investigations.

\begin{itemize}
	\item Remaining completely within the scope of methods developed in the current paper, it should be possible to find several more gluing formulas by identifying appropriate supersymmetric polarizations. One such example is 2D $\cN=(2,2)$ theories quantized on $S^1$: two supersymmetric polarizations in such theories are described in the Appendix \ref{sec:2D22}. Another example is the 3D uplift of this, namely the 3D $\cN=2$ theories glued along $S^1\times S^1$. The latter is especially interesting due to the study of half-indices \cite{Gadde:2013wq} (and holomorphic blocks \cite{Beem:2012mb}) and $(0,2)$--preserving boundary conditions  \cite{Gadde:2013wq,Okazaki:2013kaa,Yoshida:2014ssa,Dimofte:2017tpi}. In both examples, localizing the gluing theory (which is the SUSY quantum mechanics on $S^1$ in the former case, and $(0,2)$ theory on $T^2$ in the latter case) is somewhat subtle and hence is left for the future work.
	
	\item There certainly exists a possibility to find more gluing formulas that are not completely addressed by the methods of this paper. As emphasized both here and in \cite{glue1}, one should be able to extend our approach to the case of complex polarizations. There are some indications that there exist new supersymmetric complex polarizations (possibly involving Neumann boundary conditions for the gauge fields) that would result in new types of gluing formulas.
	
	\item Besides considering complex polarizations, as already mentioned in the Discussions section of \cite{glue1}, one should be able to generalize gluing to the case of manifolds with corners. Somewhat relatedly, one could hope to use such techniques to explain triple factorizations known to hold for $S^5$ partition function in 5D $\cN=1$ gauge theories \cite{Kim:2012qf,Lockhart:2012vp,Pasquetti:2016dyl,Chang:2017cdx}.
	
	\item The study of supersymmetric boundary conditions is an obvious area where gluing formulas can be applied. As we pointed out in Section \ref{sec:Mirror3d}, our 3D $\cN=4$ gluing formulas provide interesting tools to study modules over $\cA_C$ and $\cA_H$, the quantized Coulomb and Higgs branch chiral rings of these theories. Moreover, we found that boundary conditions at $S^2$ generate bimodules over such algebras. This provides a slight enrichment of the construction studied in \cite{Bullimore:2016nji}. It would be interesting to further investigate this direction. One curious question is to understand whether Hochschild homology of $\cA_C$ and $\cA_H$ with coefficients in such bimodules carry any interesting data.
	
	\item The $\mathfrak{su}(2|1)_A$--invariant gluing formula for 3D $\cN=4$ theories was used in \cite{Dedushenko:2017avn} as a tool in the computation of correlation functions of the Coulomb branch operators. Combining the $\mathfrak{su}(2|1)_A$ and $\mathfrak{su}(2|1)_B$ invariant gluing formulas should provide a similar tool and extend such computations to a more general class of 3D $\cN=4$ theories, including those not studied in \cite{Dedushenko:2016jxl,Dedushenko:2017avn} (such as theories containing both regular and twisted multiplets).
	
	\item One very interesting and important direction of future studies would be to develop gluing outside the realm of Lagrangian theories. Techniques used here and in the companion paper \cite{glue1} are very much quasi-classical in nature, and rely a lot on Lagrangian descriptions of theories. This certainly does not cover many interesting examples, such as those originating from the 6D $(2,0)$ theory by various compactifications (even though some examples of this kind are Lagrangian). One example is \cite{Gukov:2017kmk} where half-indices of $T[M_3]$ (a theory obtained by reduction of the $(2,0)$ theory on $M_3$) play a central role. Another one is the large class of Argyres-Douglas (AD) theories \cite{Argyres:1995jj,Argyres:1995xn,Wang:2015mra}, which nevertheless might be accessible through the $\cN=1$ Lagrangians that flow to AD fixed points \cite{Maruyoshi:2016tqk,Maruyoshi:2016aim,Agarwal:2016pjo}.
	
\end{itemize}

\acknowledgments

The author thanks Tudor Dimofte, Yale Fan, Bruno Le Floch, Davide Gaiotto, Sergei Gukov, Victor Mikhaylov, Alexei Morozov, Natalie Paquette, Silviu Pufu, Mauricio Romo, David Simmons-Duffin, Gustavo J. Turiaci, Ran Yacoby for comments and discussions, and in particular Tudor Dimofte and Sergei Gukov for comments on the draft. This work was supported by the Walter Burke Institute for Theoretical Physics and the U.S. Department of Energy, Office of Science, Office of High Energy Physics, under Award No.\ DE-SC0011632, as well as the Sherman Fairchild Foundation.

\appendix

\section{Open-ended examples}\label{openended}
Here we would like to present a few more examples where we can explicitly describe the supersymmetric gluing theory. However, due to additional subtleties, we do not attempt localizing it, and leave it for the future work, hence the name ``open-ended''. One such example is on gluing 2D $\cN=(2,2)$ theories quantized on $S^1$, e.g. how to glue a sphere from two hemispheres. Another example is about gluing 3D $\cN=2$ indices from half-indices.

\subsection{2D $\cN=(2,2)$ theories quantized on a circle}\label{sec:2D22}
Let us consider 2D $\cN=(2,2)$ theories quantized on a circle $S^1$. In this case, it would be straightforward to simply consider a flat space theory on $S^1\times \R$ and define a polarization in the phase space on $S^1$. Alternatively, we could consider a theory on the sphere $S^2$ in the vicinity of the equator (or on hemisphere $HS^2$ close to the boundary). The latter approach is slightly more technical than the $S^1\times \R$, however we will follow it to point out a detail. We will comment on the $S^1\times \R$ case afterwards.

Theories with $\cN=(2,2)$ SUSY on $S^2$ are based on the algebra $\mathfrak{su}(2|1)$. We have already encountered them before as the gluing theories in 3D. Here we consider them as standalone theories, while the gluing will be represented by certain quantum mechanics on the boundary circle. We adhere to conventions of \cite{Doroud:2012xw}. The sphere is cut into two hemispheres at $\theta=\pi/2$. The Killing spinor equations on $S^2$, as usual, are:
\begin{equation}
\nabla_i\epsilon=+\frac1{2r} \gamma_i \gamma^{\hat 3}\epsilon,\quad \nabla_i\bar\epsilon=-\frac1{2r}\gamma_i \gamma^{\hat 3}\bar\epsilon.
\end{equation}
The supercharges on $S^2$, following \cite{Doroud:2012xw}, are denoted by $Q_{1,2}$ and $S_{1,2}$. We choose to preserve $Q_2$ and $S_1$ at the boundary. They correspond to Killing spinors:
\begin{align}
\label{HS2kill1}
\epsilon=\exp\left( -\frac{i\theta}{2}\gamma_{\hat 2}\right)\epsilon_o,\cr
\bar\epsilon=\exp\left(+\frac{i\theta}{2}\gamma_{\hat 2}\right)\bar\epsilon_o,
\end{align}
where:
\begin{align}
\label{HS2kill2}
Q_2: \epsilon_o=\left(\begin{matrix} \alpha(\varphi)\cr 0 \end{matrix} \right)=e^{i\varphi/2}\left(\begin{matrix} a\cr 0 \end{matrix} \right),\quad \bar\epsilon_o=0,\cr
S_1: \epsilon_o=0,\quad \bar\epsilon_o=\left(\begin{matrix}0\cr \bar\alpha(\varphi) \end{matrix} \right)=e^{-i\varphi/2}\left(\begin{matrix}0\cr \bar{a} \end{matrix} \right).
\end{align}
Notice that we keep the $\varphi$-dependent factor $e^{\pm i\varphi/2}$ as a part of $\epsilon_o$, $\bar\epsilon_o$: with such conventions, the 1D SUSY parameter $\alpha$ is anti-periodic in $\varphi$. We will see that it is natural in a moment.

We construct a Dirichlet polarization designed for cutting and gluing along the equator $\theta=\pi/2$. For the 2D chiral multiplet $(\phi, \bar\phi, \psi, \bar\psi, F, \bar{F})$, we define the boundary fields as follows:
\begin{align}
u&=\phi\big|,\quad \bar{u}=\bar\phi\big|,\cr
\bar\alpha\chi&=\bar\epsilon\psi\big|,\quad \alpha\bar\chi=\epsilon\bar\psi\big|,\cr
\end{align}
and for the 2D vector multiplet $(A, \sigma_1, \sigma_2, \lambda, \bar\lambda, D)$, the boundary fields are (note that we use the same letter for the 2D and 1D gaugini):
\begin{align}
a_\varphi&=A_\varphi\big|,\quad s=\sigma_2\big|,\cr
D_{\rm 1d}&=-\frac{i}{r}\cD_\theta \sigma_1\big| + D\big|,\cr
\bar\alpha\lambda&=\bar\epsilon\gamma_{\hat 3}\lambda\big|,\quad
\alpha\bar\lambda=\epsilon\gamma_{\hat 3}\bar\lambda\big|.
\end{align}
One can easily check that these fields form a polarization, which is also supersymmetric as manifested by the following SUSY transformations. We find that the boundary values of the chiral multiplet transform under $Q_2, S_1$ as follows:
\begin{align}
\label{bS1chir}
\delta u&=\bar\alpha\chi,\quad \delta\bar{u}=\alpha\bar\chi,\cr
\delta\chi&=\alpha(\frac1r D_\varphi u + s u + i \frac{q}{2r} u),\cr
\delta\bar\chi&=\bar\alpha(\frac1r D_\varphi \bar{u} - s\bar{u} -i\frac{q}{2r}\bar{u}).
\end{align}
The vector multiplet transformations are:
\begin{align}
\label{bS1vec}
\delta s&= -\frac{i}2 (\bar\alpha\lambda +\alpha\bar\lambda),\cr
\delta a_\varphi &= -\frac{r}2 (\bar\alpha\lambda +\alpha\bar\lambda),\cr
\delta \lambda &=\alpha\left( \frac{i}{r} D_\varphi s - D_{\rm 1d} \right),\cr
\delta\bar\lambda &= \bar\alpha\left(\frac{i}{r} D_\varphi s + D_{\rm 1d} \right),\cr
\delta D_{\rm 1d} &= -\bar\alpha \left( \frac{1}{2r} \cD_\varphi \lambda - \frac{i}{4r}\lambda +\frac{1}2 [s,\lambda] \right) + \alpha \left( \frac{1}{2r} \cD_\varphi \bar\lambda + \frac{i}{4r}\bar\lambda + \frac{1}{2} [s, \bar\lambda] \right).
\end{align}
One can immediately recognize these as SUSY transformations of the 1D $\cN=2$ chiral and vector multiplets as described in \cite{Hori:2014tda}, which already appeared in \eqref{chir1D} and \eqref{vec1D}, with a small difference: here we have an R-symmetry holonomy $A_\varphi^{(R)}=1/2$ turned on. This is manifested by the terms $i\frac{q}{2r}u$ and $-i\frac{q}{2r}\bar{u}$ in $\delta\chi$ and $\delta\bar\chi$, as well as $\frac{i}{4r}\bar\lambda$ and $-\frac{i}{4r}\bar\lambda$ in $\delta D_{\rm 1d}$.

To elaborate further on this point, notice that all fermions on $S^2$ take values in the corresponding spinor bundle $\cO(-1)$. When restricted to the boundary $S^1 = \partial HS^2$, they become anti-periodic. So does the SUSY parameter $\epsilon$, and this is the reason we defined $\alpha$, the 1D SUSY parameter, to be anti-periodic as well. This also manifests, in yet another way, the fact that there is an R-charge holonomy $1/2$ preset on $S^1$: since $\alpha$ has R-charge $1$, it becomes anti-periodic in the presence of such a holonomy.

In fact, if we started with the flat space $\cN=(2,2)$ SUSY on $S^1\times \R$ rather than the sphere, we would get the same boundary SUSY on $S^1$, except that these R-holonomy shifts would not be there. In this case, we would have to explicitly go to the twisted sector on $S^1\times \R$, in which the periodicity of R-charge-$q$ fields is given by $e^{i\pi q}$, in order to the get the same boundary SUSY as in \eqref{bS1chir}, \eqref{bS1vec}. This observation means that if we want to glue a half-infinite cylinder $S^1\times \R_+$ to the hemisphere $HS^2$ along their $S^1$ boundary, the $\cN=(2,2)$ theory on $S^1\times \R_+$ has to be in the twisted sector determined by the R-symmetry holonomy\footnote{Since the flat space $\cN=(2,2)$ SUSY has two R-symmetries, the vector and the axial one, we can turn on holonomy for either of them. This corresponds to the $\mathfrak{su}(2|1)_A$ or $\mathfrak{su}(2|1)_B$-preserving background on the sphere. The cylinder with the vector R-symmetry holonomy can be glued to the hemisphere with the $\mathfrak{su}(2|1)_A$-background on it, while the cylinder with the axial holonomy can be glued to the hemisphere with the $\mathfrak{su}(2|1)_B$ background on it.} $A^{(R)}_\varphi=1/2$. Otherwise, the boundary fields simply would not match. Note that this is specific to the $\mathfrak{su}(2|1)$-preserving background on $S^2$ that we have considered. Gluing in the topological background (as in the $tt^*$-geometry \cite{Cecotti:1991me}) might work differently.

To have an even better interpretation of this, let us observe yet another fact. The algebra formed by the supercharges $Q_2$ and $S_1$ that we preserve at the boundary of $HS^2$ is $\mathfrak{su}(1|1)$, with the relation:
\begin{equation}
\{Q_2, S_1\} = P + \frac{R}{2},
\end{equation}
where $R$ is the original $U(1)$ R-charge of the $S^2$ theory. Under this R-symmetry, $s$, $a_\varphi$ and $D_{\rm 1d}$ are neutral, $\lambda$ has R-charge $1$, scalar $\varphi$ has R-charge $q$, and $\chi$ has R-charge $q+1$. However, this algebra does not look like the 1D $\cN=2$ algebra of \cite{Hori:2014tda}. In order to get their algebra, we have to redefine the translation generator as:
\begin{equation}
\tilde{P} = P + \frac{R}{2},
\end{equation}
so that the algebra becomes $\{Q_2, S_1\}=\tilde{P}$, the usual $\cN=2$ SUSY in 1D. This shift of the translation generator by $R/2$ is precisely due to the R-symmetry holonomy. 

To recapitulate, the $\mathfrak{su}(2|1)$-preserving SUSY background on $S^2$ in the vicinity of the equator looks like a twisted sector of the flat space SUSY on $S^1\times \R$, with the R-symmetry holonomy turned on. The vector R-symmetry corresponds to $\mathfrak{su}(2|1)_A$ on the sphere, while the axial R-symmetry corresponds to $\mathfrak{su}(2|1)_B$. The gluing is possible when the two pieces have the same R-symmetry holonomy.

\subsubsection{Quarter-BPS polarization}
We can describe one more potentially useful polarization at the boundary of $HS^2$: a quarter-BPS polarization that only preserves $Q_2 + S_1$. In terms of SUSY parameters, it means that $a=\bar{a}$, so $\alpha=e^{i\varphi/2}a$, $\bar\alpha=e^{-\varphi/2}\bar{a}$. For chiral multiplets, the boundary fields are defined as:
\begin{align}
u&=\phi\big|,\quad \bar{u}=\bar\phi\big|,\cr
a\chi&=\bar\epsilon\psi\big|,\quad a\bar\chi=\epsilon\bar\psi\big|.
\end{align}
For the vector multiplet, we use the same names for the boundary scalars as for their bulk counterparts:
\begin{align}
\sigma_1&=\sigma_1\big|,\quad \sigma_2=\sigma_2\big|,\quad A_\varphi=A_\varphi\big|.\cr
\end{align}
By acting with $\cQ=Q_2+S_1$ on these, we define the corresponding fermions:
\begin{align}
[\cQ,\sigma_1]&=\mu_1,\quad [\cQ,\sigma_2]=\mu_2,\quad [\cQ,A_\varphi]=-ir\mu_2,
\end{align}
whose explicit expressions in terms of $\lambda$'s are:
\begin{align}
\mu_1&=\frac1{2\sqrt{2}}\left[e^{-i\varphi/2}(\lambda_1-\lambda_2) - e^{i\varphi/2}(\bar\lambda_1 - \bar\lambda_2) \right],\cr
\mu_2&=-\frac{i}{2\sqrt{2}}\left[e^{-i\varphi/2}(\lambda_1+\lambda_2) + e^{i\varphi/2}(\bar\lambda_1 + \bar\lambda_2) \right].
\end{align}
A trivial computation shows that their Poisson bracket is zero, i.e., they define a good polarization for the fermions.

Below we summarize SUSY of the boundary fields:
\begin{align}
\delta u&=a\chi,\quad \delta\bar{u}=a\bar\chi,\cr
\delta\chi&=a\left( \frac1{r}\cD_\varphi u +\sigma_2 u + i\frac{q}{2r}u \right),\cr
\delta\bar\chi&=a\left( \frac1{r}\cD_\varphi\bar{u}-\bar{u}\sigma_2 -i\frac{q}{2r}\bar{u} \right),\cr
\delta\sigma_1&=a\mu_1,\quad \delta\sigma_2=a\mu_2,\quad \delta A_\varphi = -i\alpha r\mu_2,\cr
\delta\mu_1&=\alpha\left(\frac1{r}\cD_\varphi\sigma_1 + [\sigma_2,\sigma_1] \right),\cr
\delta\mu_2&=\alpha \cdot \frac1{r}\cD_\varphi\sigma_2,
\end{align}
where $q$ is the R-charge of $\phi$. These fields form a $\frac14$-BPS polarization which can be used to glue $\cQ$-closed states in the usual way, i.e. $\langle\Psi_2|\Psi_1\rangle=\int\pD\bB\, \langle\Psi_2|\bB\rangle \langle\bB|\Psi_1\rangle$, with $\bB=(u, \bar{u}, \chi, \bar\chi, \sigma_1, \sigma_2, A_\varphi, \mu_1, \mu_2)$.

\subsection{3D $\cN=2$ index and half-index}
Discussion of the previous subsection can be uplifted to three dimensions, for $\cN=2$ theories quantized on $S^1\times S^1$. The most important application would be to understanding half-indices \cite{Gadde:2013wq}, that is partition functions on $S^1\times HS^2$, and how they are glued into full indices on $S^1\times S^2$. %Here we present some preliminary facts necessary for the future investigation of this question, namely the basic half-BPS Dirichlet polarization.

In complete analogy with the 2D $\cN=(2,2)$ case, the vector and chiral multiplets of 3D $\cN=2$ theory quantized on $T^2$ give vector and chiral multiplets in the gluing theory. The gluing theory is an $\cN=(0,2)$ gauge theory on $T^2$, and we mostly think of it as the device to glue two half-indices on $S^1\times HS^2$. Fields of this theory are also defined in the twisted sector. The origin of this twisted sector is twofold: on the one hand, like in the $S^1$ case of previous subsection, it comes from the supersymmetric background on $S^2$; on the other hand, when we study indices on $S^1\times S^2$, $\epsilon$ is not periodic in the $S^1$ direction. Following \cite{Imamura:2011su}, and as already discussed in Section \ref{sec:4dindex}, one has to turn on proper holonomies in the $S^1$ direction that would generate appropriate fugacities in the index.

As usual, denoting the boundary multiplets collectively by $\bB$, the gluing is performed by:
\begin{equation}
\int \pD \bB \langle\Psi_1|\bB\rangle \langle\bB|\Psi_2\rangle.
\end{equation}

This is a 2D $\cN=(0,2)$ theory on the torus with appropriate holonomies turned on. The matter content of this 2D theory is uniquely determined by the matter content of the 3D theory: each 3D $\cN=2$ vector multiplet gives a 2D $\cN=(0,2)$ vector multiplet, and each 3D $\cN=2$ chiral of R-charge $\Delta$ gives a 2D $\cN=(0,2)$ chiral of the same R-charge. Notice that unlike in theories with twice as many SUSY (which were discussed earlier in this paper), the R-charges of matter multiplets are not canonically fixed. The 2D theory by itself appears anomalous due to the unbalanced chiral matter, however as a gluing theory, it is non-anomalous thanks to the bulk contribution $\langle\Psi_1|\bB\rangle \langle\bB|\Psi_2\rangle$ providing the necessary anomaly inflow. This was discussed from the general point of view in \cite{glue1}.

One should further localize this theory \cite{Benini:2013nda,Benini:2013xpa,Gadde:2013ftv}. The Coulomb branch localization locus is parametrized by holonomies of the gauge fields around the two 1-cycles of $T^2$. They are combined into a single complex variable $u$ valued in the complexified maximal torus of the gauge group. In addition to that, there are holonomies for the flavor symmetries and the R-symmetry turned on. The wavefunctions $\langle\Psi_1|\bB\rangle$ and $\langle\bB|\Psi_2\rangle$ evaluated at this localization locus are half-indices \cite{Gadde:2013wq} with Dirichlet boundary conditions:
\begin{equation}
\langle\Psi_1|\bB\rangle\big|_{L.L.} = \langle\bB|\Psi_2\rangle\big|_{L.L.} = I_D(x;q;u).
\end{equation}
We leave detailed treatment of this problem for the future work.

\newpage

\bibliographystyle{JHEP}
\bibliography{gluing}

\end{document}